\pdfoutput=1
\documentclass[PAPER, cernpreprint, atlasdraft=false, texlive=2023, UKenglish, texmf, orcidlogo]{atlasdoc}
 
\usepackage[subfigure=true]{atlaspackage}
\usepackage{atlasbiblatex}
 
\usepackage{atlasphysics}

\graphicspath{{logos/}{figures/}}
 
\usepackage{ANA-HDBS-2021-02-PAPER-defs}
 
\usepackage{multirow}
\AtlasTitle{Search for a CP-odd Higgs boson decaying into a heavy CP-even Higgs boson and a $Z$ boson in the $\ell^+\ell^- t\bar{t}$ and $\nu\bar{\nu}b\bar{b}$ final states using 140~fb$^{-1}$ of data collected with the ATLAS detector}

\AtlasAbstract{
A search for a heavy CP-odd Higgs boson, $A$, decaying into a $Z$ boson and a heavy CP-even Higgs boson, $H$, is presented. It uses the full LHC Run~2 dataset of $pp$ collisions at $\sqrt{s}=13$~TeV collected with the ATLAS detector, corresponding to an integrated luminosity of $140$~fb$^{-1}$.
The search for $A\to ZH$ is performed in the $\ell^+\ell^- t\bar{t}$ and $\nu\bar{\nu}b\bar{b}$ final states and surpasses the reach of previous searches in different final states in the region with $m_H>350$~GeV and $m_A>800$~GeV. No significant deviation from the Standard Model expectation is found. Upper limits are placed on the production cross-section times the decay branching ratios.
Limits with less model dependence are also presented as functions of the reconstructed $m(t\bar{t})$ and $m(b\bar{b})$ distributions in the $\ell^+\ell^- t\bar{t}$ and $\nu\bar{\nu}b\bar{b}$ channels, respectively.
In addition, the results are interpreted in the context of two-Higgs-doublet models.
}

\AtlasRefCode{HDBS-2021-02}
 
\PreprintIdNumber{CERN-EP-2023-210}

\AtlasJournal{JHEP}
\AtlasJournalRef{JHEP 02 (2024) 197}
\AtlasDOI{10.1007/JHEP02(2024)197}

\hypersetup{pdftitle={ATLAS document},pdfauthor={The ATLAS Collaboration}}
 
\begin{document}
 
\maketitle
 
\tableofcontents

\section{Introduction}
\label{sec:intro}

The discovery of a Higgs boson at the Large Hadron Collider
(LHC)~\cite{HIGG-2012-27,CMS-HIG-12-028} raised the question of
whether this particle is part of an extended scalar sector, a scenario which arises
in several models that attempt to explain the shortcomings of the Standard Model (SM),
such as the hierarchy problem, the excess of matter over antimatter in the observable universe and
the existence of dark matter.
This question has motivated experimental
searches for extended scalar sectors at the LHC~\cite{Ferrari:2021hqc}. A simple
such extension that has received a lot of attention
is the two-Higgs-doublet model
(2HDM)~\cite{Lee:1973iz,Branco:2011iw}. The 2HDM is popular due to its
very rich phenomenology and the fact that it is motivated
by several new physics scenarios such as
supersymmetry~\cite{Djouadi:2005gj}, dark
matter~\cite{Abdallah:2015ter,Bauer:2017ota} and axion~\cite{Kim:2008hd}
models, electroweak baryogenesis~\cite{Cohen:1993nk}  and
neutrino mass models~\cite{King:2003jb}.
 
A second Higgs doublet leads to five Higgs bosons after electroweak
symmetry breaking. The phenomenology depends on many parameters, including the masses of the Higgs bosons
and the parameters of the Higgs potential. Phenomenological studies often assume
CP-conservation and discrete symmetries that eliminate quartic terms odd
in either of the doublets~\cite{Gunion:2002zf}. This leads to two CP-even Higgs bosons,
$h$ and $H$, with
$m_h < m_H $, one CP-odd boson, $A$, and two charged scalars, $H^{\pm}$.
This model has seven free parameters, usually taken to be the masses
of the Higgs bosons ($m_h$, $m_H$, $m_A$, $m_{H^{\pm}}$), the ratio of the
vacuum expectation values of the two doublets ($\tan\beta$), the
potential parameter ($m_{12}$), and the mixing angle ($\alpha$)
of the CP-even Higgs bosons. It is usually assumed that the $h$ boson is the Higgs boson that was discovered at the
LHC and has $m_h\simeq 125$~\GeV. In this scenario, $h$ has the
same couplings to fermions and vector bosons as the SM
Higgs boson at lowest order in the limit that $\cos(\beta-\alpha)=0$, known as the alignment limit.
 
Precision electroweak measurements~\cite{Haller:2018nnx} suggest that
the masses of two of the heavy Higgs bosons in the 2HDM are degenerate.
This has motivated many LHC searches, such as for the $A\to Zh$ process~\cite{ATLAS:2022enb,CMS-B2G-16-003},
where $m_A=m_H$ is assumed when interpreting the results.
The scenario $m_A\neq m_H$  has strong motivation from
electroweak baryogenesis
models~\cite{Dorsch:2013wja,Dorsch:2014qja,Turok:1990zg,Fromme:2006cm,Basler:2016obg};
in particular, $m_A > m_H$ is favoured~\cite{Dorsch:2014qja} for
a strong first-order phase transition to have occurred in the early universe.
The $A$ boson mass is also constrained to be not far above 1~\TeV~\cite{Dorsch:2013wja,Su:2020pjw},
whereas the $h$ boson is required to have properties similar to those of a SM Higgs boson and
hence it is compatible with the Higgs boson that was observed
at the LHC~\cite{Dorsch:2014qja}.
 
\begin{figure}[ht]
\centering
\subfigure[]{\includegraphics[height=3cm]{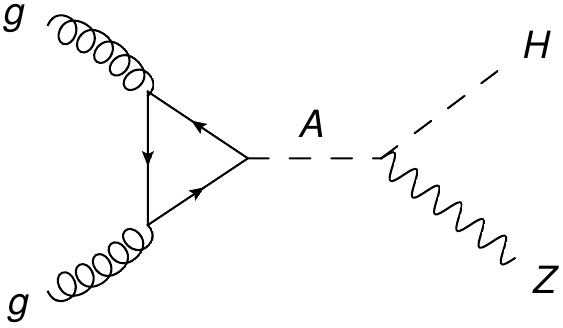}\label{fig:Feynman:ggF}}\hspace{2cm}
\subfigure[]{\includegraphics[height=3cm]{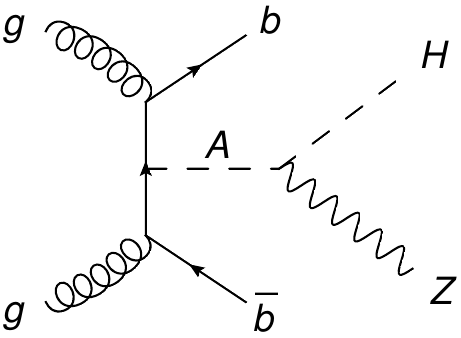}\label{fig:Feynman:bbA}}
\caption{Feynman diagrams for the ggF (a) and bbA (b) production modes. The searches presented in this paper target final states in which the \(H\) boson decays into \ttbar or \bbbar and the \(Z\) boson decays into $\ell^+\ell^-$ or $\nu\bar{\nu}$.}
\label{fig:Feynman}
\end{figure}

In such a scenario, the most promising experimental signature is an $A$ boson
produced either via gluon--gluon fusion (ggF, \Cref{fig:Feynman:ggF}) or in association
with $b$-quarks (bbA, \Cref{fig:Feynman:bbA}), with a subsequent decay into $ZH$, which dominates
when the mass difference $m_A-m_H$ becomes large.
The signature of the $A\to ZH$ process has been sought at the LHC in final states where the
$Z$ boson decays leptonically ($Z\to\ellell$) and the $H$ boson decays into
$b\bbar$, $WW$ or $\tau\tau$~\cite{HDBS-2018-13,CMS-HIG-15-001,CMS-HIG-18-012}.
These final states, although very sensitive, cannot probe the parameter space in which $m_H > 2 m_{\mathrm{top}}$, where the $H\to \ttbar$ decay becomes dominant.
This parameter space was probed recently with $A\to ZH$ and $H\to hh$~\cite{HDBS-2019-31}, but this decay chain is not sensitive in the alignment limit.
 
The first search presented in this paper aims to cover this unexplored phase space by searching for
$A\to ZH$ in the \lltt final state.
Top-quark pairs ($t\bar{t}$) in which one top-quark decays semileptonically and the other decays hadronically are considered.
This leads to a signature with three leptons (electrons or muons) and at least four jets, two of which
are expected to have originated from $b$-quarks.
 
In addition, a search for $A\to ZH$ with the $Z$ boson decaying into neutrinos and the $H$ boson decaying into a pair of $b$-quarks
is performed. This constitutes the first search at the LHC for $b\bar{b}$ resonances with a mass up to 2~TeV produced in association with  missing transverse momentum,
and complements the sensitivity of the $\ellell b\bbar$ search~\cite{HDBS-2018-13} at high $m_A$, due to the higher branching ratio of the $Z$ boson decay into neutrinos~\cite{ATLAS:2022enb}.
Moreover, this search could also be reinterpreted
in the context of hierarchical 2HDMs with extra mediators coupling to dark matter~\cite{Bauer:2017ota}, thus also complementing the
existing searches~\cite{ATLAS:2021shl} in the high \mHhypo regime.


\section{ATLAS detector}
\label{sec:detector}

\newcommand{\AtlasCoordFootnote}{
ATLAS uses a right-handed coordinate system with its origin at the nominal interaction point (IP)
in the centre of the detector and the \(z\)-axis along the beam pipe.
The \(x\)-axis points from the IP to the centre of the LHC ring,
and the \(y\)-axis points upwards.
Cylindrical coordinates \((r,\phi)\) are used in the transverse plane,
\(\phi\) being the azimuthal angle around the \(z\)-axis.
The pseudorapidity is defined in terms of the polar angle \(\theta\) as \(\eta = -\ln \tan(\theta/2)\).
Angular distance is measured in units of \(\Delta R \equiv \sqrt{(\Delta\eta)^{2} + (\Delta\phi)^{2}}\).}
 
The ATLAS detector~\cite{PERF-2007-01} at the LHC covers nearly the entire solid angle around the collision point.\footnote{\AtlasCoordFootnote}
It consists of an inner tracking detector surrounded by a thin superconducting solenoid, electromagnetic and hadron calorimeters,
and a muon spectrometer incorporating three large superconducting air-core toroidal magnets.
 
The inner-detector system (ID) is immersed in a \qty{2}{\tesla} axial magnetic field
and provides charged-particle tracking in the range \(|\eta| < 2.5\).
The high-granularity silicon pixel detector covers the vertex region and typically provides four measurements per track,
the first hit normally being in the insertable B-layer installed before Run~2~\cite{ATLAS-TDR-19,PIX-2018-001}.
It is followed by the silicon microstrip tracker, which usually provides eight measurements per track.
These silicon detectors are complemented by the transition radiation tracker (TRT),
which enables radially extended track reconstruction up to \(|\eta| = 2.0\).
The TRT also provides electron identification information
based on the fraction of hits (typically 30 in total) above a higher energy-deposit threshold corresponding to transition radiation.
 
The calorimeter system covers the pseudorapidity range \(|\eta| < 4.9\).
Within the region \(|\eta|< 3.2\), electromagnetic calorimetry is provided by barrel and
endcap high-granularity lead/liquid-argon (LAr) calorimeters,
with an additional thin LAr presampler covering \(|\eta| < 1.8\)
to correct for energy loss in material upstream of the calorimeters.
Hadron calorimetry is provided by the steel/scintillator-tile calorimeter,
segmented into three barrel structures within \(|\eta| < 1.7\), and two copper/LAr hadron endcap calorimeters.
The solid angle coverage is completed with forward copper/LAr and tungsten/LAr calorimeter modules
optimised for electromagnetic and hadronic energy measurements respectively.
 
The muon spectrometer (MS) comprises separate trigger and
high-precision tracking chambers measuring the deflection of muons in a magnetic field generated by the superconducting air-core toroidal magnets.
The field integral of the toroids ranges between \num{2.0} and \qty{6.0}{\tesla\metre}
across most of the detector.
Three layers of precision chambers, each consisting of layers of monitored drift tubes, cover the region \(|\eta| < 2.7\),
complemented by cathode-strip chambers in the forward region, where the background is highest.
The muon trigger system covers the range \(|\eta| < 2.4\) with resistive-plate chambers in the barrel, and thin-gap chambers in the endcap regions.
 
Interesting events are selected by the first-level trigger system implemented in custom hardware,
followed by selections made by algorithms implemented in software in the high-level trigger~\cite{TRIG-2016-01}.
The first-level trigger accepts events from the \qty{40}{\MHz} bunch crossings at a rate below \qty{100}{\kHz},
which the high-level trigger reduces in order to record events to disk at about \qty{1}{\kHz}.
 
An extensive software suite~\cite{ATL-SOFT-PUB-2021-001} is used in data simulation, in the reconstruction
and analysis of real and simulated data, in detector operations, and in the trigger and data acquisition
systems of the experiment.


\section{Data and simulated event samples}
\label{sec:dataMC}

The search presented here uses 140~fb$^{-1}$ of proton--proton collision data at a centre-of-mass energy (\(\sqrt{s}\)) of 13~\TeV\ recorded by the ATLAS detector during the years 2015--2018 (Run~2). All events were required to pass basic data-quality requirements which ensure that all components of the ATLAS detector were functioning correctly~\cite{DAPR-2018-01}.
 
For the analysis targeting the \lltt final state, events were recorded using a logical OR of single-electron triggers with transverse momentum (\pt) thresholds varying from 24 to 26~\GeV\ or single-muon triggers with \pt thresholds varying from 20 to 26~\GeV\ and a combination of quality and isolation requirements.
Offline, the leptons were required to be geometrically matched to the corresponding trigger object and to have a \pt threshold 1--2~\GeV\ above the high-level trigger threshold to operate in the region where the trigger reaches its maximum efficiency.
For the analysis targeting the \vvbb final state, events were recorded by the missing transverse momentum (\met) triggers with thresholds varying between 70 and 110~\GeV, which become fully efficient for an offline \met value of approximately 200~\GeV.
The trigger efficiencies in simulation are corrected to match those observed in data.
For the \met triggers, this is done following the procedure in Ref.~\cite{ATLAS:2020fcp}.
Events with one or two leptons that are used to define background-enriched control regions in the \vvbb channel are selected with the same single-lepton triggers as in the \lltt channel.
 
Simulated signal events were generated with \textsc{MadGraph5\_aMC@NLO} (\MGaMC)~2.3.3~\cite{Alwall:2014hca}, requiring an $s$-channel $A$ boson that decays into a $Z$ boson and $H$ boson, and using the UFO model provided in Ref.~\cite{Bauer:2017ota} to calculate the loop-induced ggF process with a finite $m_\mathrm{top}$ value.\footnote{This UFO model contains a priori an extra CP-odd Higgs boson $a$ that mixes with the CP-odd Higgs boson $A$ of the 2HDM with a mixing angle $\sin\theta$. This is completely decoupled in the signal generation by setting $\sin\theta=0$, thereby effectively yielding a 2HDM with five physical Higgs bosons.}
Both the ggF and bbA production modes (shown in \Cref{fig:Feynman}) were generated at leading order (LO) in QCD for various combinations of $(m_A,m_H)$ using the \NNPDF[3.0nlo] set of parton distribution functions (PDF)~\cite{Ball:2014uwa}, with the former including contributions from top-quark loop-induced processes but neglecting contributions from bottom-loop induced processes, which have a negligible impact on the kinematic distributions. The ggF samples were generated at $\tan\beta=1$ and the bbA production samples were generated at $\tan\beta=5$. Simulated events with different values of $\tan\beta$ were obtained via matrix-element reweighting~\cite{Artoisenet:2010cn}. The decay widths of the $A$ and $H$ bosons were calculated by \MGaMC at LO and finite-width effects were included in the simulation. The decays of the $Z$ and $H$ bosons were simulated using \MADSPIN~\cite{Frixione:2007zp,Artoisenet:2012st}. \PYTHIA[8.244]~\cite{Sjostrand:2014zea} was used to model the parton shower and hadronisation, using the A14~\cite{ATL-PHYS-PUB-2014-021} set of tuned parameters (tune). Non-resonant diagrams, in which the $ZH$ final state is produced through a top-quark box were found to have a negligible impact and thus were not included in the simulation. The interference between the resonant diagram shown in \Cref{fig:Feynman} and the non-resonant box diagrams, and also the SM $t\bar{t}Z$ process, was studied using the UFO model provided in Ref.~\cite{Haisch:2018djm} and found to be negligible.
The generation of events in the \vvbb final state required \(\met > 100\)~\GeV,  which increased the efficiency of the simulation by improving the acceptance.
Simulated events in the \lltt final state were filtered to select events with at least one top-quark decaying semileptonically, with no kinematic requirements on the generator-level leptons.
 
Simulated events for the background processes were generated as shown in \Cref{tab:MC}. The decays of bottom and charm hadrons were simulated using \EVTGEN[1.6.0]~\cite{Lange:2001uf}, except for the samples generated using \SHERPA~\cite{Schumann:2007mg}, in which case these decays were generated within \SHERPA.
 
The effect of multiple interactions in the same and neighbouring bunch crossings (pile-up) in all simulated samples was modelled by overlaying each hard-scattering event with simulated inelastic proton--proton events generated with \PYTHIA[8.186]~\cite{Sjostrand:2007gs} using the \NNPDF[2.3lo] PDF set and the A3 tune~\cite{ATL-PHYS-PUB-2016-017}. The simulated events were weighted to reproduce the distribution of the average number of interactions per bunch crossing observed in the data. All generated background samples were passed through the \GEANT-based~\cite{Agostinelli:2002hh} detector simulation~\cite{SOFT-2010-01} of the ATLAS detector. The ATLFAST-II simulation~\cite{SOFT-2010-01} was used for the signal samples. The simulated events were reconstructed in the same way as the data (see \cref{sec:objects}).
 
\begin{table}
\caption{List of generators used for the simulation of different processes. (N)(N)LO or (N)NLL refers to (next-to-)(next-to-)leading order or (next-to-)next-to-leading-logarithm accuracy in the perturbative expansion of the cross-sections, while EW refers to the inclusion of electroweak corrections at the order indicated. For all samples using the A14 or AZNLO tune, the \NNPDF[2.3lo]~\cite{Ball:2012cx} or CTEQ6L1~\cite{Pumplin:2002vw} PDF set was used, respectively. The diagram removal (DR) and diagram subtraction (DS) schemes~\cite{Frixione:2008yi} were used to remove the overlap between the $t\bar{t}Z/tWZ$ and \ttbar/$tW$ processes, respectively. $V$ denotes a $W$ or $Z$ boson.}\label{tab:MC}
\resizebox{\textwidth}{!}{
\begin{tabular}{|c|c|c|c|c|c|c|c|}
\hline
\textbf{Process} & \textbf{Matrix element} & \textbf{QCD order} & \textbf{PDF} & \textbf{Parton shower} & \textbf{Tune} & \textbf{Cross-section} & \textbf{Special settings}  \\\hline
\multirow{2}{*}{\ttbar} & \multirow{2}{*}{\PWHG[v2]~\cite{Frixione:2007nw,Nason:2004rx,Frixione:2007vw,Alioli:2010xd}} & \multirow{2}{*}{NLO} & \multirow{2}{*}{\NNPDF[3.0nlo]~\cite{Ball:2014uwa}} & \multirow{2}{*}{\PYTHIA[8.230]~\cite{Sjostrand:2014zea}} & \multirow{2}{*}{A14~\cite{ATL-PHYS-PUB-2014-021}} & NNLO+NNLL
& \multirow{2}{*}{$\hdamp=1.5\mtop$}  \\
& & & & & &  QCD~\cite{Beneke:2011mq,Cacciari:2011hy,Baernreuther:2012ws,Czakon:2012zr,Czakon:2012pz,Czakon:2013goa,Czakon:2011xx} & \\\hline
\multirow{2}{*}{$tW$} & \multirow{2}{*}{\PWHG[v2]~\cite{Re:2010bp,Nason:2004rx,Frixione:2007vw,Alioli:2010xd}} & \multirow{2}{*}{NLO} & \multirow{2}{*}{\NNPDF[3.0nlo]~\cite{Ball:2014uwa}} & \multirow{2}{*}{\PYTHIA[8.230]~\cite{Sjostrand:2014zea}} & \multirow{2}{*}{A14~\cite{ATL-PHYS-PUB-2014-021}} & NLO+NNLL & \multirow{2}{*}{DS scheme~\cite{Frixione:2008yi}} \\
& & & & & &  QCD~\cite{Kidonakis:2010ux,Kidonakis:2013zqa} & \\\hline
Single top & \multirow{2}{*}{\PWHG[v2]~\cite{Frederix:2012dh,Nason:2004rx,Frixione:2007vw,Alioli:2010xd}} &  \multirow{2}{*}{NLO} &  \multirow{2}{*}{\NNPDF[3.0nlo]~\cite{Ball:2014uwa}} &  \multirow{2}{*}{\PYTHIA[8.230]~\cite{Sjostrand:2014zea}} &  \multirow{2}{*}{A14~\cite{ATL-PHYS-PUB-2014-021}} & \multirow{2}{*}{NLO QCD~\cite{Aliev:2010zk,Kant:2014oha}} &  \multirow{2}{*}{4-flavour scheme}  \\
$t$-channel & & & & & & &  \\\hline
Single top & \multirow{2}{*}{\PWHG[v2]~\cite{Frederix:2012dh,Nason:2004rx,Frixione:2007vw,Alioli:2010xd}} &  \multirow{2}{*}{NLO} &  \multirow{2}{*}{\NNPDF[3.0nlo]~\cite{Ball:2014uwa}} &  \multirow{2}{*}{\PYTHIA[8.230]~\cite{Sjostrand:2014zea}} &  \multirow{2}{*}{A14~\cite{ATL-PHYS-PUB-2014-021}} & \multirow{2}{*}{NLO QCD~\cite{Aliev:2010zk,Kant:2014oha}} &  \multirow{2}{*}{5-flavour scheme} \\
$s$-channel & & & & & & & \\\hline
\multirow{2}{*}{$V$+\;\!jets} & \multirow{2}{*}{\SHERPA[2.2.1]~\cite{Bothmann:2019yzt}} & 0,1 jets @ NLO & \multirow{2}{*}{\NNPDF[3.0nnlo]~\cite{Ball:2014uwa}} & \multirow{2}{*}{\SHERPA~\cite{Schumann:2007mg}} & \multirow{2}{*}{default} & \multirow{2}{*}{NNLO~\cite{Anastasiou:2003ds}} & \multirow{2}{*}{\MEPSatNLO~\cite{Hoeche:2011fd,Hoeche:2012yf,Catani:2001cc,Hoeche:2009rj}} \\
&  & 2,3,4 jets @ LO & & & & &  \\\hline
Diboson & \multirow{2}{*}{\SHERPA[2.2.1]~\cite{Bothmann:2019yzt}} & 0,1 jets @ NLO & \multirow{2}{*}{\NNPDF[3.0nnlo]~\cite{Ball:2014uwa}} & \multirow{2}{*}{\SHERPA~\cite{Schumann:2007mg}} & \multirow{2}{*}{default} & \multirow{2}{*}{NLO QCD} & \multirow{2}{*}{\MEPSatNLO~\cite{Hoeche:2011fd,Hoeche:2012yf,Catani:2001cc,Hoeche:2009rj}} \\
$q\bar{q}\to VV$ &  & 2,3 jets @ LO & & & & &  \\\hline
Diboson & \multirow{2}{*}{\SHERPA[2.2.2]~\cite{Bothmann:2019yzt}} & \multirow{2}{*}{0,1 jets @ LO} & \multirow{2}{*}{\NNPDF[3.0nnlo]~\cite{Ball:2014uwa}} & \multirow{2}{*}{\SHERPA~\cite{Schumann:2007mg}} & \multirow{2}{*}{default} & \multirow{2}{*}{NLO QCD} & \multirow{2}{*}{\MEPSatNLO~\cite{Hoeche:2011fd,Hoeche:2012yf,Catani:2001cc,Hoeche:2009rj}} \\
$gg\to VV$ &  &  & & & & &  \\\hline
\multirow{2}{*}{$q\bar{q}\to Vh$} & \multirow{2}{*}{\PWHG[v2]~\cite{Alioli:2010xd}} & \multirow{2}{*}{NLO} & \multirow{2}{*}{\NNPDF[3.0nlo]~\cite{Ball:2014uwa}} & \multirow{2}{*}{\PYTHIA[8.212]~\cite{Sjostrand:2014zea}} & \multirow{2}{*}{AZNLO~\cite{Aad:2014xaa}} & NNLO QCD + & \multirow{2}{*}{\MINLO~\cite{Hamilton:2012np,Luisoni:2013kna}}\\
& & & & & & NLO EW~\cite{Ciccolini:2003jy,Brein:2003wg,Ferrera:2011bk,Brein:2011vx,Ferrera:2013yga,Ferrera:2014lca,Campbell:2016jau}  & \\\hline
\multirow{2}{*}{$gg\to Zh$} & \multirow{2}{*}{\PWHG[v2]~\cite{Alioli:2010xd}} & \multirow{2}{*}{NLO} & \multirow{2}{*}{\NNPDF[3.0nlo]~\cite{Ball:2014uwa}} & \multirow{2}{*}{\PYTHIA[8.212]~\cite{Sjostrand:2014zea}} & \multirow{2}{*}{AZNLO~\cite{Aad:2014xaa}} & NLO+NLL  & \multirow{2}{*}{-} \\
& & & & & & QCD~\cite{Altenkamp:2012sx,Harlander:2014wda,Brein:2012ne} & \\\hline
\multirow{2}{*}{\ttV} & \multirow{2}{*}{\MGaMC[2.3.3]~\cite{Alwall:2014hca}} & \multirow{2}{*}{NLO} & \multirow{2}{*}{\NNPDF[3.0nlo]~\cite{Ball:2014uwa}}  & \multirow{2}{*}{\PYTHIA[8.210]~\cite{Sjostrand:2014zea}} & \multirow{2}{*}{A14~\cite{ATL-PHYS-PUB-2014-021}} & NLO QCD + & \ttll scaled by $\mathcal{O}(\alphas^3\alpha)$ \\
& & & & & & NLO EW~\cite{deFlorian:2016spz} & off-shell correction \\\hline
\multirow{2}{*}{\ttH} & \multirow{2}{*}{\PWHG[v2]~\cite{Frixione:2007nw,Nason:2004rx,Frixione:2007vw,Alioli:2010xd,Hartanto:2015uka}} & \multirow{2}{*}{NLO} & \multirow{2}{*}{\NNPDF[3.0nlo]~\cite{Ball:2014uwa}}  & \multirow{2}{*}{\PYTHIA[8.230]~\cite{Sjostrand:2014zea}} & \multirow{2}{*}{A14~\cite{ATL-PHYS-PUB-2014-021}} & NLO QCD + & \multirow{2}{*}{-}
\\
& & & & & & NLO EW~\cite{deFlorian:2016spz} & \\\hline
\multirow{4}{*}{\tZq} & \multirow{4}{*}{\MGaMC[2.3.3]~\cite{Alwall:2014hca}} & \multirow{4}{*}{NLO} & \multirow{4}{*}{\NNPDF[3.0nlo]~\cite{Ball:2014uwa}} & \multirow{4}{*}{\PYTHIA[8.230]~\cite{Sjostrand:2014zea}} & \multirow{4}{*}{A14~\cite{ATL-PHYS-PUB-2014-021}} & \multirow{4}{*}{NLO QCD} & 4-flavour scheme,  \\
& & & & & & & off-shell effects,\\
& & & & & & & top decays in\\
& & & & & & & \MADSPIN~\cite{Frixione:2007zp,Artoisenet:2012st} \\\hline
\tttt & \multirow{2}{*}{\MGaMC[2.3.2]~\cite{Alwall:2014hca}} &\multirow{2}{*}{LO} & \multirow{2}{*}{\NNPDF[3.1lo]~\cite{Ball:2014uwa}} & \multirow{2}{*}{\PYTHIA[8.186]~\cite{Sjostrand:2007gs}} & \multirow{2}{*}{A14~\cite{ATL-PHYS-PUB-2014-021}} & \multirow{2}{*}{LO QCD} & top decays in \\
\ttbar$WW$ & & & & & & & \MADSPIN~\cite{Frixione:2007zp,Artoisenet:2012st} \\\hline
\multirow{3}{*}{$tWZ$} & \multirow{3}{*}{\MGaMC[2.6.2]~\cite{Alwall:2014hca}} & \multirow{3}{*}{LO} & \multirow{3}{*}{\NNPDF[3.1lo]~\cite{Ball:2014uwa}} & \multirow{3}{*}{\PYTHIA[8.235]~\cite{Sjostrand:2014zea}} & \multirow{3}{*}{A14~\cite{ATL-PHYS-PUB-2014-021}} & \multirow{3}{*}{LO QCD} & top decays in  \\
& & & & & & & \MADSPIN~\cite{Frixione:2007zp,Artoisenet:2012st}, \\
& & & & & & & DR scheme~\cite{Frixione:2008yi} \\\hline
\multirow{2}{*}{$VVV$} & \multirow{2}{*}{\SHERPA[2.2.2]~\cite{Bothmann:2019yzt}} & 0 jets @ NLO & \multirow{2}{*}{\NNPDF[3.0nnlo]~\cite{Ball:2014uwa}} & \multirow{2}{*}{\SHERPA~\cite{Schumann:2007mg}} & \multirow{2}{*}{default} & \multirow{2}{*}{NLO QCD} & \multirow{2}{*}{-} \\
& & 1.2 jets @ LO & & & & & \\
\hline
\end{tabular}
}
\end{table}


\section{Object reconstruction}
\label{sec:objects}

Tracks measured in the ID are used to reconstruct interaction vertices~\cite{ATL-PHYS-PUB-2015-026}. The one with the highest sum of squared transverse momenta of associated tracks is selected as the primary vertex and its position is taken as the proton--proton collision point in the reconstruction of physics objects.
 
Electrons are reconstructed from a track matched to a cluster built from energy deposits in the calorimeter~\cite{EGAM-2018-01}. They are identified using a multivariate likelihood technique~\cite{PERF-2017-01}, using the `loose' working point (WP) for the \vvbb channel and the `medium' WP for the \lltt channel, and they are required to fulfil loose calorimeter isolation criteria. Electrons must have $\pt > 7$~\GeV\ and $|\eta| < 2.47$. To ensure that they are compatible with the primary vertex, the track associated with the electron candidate is required to have $\sigma(d_0)<5$ and $|z_0\sin\theta|<0.5$~mm, where $\sigma(d_0)$ is the significance of the transverse impact parameter, $z_0$ is the longitudinal impact parameter and $\theta$ is the polar angle of the track.
 
Muons are reconstructed by matching track segments in the MS to a track in the ID~\cite{MUON-2018-03}. They are identified by using selections on the quality of the tracks and the compatibility of the ID and MS measurements, and they are required to satisfy the `loose' identification WP and loose isolation criteria combining calorimeter and track information~\cite{MUON-2018-03}. Muons are required to have $\pt>7$~\GeV, $|\eta|<2.5$, $\sigma(d_0)<3$ and $|z_0\sin\theta|<0.5$~mm.
 
Hadronically decaying $\tau$-leptons are reconstructed from calorimeter-cell energy clusters~\cite{ATL-PHYS-PUB-2015-045} formed by the \antikt algorithm~\cite{Cacciari:2008gp,Fastjet} with a radius parameter of $R=0.4$. Either one or three tracks must lie within a cone of size $\Delta R=0.2$ around the direction of the hadronically decaying $\tau$ candidate, which is identified using a recurrent neural network~\cite{ATL-PHYS-PUB-2019-033} and the `loose' WP. The $\tau$-lepton candidates must have $\pt>15$~\GeV\ and $|\eta|<2.5$, excluding the calorimeter barrel/endcap transition region ($1.37\leq |\eta|\leq 1.52$).
 
Jets are reconstructed from particle-flow objects formed from ID tracks and calorimeter energy clusters~\cite{PERF-2015-09} by using the \antikt algorithm with a radius parameter of $R=0.4$. Jets with $|\eta|<2.5$ ($2.5 < |\eta| < 4.5$) are classified as central (forward) jets and are required to have $\pt>20~(30)$~\GeV. Central jets with $20~\GeV <\pt<60$~\GeV\ and $|\eta|<2.4$ are required to pass the `tight' jet vertex tagger (JVT)~\cite{PERF-2014-03} WP that suppresses jets originating from pile-up interactions.
 
Jets containing $b$-hadrons, hereafter referred to as $b$-jets, are identified using the DL1r tagger~\cite{FTAG-2018-01}. A WP corresponding to a 77\% efficiency in simulated inclusive \ttbar events is used for the \lltt channel, while a WP corresponding to a 70\% efficiency is used for the \vvbb channel. The decays of the $b$-hadrons can produce muons which are vetoed when building particle-flow objects and therefore are not included in the energy of the reconstructed jets. To correct for this effect, the four-momentum of the closest non-isolated muon that satisfies $\Delta R(b$-jet$,\mu)<\min(0.4,0.04+10/\pt(\mu)[\GeV])$ is added to the four-momentum of the $b$-jet. The corrected four-momentum is used when defining the event selection criteria described in the following section.
 
The event's missing transverse momentum, \metvec (or \met for its modulus), is defined as the negative vector sum of the transverse momenta of all the observable electron, muon and jet objects described above, plus a soft term comprising ID tracks that are matched to the primary vertex but not to any of the already included objects~\cite{ATLAS-CONF-2018-023}.  An \met significance variable (\metsig) which is sensitive to fake-\met effects is defined using the expected resolutions of all objects used in the \met reconstruction and the correlations between them~\cite{ATLAS-CONF-2018-038}.
 
An overlap removal procedure is applied to avoid any double-counting between the reconstructed leptons, including the hadronically decaying $\tau$-leptons, and jets~\cite{ATLAS:2020fcp}.


\section{Event selection and background estimation}
\label{sec:selection}

The final state resulting from the $A\to Z(\to\ellell)H(\to\ttbar)$ signal process, where one top-quark decays semileptonically and the other decays hadronically, is expected to contain three high-\pt leptons, two of which should have an invariant mass close to the $Z$~boson mass, $m_Z$, and a resonant \ttbar pair. The main backgrounds expected in the \lltt channel consist of $\ttbar Z$ events, which have a non-resonant $m(\ttbar)$ spectrum, and events with a jet misidentified as a lepton which mostly arise from the \ttbar process with both top quarks decaying semileptonically.
 
The $A\to Z(\to\nu\bar{\nu})H(\to b\bar{b})$ signal process leads to a final state with large \met, no leptons, at least two $b$-jets and a resonant $m(b\bar{b})$ spectrum. The main backgrounds in the \vvbb channel arise from $Z$\;\!+\;\!heavy-flavour (denoted by Zhf)\footnote{Jets in simulated events are labelled as $b/c$-jets if a $b/c$-hadron with $\pt>5$~\GeV\ is found within a cone of size $\Delta R=0.3$ around the jet axis, or as light jets ($l$-jets) otherwise. In the \vvbb channel the $W/Z$+jets events are divided according to the true flavour of the jets that constitute the Higgs boson candidate, into heavy flavour, consisting of $bb$, $bc$, $bl$ and $cc$, and light flavour, consisting of $cl$ and $ll$. These components are denoted in the following by $V$hf and $V$lf, respectively.} and \ttbar processes.
 
Differences between the signal and background processes in lepton multiplicity, flavour, charge and kinematics are exploited to define background-enriched control regions that can be used to constrain the main backgrounds, as described in the following sections.
 
A common preselection is applied to the \lltt and \vvbb channels to reject events without a reconstructed primary vertex or events containing jets with properties consistent with beam-induced background processes, cosmic-ray showers or noisy calorimeter cells~\cite{ATLAS-CONF-2015-029}. The subsequent channel-specific selections are described in the following sections.
 
\subsection{\lltt selection}
 
In the \lltt channel, the dominant background consists of $\ttbar Z$ events, which, unlike signal events, produce a non-resonant $m(\ttbar)$ spectrum. Another major background consists of \ttbar events with two prompt leptons from the top-quark decays and an extra lepton which is expected to originate from $b$-hadron decays in 60\% of cases or a jet misidentified as a lepton in the remaining 40\% of cases.
Other backgrounds arise from multi-boson events and events with a single top quark produced in association with vector bosons; these backgrounds generally have lower lepton \pT, a non-resonant $m(\ttbar)$ spectrum, and can be accompanied by $Z$ bosons. Events are therefore separated into signal (SR), control (CR) and validation (VR) regions using a combination of requirements on the following three kinematic quantities: the \pt of the third-highest-\pt lepton (\ptlep), the mass of the $Z$ boson candidate (\mZcand) and the invariant mass of the $H$ boson candidate (\mHcand), as defined below.
 
Events in all regions are required to have exactly three leptons (electrons or muons), leading to the four flavour combinations $eee$, $ee\mu$, $\mu\mu e$, and $\mu\mu\mu$. The leptons must have $\pt>7$~\GeV, with the highest-\pt lepton having $\pt>27$~\GeV. Furthermore, only events with at least four jets and exactly two $b$-tagged jets are retained.\footnote{For the bbA production mode, it was found that the majority of events with three or more $b$-jets ($\gtrsim 60\%$) are reconstructed in the \twotag region, since the additional $b$-jets are soft and forward.}
The events which do not contain any pairs of leptons with opposite-sign charges and the same flavour (OSSF), namely the $e^\pm e^\pm\mu^\mp$ and $\mu^\pm\mu^\pm e^\mp$ combinations, are selected for the same-sign (\fake) region, which serves as the \ttbar CR.
Events with at least one OSSF lepton pair are considered further in the selection of the SR and other CRs and VRs.
 
The following regions are defined by requirements on \ptlep: the region with $\ptlep > 13$~\GeV\ (denoted by \Lhi) is enriched in signal events, whereas the region with $7~\GeV<\ptlep<13~\GeV$ (denoted by \Llo) is enriched in background events.
 
The $Z$ candidate is defined as the OSSF lepton pair whose invariant mass is closest to $m_Z$~\cite{Zyla:2020zbs}, and only events with $|\mZcand-\mZ|< 20$~\GeV\ are retained. Events that satisfy $|\mZcand-\mZ|< 10$~\GeV\ define the \Zin region, where more signal events are expected, and the remaining events define the \Zout region.
In the \fake region, the $Z$ candidate is reconstructed from the pair of leptons with same-sign charges and the same flavour, and only events with $|\mZcand-\mZ|< 20$~\GeV\ are selected.
 
A combination of the \ptlep and \mZcand requirements allows the definition of several regions, enriched in either signal or background events. The signal events generally populate the \Lhi\!\!\_\Zin region.
Two signal-depleted regions are also defined: \Llo\!\!\_\Zin, with roughly equal contributions from Zhf, \ttbar and \ttZ background processes, and \Lhi\!\!\_\Zout, with a relatively large contribution from \ttbar and \ttZ background processes.
These signal-depleted regions cannot be used as CRs to simultaneously constrain the normalisation of the Zhf, \ttbar and \ttZ backgrounds because they receive fairly similar background contributions and have a limited number of events.
They are therefore only used as VRs to verify that the fit model (described in \cref{sec:fit}) can describe the data in regions that are kinematically close to the SR.
 
The semileptonically decaying top-quark candidate (\tlep) is reconstructed from the lepton that is not used in the reconstruction of the $Z$ candidate, the $b$-jet which is the closest in $\Delta R$ to this lepton, and \metvec. To improve the resolution in \mtlep, the longitudinal-momentum component of the neutrino from the \tlep decay is calculated by constraining the mass of the lepton--neutrino system to be equal to the $W$ boson mass, $m_W$.\footnote{In the resulting quadratic equation, the neutrino $p_z$ is taken from the real component in the case of complex solutions or the smaller component of the two solutions if both solutions are real.} The hadronically decaying top-quark candidate (\thad) is reconstructed from the light-jet pair with mass $m_{\mathrm{jj}}$ closest to $m_W$ and the $b$-jet that is not used in the \tlep reconstruction. To improve the resolution in \mthad, the four-momenta of the light-jet pair that constitutes the hadronic $W$ candidate are rescaled by $m_W/m_{\mathrm{jj}}$.
 
The $H$ candidate is defined as the sum of the four-momenta of \tlep and \thad, while the $A$ candidate is reconstructed as the sum of the four-momenta of the $H$ candidate and the $Z$ candidate.
The fact that the $H$ and $Z$ candidates for signal events are produced by the decay of a resonance constrains the kinematic properties of these candidates; the constraints depend on the $m_A$ and $m_H$ values of the signal hypothesis. In particular, the $H$ candidate is expected to be produced more centrally than background events. It is found that requiring the pseudorapidity of the $H$ candidate in the rest frame of the $A$ candidate to satisfy $\left|\etaRZH\right| < 2.2 + 0.0004 \cdot m(\ttbar)~[\GeV] - 0.0011 \cdot m(\lltt)~[\GeV]$ provides the optimal sensitivity across the whole \((m_A,m_H)\) plane. The parameters of the linear function defining this requirement are determined by a fit to the values of the $\left|\etaRZH\right|$ cut that maximise the expected significance for each \((m_A,m_H)\) hypothesis.
 
The presence of a signal would manifest itself as a resonance in the $m(\ttbar)$ and $m(\lltt)$ distributions, as well as in the distribution of the mass difference $\Delta m=m(\lltt)-m(\ttbar)$ \cite{Haisch:2018djm}. The region that is expected to contain most of the signal events for a given mass hypothesis \mHhypo is constructed using a sliding window defined by the condition |$m(\ttbar) - \mHhypo| < N\cdot\sigma$, where $\sigma\approx 0.16\cdot\mHhypo$ is the resolution in $m(\ttbar)$ and $N=2~(1.5)$ for $\mHhypo<(\geq)500$~\GeV, and this region is referred to as the \Hin{} SR. The sideband regions with a lower or higher $m(\ttbar)$ value define the \Hlo{} and \Hhi{} CRs, which are used to constrain the normalisation of the simulated $\ttbar Z$ sample. The $N$ factor, which defines the width of the signal region, is optimised to achieve the highest signal significance.
 
The four-momentum vector of the \(H\) candidate is rescaled by $\mHhypo/m(\ttbar)$ to improve the resolution in $m(\lltt)$. The rescaling is performed only in the SR, where the resonance is expected, and is applied to both the simulated events and data events. After this rescaling, the resolution in $m(\lltt)$ improves by as much as a factor of three, particularly for signal hypotheses with small $\mAhypo-\mHhypo$ values, and ranges from 3\% to 20\% for small and large $\mAhypo-\mHhypo$ values, respectively.
 
The fraction of signal events passing the full event selection varies from 2\% to 3.5\%, depending on the mass hypothesis, and the fraction increases slightly with increasing \(m_H\).
 
A summary of the selection criteria that define the different regions considered in the statistical analysis is given in \Cref{tab:llttSelection}.
 
\begin{table}[ht]
\caption{Event selection for the \lltt channel. The SR, CR and VR symbols next to the region name indicate that this region is used as a signal, control or validation region in the fit, respectively.}\label{tab:llttSelection}
\centering\scriptsize
\resizebox{\textwidth}{!}{
\begin{tabular}{|c||c|c|c|c|c|c|}
\hline
\multicolumn{2}{|c|}{\multirow{3}{*}{\textbf{Requirement}}} & \multicolumn{5}{c|}{\textbf{Regions}} \\\cline{3-7}
\multicolumn{2}{|c|}{} & \multirow{2}{*}{\textbf{\fake (CR)}} & \multirow{2}{*}{\textbf{\Lhi\!\!\_\Zout (VR)}} & \multicolumn{2}{c|}{\textbf{\Lhi\!\!\_\Zin}} & \multirow{2}{*}{\textbf{\Llo\!\!\_\Zin (VR)}} \\\cline{5-6}
\multicolumn{2}{|c|}{} & & & \textbf{\Hlo{} / \Hhi{} (CR)} & \textbf{\Hin{} (SR)} & \\\hline
\multicolumn{2}{|c|}{Number of leptons} & \multicolumn{5}{c|}{3} \\
\multicolumn{2}{|c|}{$\pt(\ell_1)$}  & \multicolumn{5}{c|}{$>  27$~\GeV} \\
\multicolumn{2}{|c|}{Number of jets} & \multicolumn{5}{c|}{$\geq 4$} \\
\multicolumn{2}{|c|}{Number of $b$-jets} & \multicolumn{5}{c|}{2} \\[3pt]
\multicolumn{2}{|c|}{$\left|\etaRZH\right|$} & \multicolumn{5}{c|}{$< 2.2 + 0.0004 \cdot m(\ttbar) [\GeV] - 0.0011 \cdot m(\lltt) [\GeV]$} \\[2pt]\hline
\multicolumn{2}{|c|}{$\pt(\ell_3)$}  & \multicolumn{4}{c|}{$> 13$~\GeV} & \multicolumn{1}{c|}{$>  7$~\GeV\ \& $< 13$~\GeV} \\\hline
\multicolumn{2}{|c|}{Lepton flavour} & $ee\mu/\mu\mu e$ & \multicolumn{4}{c|}{$eee/ee\mu/\mu\mu e/\mu\mu\mu$} \\\hline
\multicolumn{2}{|c|}{OSSF lepton pairs} & 0 & \multicolumn{4}{c|}{$\geq 1$} \\[1pt]\hline
\multicolumn{2}{|c|}{$|\mZcand - m_Z|$} & $< 20$~\GeV\ & $>  10$~\GeV\ \& $<  20$~\GeV & \multicolumn{3}{c|}{$<  10$~\GeV} \\[1pt]\hline
\multicolumn{1}{|c}{\multirow{2}{*}{$|m(\ttbar) - \mHhypo|$}} & \mHhypo < 500~\GeV & \multicolumn{2}{c|}{\multirow{2}{*}{-}} & $>0.32\cdot\mHhypo$ & $<0.32\cdot\mHhypo$ & \multirow{2}{*}{-} \\[4pt]
\multicolumn{1}{|c}{} & $\mHhypo \geq 500~\GeV$ & \multicolumn{2}{c|}{} & $>0.24\cdot\mHhypo$ & $<0.24\cdot\mHhypo$ &  \\
\hline
\end{tabular}
}
\end{table}
 
\subsection{\vvbb selection}
 
In the \vvbb channel, the events are split into regions with different lepton multiplicities. The signal is expected to manifest in the region with no leptons. A region consisting of exactly two leptons of the same flavour (\twoL),\footnote{In principle, the \twoL region could also contain signal from the $A\to ZH\to \ell^+\ell^- b\bar{b}$ process. Based on the constraints on the cross-section for this process derived in Ref.~\cite{HDBS-2018-13} and given that the \twoL region is included in the statistical analysis as a single bin (see Section~\ref{sec:fit}) it has been estimated that the impact of such a signal contamination in the \twoL control region would be smaller than $3\%$, with a negligible impact on the analysis, and is therefore neglected.} enriched in Zhf events, and a region with one electron and one muon (\emu), enriched in \ttbar events, are used to constrain the corresponding background normalisations. Finally, a region with exactly one lepton (\oneL) is used as a VR. The SR, CR and VR regions are further divided into regions with exactly two and at least three $b$-jets, which target the ggF and bbA production modes, respectively.
 
In events with at least one lepton, the highest-\pt lepton is required to have $\pt(\ell_1)>27$~\GeV. Events in all regions are required to have $\ptv > 150~\GeV$ ($V$ denotes a $Z$ or $W$ boson), where $\ptv=\met$ in the region with no leptons, $\ptv=|\vec{\pt}(\ell)+\metvec|$ in the \oneL region and $\ptv=|\vec{\pt}(\ell_1)+\vec{\pt}(\ell_2)|$  in the \twoL and \emu regions. Events are also required to have at least two \(b\)-jets. A veto on events with more than five jets or events containing any hadronically decaying $\tau$-lepton (\(\tau^\mathrm{had}\)) candidates is applied to suppress the \ttbar background.
 
To suppress the background from multi-jet events, only events in which the smallest
azimuthal angle between \metvec and any jet, \mindPhiMetJet, is larger than $\pi/10$ are selected.
This background is further suppressed by selecting events with $\metsig > 3$ ($\metsig > 10$) in the region with one (zero) lepton(s). These selection criteria are found to reduce the multi-jet contamination to a negligible level. In contrast, a selection $\metsig < 5$ is applied in the \twoL region to reduce the contamination from the \ttbar background and maximise the purity in the Zhf background in this CR.
The purity of the Zhf background in the \twoL region is further increased by retaining only events that satisfy $|\mZcand - m_Z|<10$~\GeV.
 
The $H$ candidate is reconstructed from the two highest-\pt $b$-jets and only events with $m(\bbbar) > 50$~\GeV\ are retained. The $\Delta R$ between the $b$-jets forming the $H$ candidate is required to be smaller than 3.3 (3.5) for events with exactly two (at least three) $b$-jets.
 
To further suppress the \ttbar background in events with no leptons, two top-quark-mass proxy variables are defined as follows \cite{ATLAS:2021shl}:
\begin{equation*}
\label{eqn:mtop}
m_{\mathrm{top}}^{\mathrm{near/far}} = \sqrt{2p_{\mathrm{T},b_\mathrm{near/far}}\met\left[1-\cos\Delta\phi\left(\vec{p}_{\mathrm{T},b_{\mathrm{near/far}}},\metvec\right)\right]},
\end{equation*}
where near (far) refers to the $H$ candidate's $b$-jet that is nearer to (farther from) the \metvec in azimuthal angle.
The $b$-jet closer to \metvec in azimuthal angle is used for the calculation of
$m_{\mathrm{top}}^{\mathrm{near}}$, whereas the $b$-jet farther from \metvec is used for $m_{\mathrm{top}}^{\mathrm{far}}$.
Events are retained only if $\mtopnear>180$~\GeV\ and $\mtopfar > 200$~\GeV.
 
The presence of a signal would manifest as a broad resonance in the distribution of the $A$ candidate transverse mass (\mtvh) in the final state with no leptons. The region which is expected to contain most of the signal events for a given mass hypothesis \mHhypo is constructed using a sliding window defined by the condition |$m(\bbbar) - \mHhypo| < 2\cdot\sigma$, where $\sigma= 0.1\cdot\mHhypo$ is the  resolution in $m(\bbbar)$. The adjacent regions with a lower or higher $m(\bbbar)$ define the \Hlo{} and \Hhi{} regions, which are used as CRs in the statistical analysis. The \Hlo{} and \Hhi{} regions contain events from a mix of background processes, so they constrain all the background processes present there, rather than a specific one.
 
The four-momentum vector of the \(H\) candidate is rescaled by $\mHhypo/m(\bbbar)$ to improve the resolution in \mtvh. The rescaling is performed only in the SR, where the resonance is expected, and is applied to both the simulated events and data events. The  resolution in \mtvh after this rescaling ranges from 8\% for signal hypotheses with high $m_H$ and low $m_A-m_H$ values, to 27\% for signal hypotheses with low $m_H$ and high $m_A-m_H$ values.
 
The fraction of signal events passing the full event selection varies from less than 1\% for signal events with low $m_A-m_H$, which also have low \met, to about $21\%$ for signal events with high $m_A-m_H$.
 
A summary of the selection requirements applied in the SR, VRs and CRs is shown in \Cref{tab:vvbbSelection}.
 
\begin{table}[ht]
\centering
\caption{Event selection for the \vvbb channel. The SR, CR and VR symbols next to the region name indicate that this region is used as a signal, control or validation region in the fit, respectively.}\label{tab:vvbbSelection}
\resizebox{0.7\textwidth}{!}{
\begin{tabular}{|c|c|c|c|c|c|}
\hline
\multirow{3}{*}{\textbf{Requirement}} & \multicolumn{5}{c|}{\textbf{Regions}} \\\cline{2-6}
& \multirow{2}{*}{\textbf{\twoL (CR)}} & \multirow{2}{*}{\textbf{\emu (CR)}} & \multirow{2}{*}{\textbf{\oneL (VR)}} & \multicolumn{2}{c|}{\textbf{\zeroL}} \\\cline{5-6}
&  &  & & \textbf{\Hlo{} / \Hhi{} (CR)} & \textbf{\Hin{} (SR)} \\\hline
Number of jets & \multicolumn{5}{c|}{2--5}\\
Number of $b$-jets & \multicolumn{5}{c|}{$\geq  2$}\\
$m(\bbbar)$ & \multicolumn{5}{c|}{$>  50$~\GeV}\\
Number of \(\tau^\mathrm{had}\) & \multicolumn{5}{c|}{0}\\
$\pt(V)$ & \multicolumn{5}{c|}{$>  150$~\GeV}\\[2pt]
\mindPhiMetJet & \multicolumn{5}{c|}{$>\pi/10$}\\
\multirow{2}{*}{\dRbb} & \multicolumn{5}{c|}{$<3.3$ (2 $b$-jets)}  \\[1pt]
\multirow{2}{*}{} & \multicolumn{5}{c|}{$<3.5$ ($\geq  3$ $b$-jets)} \\[1pt]\hline
Number of leptons & \multicolumn{2}{c|}{2} & 1 & \multicolumn{2}{c|}{0}\\\hline
Lepton flavour & $ee/\mu\mu$ & $e\mu$ & $e/\mu$ & \multicolumn{2}{c|}{-} \\\hline
$\pt(\ell_1)$ & \multicolumn{3}{c|}{$> 27$~\GeV} & \multicolumn{2}{c|}{-} \\[2pt]\hline
$|\mZcand - m_Z|$ &  $< 10$~\GeV & \multicolumn{4}{c|}{-} \\\hline
\metsig & $< 5$ & - & $> 3$ & \multicolumn{2}{c|}{$> 10$} \\[2pt]\hline
\mtopnear  & \multicolumn{3}{c|}{-} & \multicolumn{2}{c|}{$> 180$~\GeV} \\[2pt]\hline
\mtopfar & \multicolumn{3}{c|}{-} & \multicolumn{2}{c|}{$> 200$~\GeV} \\[2pt]\hline
$|m(\bbbar) - \mHhypo|$ & \multicolumn{3}{c|}{-}  & $>0.2\cdot\mHhypo$ & $<0.2\cdot\mHhypo$  \\
\hline
\end{tabular}
}
\end{table}
 
\FloatBarrier


\section{Systematic uncertainties}
\label{sec:systematics}

Experimental uncertainties from detector effects and theoretical uncertainties related to the modelling of the simulated signal and background processes affect the normalisation and shape of the final discriminating variable (see \cref{sec:fit}) as well as the relative normalisation of events in the SR and CRs.
 
Uncertainties arising from the calibration of the jet energy scale and resolution are derived as functions of the jet $\pt$ and $\eta$~\cite{ATLAS:2020cli}. Uncertainties arising from the identification of $b$-jets are derived in Refs.~\cite{FTAG-2018-01,FTAG-2020-08,ATLAS-CONF-2018-006}. Uncertainties related to the reconstruction and identification of leptons are derived in Refs.~\cite{EGAM-2018-01,MUON-2018-03}.
Uncertainties in the reconstruction of \met are taken into account as described in Ref.~\cite{ATLAS:2018txj}.
The number of pile-up collisions in simulation is reweighted to match the data and a 4\% variation of this reweighting factor is assigned as an uncertainty. The uncertainty in the total integrated luminosity is 0.83\%~\cite{DAPR-2021-01}, obtained using the LUCID-2 detector~\cite{LUCID2} for the primary luminosity measurements.
 
For the $V$+\;\!jets and diboson processes, uncertainties are estimated by comparing the nominal \SHERPA samples with alternative samples simulated with \MGaMC[2.2.2] interfaced to \PYTHIA[8.186] and including up to four jets in the matrix element calculated at LO accuracy in QCD. Uncertainties related to multi-jet merging and resummation are calculated by varying the associated scales in \SHERPA.
 
For the \ttZ process, the uncertainty related to the modelling of the parton shower is obtained by comparing the nominal sample with a sample simulated with \MGaMC[2.3.3] interfaced to \HERWIG[7.13]~\cite{Bellm:2017jjp}. An additional uncertainty related to the modelling of initial- and final-state radiation is obtained using the associated eigentune of the A14 tune~\cite{ATL-PHYS-PUB-2014-021}.
 
For the \ttbar process, an uncertainty related to the NLO matching is obtained by comparing the nominal sample with a sample simulated with \MGaMC[2.6.0] interfaced to \PYTHIA[8.230], while the parton shower uncertainty is obtained by comparison with a sample simulated with \POWHER[7.04]~\cite{Bahr:2008pv,Bellm:2015jjp}.
The initial- and final-state radiation uncertainties are obtained by varying the renormalisation scale in the initial- and final-state emissions by a factor of two.
The downward variation of the renormalisation scale in initial-state emissions is accompanied by an upward variation of the $h_{\text{damp}}$ parameter by a factor of two.
 
For the $tW$ process, an additional uncertainty related to the removal of the overlap with the \ttbar process is obtained by comparing the nominal sample evaluated using the diagram subtraction technique with a sample which uses the diagram removal technique.
 
For all simulated samples, the uncertainties due to missing higher orders in QCD are estimated by independent variations of the renormalisation and factorisation scales by a factor of two, while the PDF and \alphas uncertainties are calculated using the PDF4LHC prescription~\cite{Butterworth:2015oua}.
 
Systematic uncertainties for background processes that contribute less than 1\% to the total background are neglected.
 
Statistical uncertainties are dominant in the \lltt channel for all $(m_A,m_H)$ hypotheses, while in the \vvbb analysis the systematic uncertainties dominate. The dominant systematic uncertainties in the \vvbb channel are related to the modelling of the Zhf and Whf background processes as well as to the jet energy scale and resolution.
 
The relative contributions from the different sources of uncertainty for representative $(m_A,m_H)$ hypotheses in the two channels are shown in \Cref{tab:uncbreak}.
The fractional impact is calculated by considering the square of the uncertainty in the signal strength parameter (\(\mu\)) arising from a group of uncertainties (as listed in the left column of the table), divided by the square of the total uncertainty in \(\mu\). The value of  \(\mu\) used is determined by the fit to data.
Due to correlations, the sum of the impacts of different systematic uncertainties may not be equal to the total impact of all systematic uncertainties.
Statistical uncertainties include the uncertainties due to the size of the data sample and the uncertainties associated with the background process normalisations which are free to float in the fit.
The row ``MC sample size" refers to the systematic uncertainty due to the finite number of generated events in the simulation.

\begin{table}[ht]
\caption{Fractional squared uncertainty in \(\mu\) from the different sources of uncertainty for different $(m_A,m_H)$ values in the \lltt and \vvbb channels.}
\centering
\resizebox{0.9\textwidth}{!}{
\begin{tabular}{|l|cc|ccc|}
\hline
\multirow{3}{*}{Source of uncertainty} & \multicolumn{5}{c|}{Fractional squared uncertainty in \(\mu\)} \\\cline{2-6}
& \multicolumn{2}{c|}{\lltt signals $(m_A,m_H)$ [GeV]}  & \multicolumn{3}{c|}{\vvbb signals $(m_A,m_H)$ [GeV]} \\
& (700, 500) & (1200, 800) & (400, 130) & (700, 300) & (1200, 800) \\ \hline
Total statistical uncertainty  &   ~~~0.91  &   ~~~0.90   &   ~~~0.19  &   ~~~0.27  &   ~~~0.48   \\
Total systematic uncertainty   &   ~~~0.09  &   ~~~0.10   &   ~~~0.81  &   ~~~0.73  &   ~~~0.52   \\ \hline
\multicolumn{6}{|c|}{Statistical uncertainties}   \\ \hline
Data statistics                &   ~~~0.40  &   ~~~0.72   &   ~~~0.16  &   ~~~0.24  &   ~~~0.48   \\
\ttZ normalisation             &   ~~~0.36  &   ~~~0.14   &  \multicolumn{3}{c|}{neglected} \\
Zhf normalisation              & \multicolumn{2}{c|}{not free to float, included in `Other'}   &   ~~~0.01  &   ~~~0.05  &   ~~~0.12  \\
\ttbar normalisation           &   ~~~0.01  &   ~~~0.02   &   ~~~0.06  &   ~~~0.01  &   ~~~0.01   \\ \hline
\multicolumn{6}{|c|}{Systematic uncertainties}    \\ \hline
Jets                           &   ~~~0.02  &   ~~~0.01   &   ~~~0.15  &   ~~~0.10  &   ~~~0.10   \\
$b$-tagging                    & \(<\) 0.01\;\!  & \(<\) 0.01\;\!   &   ~~~0.02  & \(<\) 0.01\;\!  &   ~~~0.05   \\
\met soft-term and pile-up     & \(<\) 0.01\;\!  & \(<\) 0.01\;\!   &   ~~~0.01  &   ~~~0.01  & \(<\) 0.01\;\!   \\
Luminosity                     & \(<\) 0.01\;\!  & \(<\) 0.01\;\!   & \(<\) 0.01\;\!  & \(<\) 0.01\;\!  & \(<\) 0.01\;\!   \\
Other experimental sources     & \(<\) 0.01\;\!  & \(<\) 0.01\;\!   & \(<\) 0.01\;\!  & \(<\) 0.01\;\!  & \(<\) 0.01\;\!   \\ \hline
\ttZ modelling                 &   ~~~0.03  &   ~~~0.05   & \multicolumn{3}{c|}{not applied} \\
\ttbar modelling               &   ~~~0.01  &   ~~~0.02   &   ~~~0.05  &   ~~~0.01  &   ~~~0.01   \\
Zhf modelling                  &  \multicolumn{2}{c|}{included in `Other'}  &   ~~~0.21  &   ~~~0.47  &   ~~~0.30   \\
Whf modelling                  &  \multicolumn{2}{c|}{neglected}  &   ~~~0.14  &   ~~~0.04  &   ~~~0.10   \\
$tW$ modelling                 &  \multicolumn{2}{c|}{neglected}  &   ~~~0.02  &   ~~~0.03  & \(<\) 0.01\;\!   \\
Other modelling sources        &   ~~~0.01  &   ~~~0.02   &   ~~~0.08  &   ~~~0.01  & \(<\) 0.01\;\!   \\
Signal modelling               & \(<\) 0.01\;\!  & \(<\) 0.01\;\!   & \(<\) 0.01\;\!  & \(<\) 0.01\;\!  & \(<\) 0.01\;\!   \\
MC sample size                 &   ~~~0.01  &   ~~~0.01   &   ~~~0.05  &   ~~~0.05  &   ~~~0.05   \\
\hline
\end{tabular}
}
\label{tab:uncbreak}
\end{table}



\section{Statistical analysis}
\label{sec:fit}

To extract the final result of the search presented here, a binned profile likelihood fit~\cite{Cowan:2010js} to the data is used.
The likelihood function $\mathcal{L}(\mu,\theta)$ is constructed
from a product of Poisson probabilities Pois$(N_{\text{obs}}|\mu s+b)$
for observing $N_{\text{obs}}$ data events when $\mu s$ signal and $b$ background events are expected in each bin of the fitted distribution.
It contains the parameter-of-interest $\mu$, which is a factor multiplying the nominal signal cross-section and is extracted by maximising the likelihood.
Background processes whose event yields are constrained using data have freely floating normalisations in the likelihood function (see Table~\ref{tab:uncbreak}).
The uncertainties due to the limited number of simulated background events are incorporated in the likelihood as nuisance parameters (NP) using Poisson probability density functions following a simplified version of the Beeston--Barlow technique~\cite{Barlow:1993dm}.
 
The rest of the systematic uncertainties are incorporated in the fit as NP, constrained using Gaussian probability density functions.
These include (i) the NP corresponding to the normalisation uncertainty in the measured phase space, added for each background sample whose normalisation is not left to float freely in the fit, (ii) the NP corresponding to the uncertainty in the acceptance difference between the different regions considered in the fit and (iii) the NP corresponding to the uncertainty in the distribution shape of the fitted final discriminating variable.
 
In the \lltt channel, the normalisation of the \ttbar background is constrained by using the \fake region, and the normalisation of the \ttZ background is constrained by using the \Hlo{} and \Hhi{} regions, so these normalisations are allowed to float freely in the fit.
A simultaneous fit is performed to the data yields in the \fake, \Hlo{} and \Hhi{} CRs and to the $\Delta m$ distribution in the \Hin{} SR, which is the final discriminating variable for this channel.
 
In the \vvbb channel, the normalisations of the \ttbar and Zhf background processes are allowed to float freely in the likelihood function, since they are constrained by using the \emu and \twoL regions, respectively.
Due to a known mismodelling of Zhf events~\cite{ATLAS:2021shl,ATLAS:2020fcp,ATLAS:2022enb}, which is expected to be more severe in the \threetag region, two decorrelated normalisation factors are used for the Zhf background in the \twotag and \threetag regions.
The \ttbar events populating the \threetag region can arise either from events with at least three real $b$-jets, via $\ttbar+(g\to b\bar{b}$), or from the misidentification of a $c$-jet originating from a $W$ boson produced in the top-quark decays. The contributions of the two processes depend on \(m(\bbbar)\), with the former dominating in the high \(m(\bbbar)\) region. To account for the different underlying processes, two decorrelated normalisation factors are applied to the \ttbar background in the \twotag and \threetag regions.
A simultaneous fit of the \mtvh distribution, which is the final discriminating variable for this channel, is performed in the \twoL, \emu, \Hlo{} and \Hhi{} CRs and the \Hin{} SR including both the \twotag and \threetag regions. Eight bins of the \mtvh distribution are used in the \twoL \twotag region, but only two bins in the \emu region and a single bin in the \twoL \threetag,  \Hlo{} and \Hhi{} regions.
 
Since a large number of events satisfy the selection criteria, both the significance of any excess and the upper limits are calculated in the asymptotic approximation~\cite{Cowan:2010js}.
Upper limits on $\mu$ are extracted using the profile-likelihood test statistic and the CL$_\text{s}$ prescription~\cite{Read:2002hq} by performing fits for multiple $(\mAhypo, \mHhypo)$ values. All upper limits mentioned in \cref{sec:results} correspond to a confidence level (CL) of 95\%.


\section{Results}
\label{sec:results}

No significant deviation from the background-only hypothesis is observed in the likelihood fits.
The largest excess over the SM background prediction, amounting to a local significance of 2.85\,$\sigma$, is observed in the \lltt channel, for the signal hypothesis corresponding to \((m_A,m_H)=(650,450)\)~\GeV. The global significance for the \lltt channel is estimated following Refs.~\cite{Vitells:2011da,Gross:2010qma} to be 2.35\,$\sigma$.
 
\begin{figure}
\centering
\subfigure[]{\includegraphics[width=0.45\textwidth]{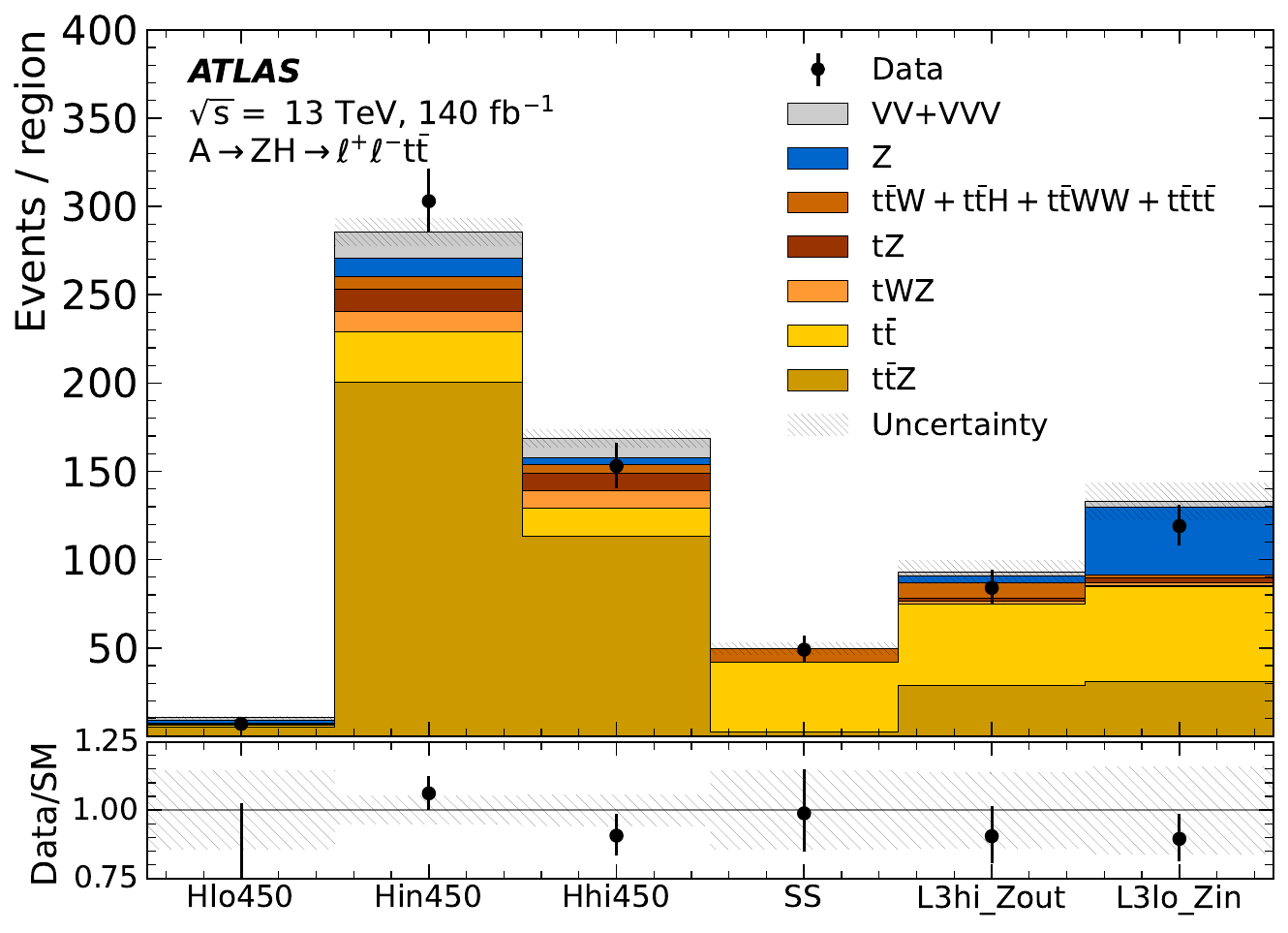}}
\subfigure[]{\includegraphics[width=0.45\textwidth]{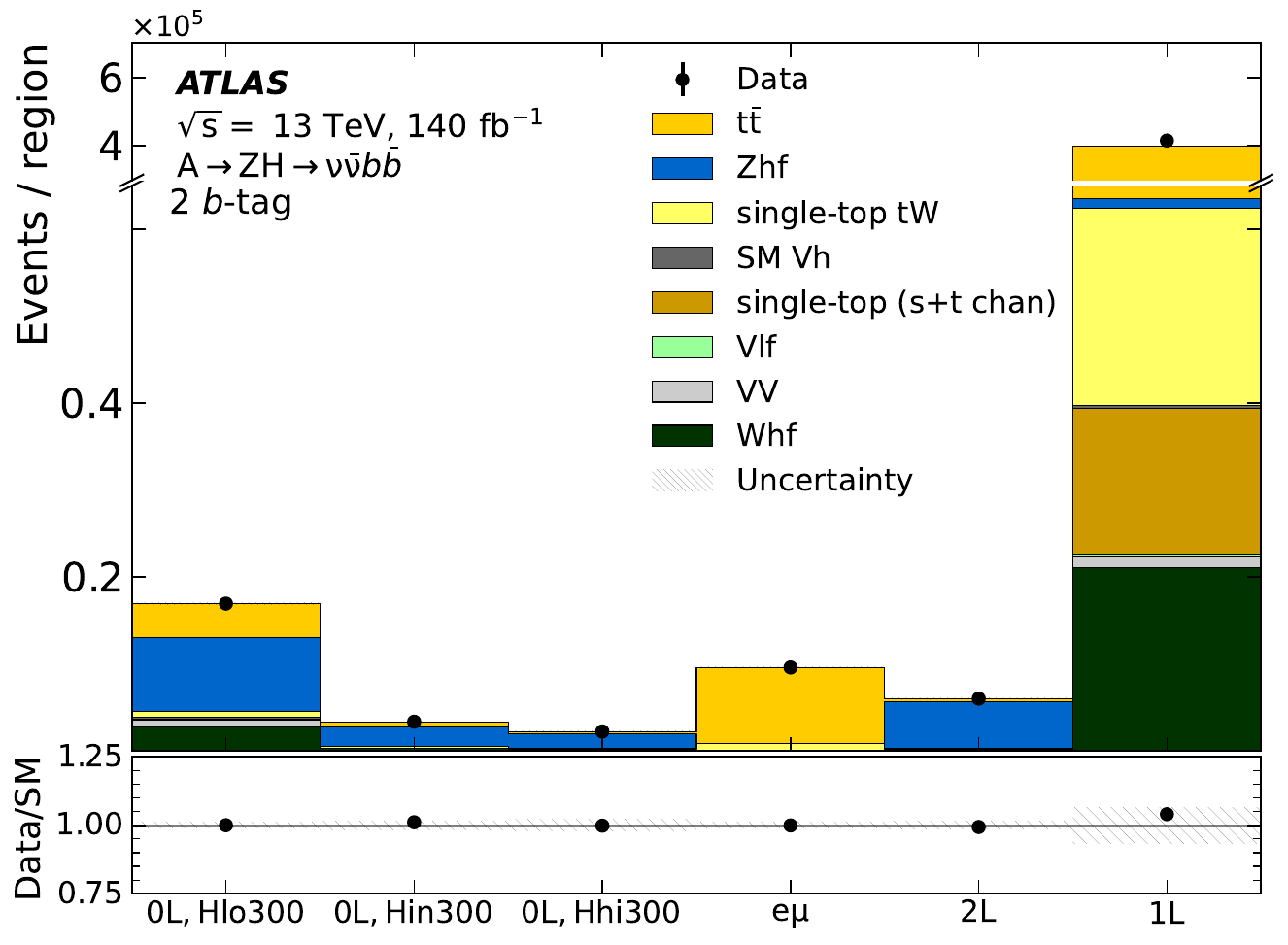}}
\subfigure[]{\includegraphics[width=0.45\textwidth]{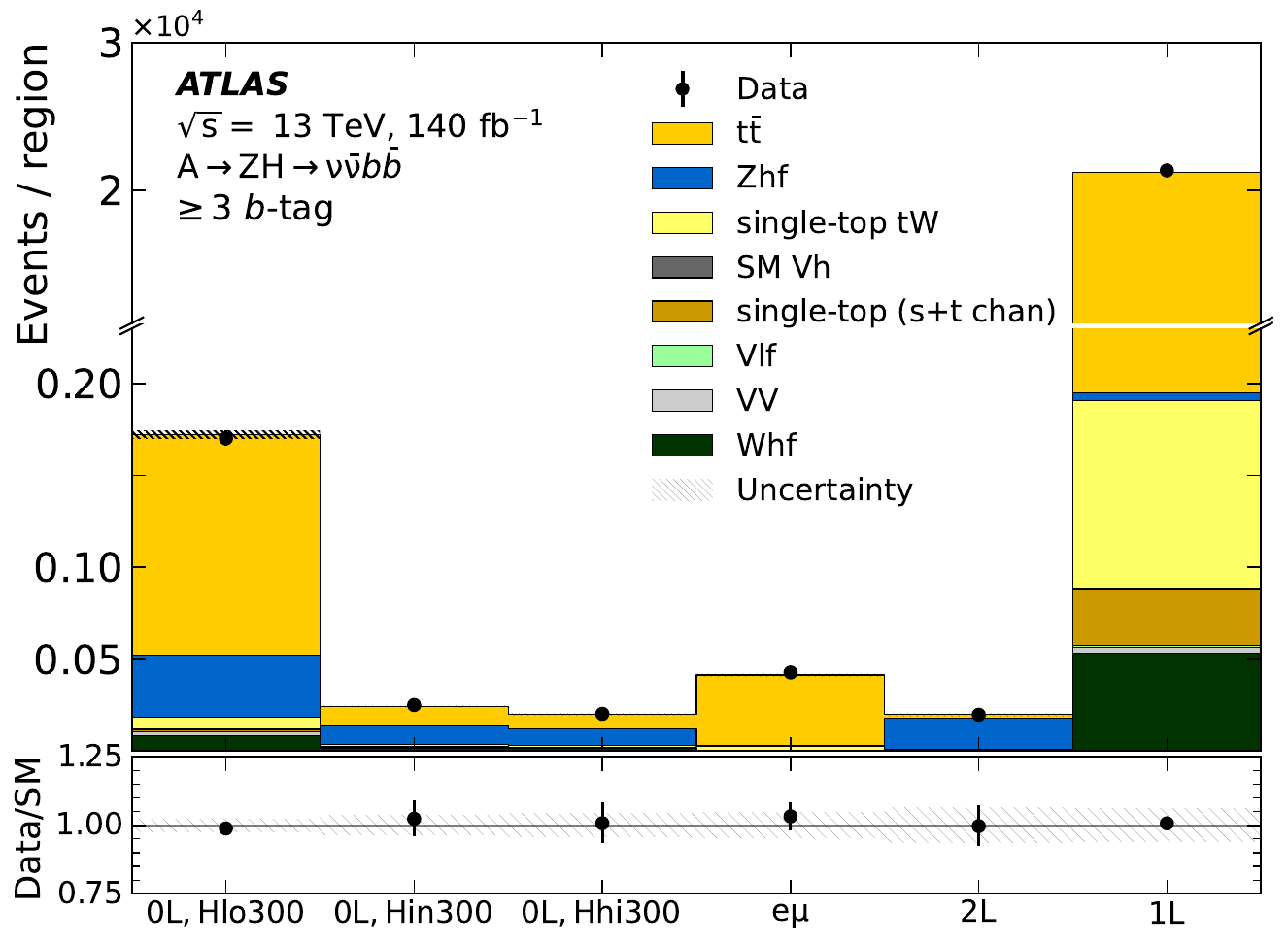}}
\caption{Yields in the SR, CRs and VRs used in the \lltt channel (a) and in the \twotag (b) and \threetag (c) regions of the \vvbb channel. The yields are obtained from a background-only fit to data. The region names are defined in \Cref{tab:llttSelection,tab:vvbbSelection}. The value next to the region name refers to the \mHhypo hypothesis. The data are represented as black points and the associated error bars represent the statistical uncertainty. The hatched band indicates the combined statistical and systematic uncertainty for the sum of backgrounds.}
\label{fig:yields}
\end{figure}

\begin{figure}
\centering
\subfigure[]{\includegraphics[width=0.329\textwidth]{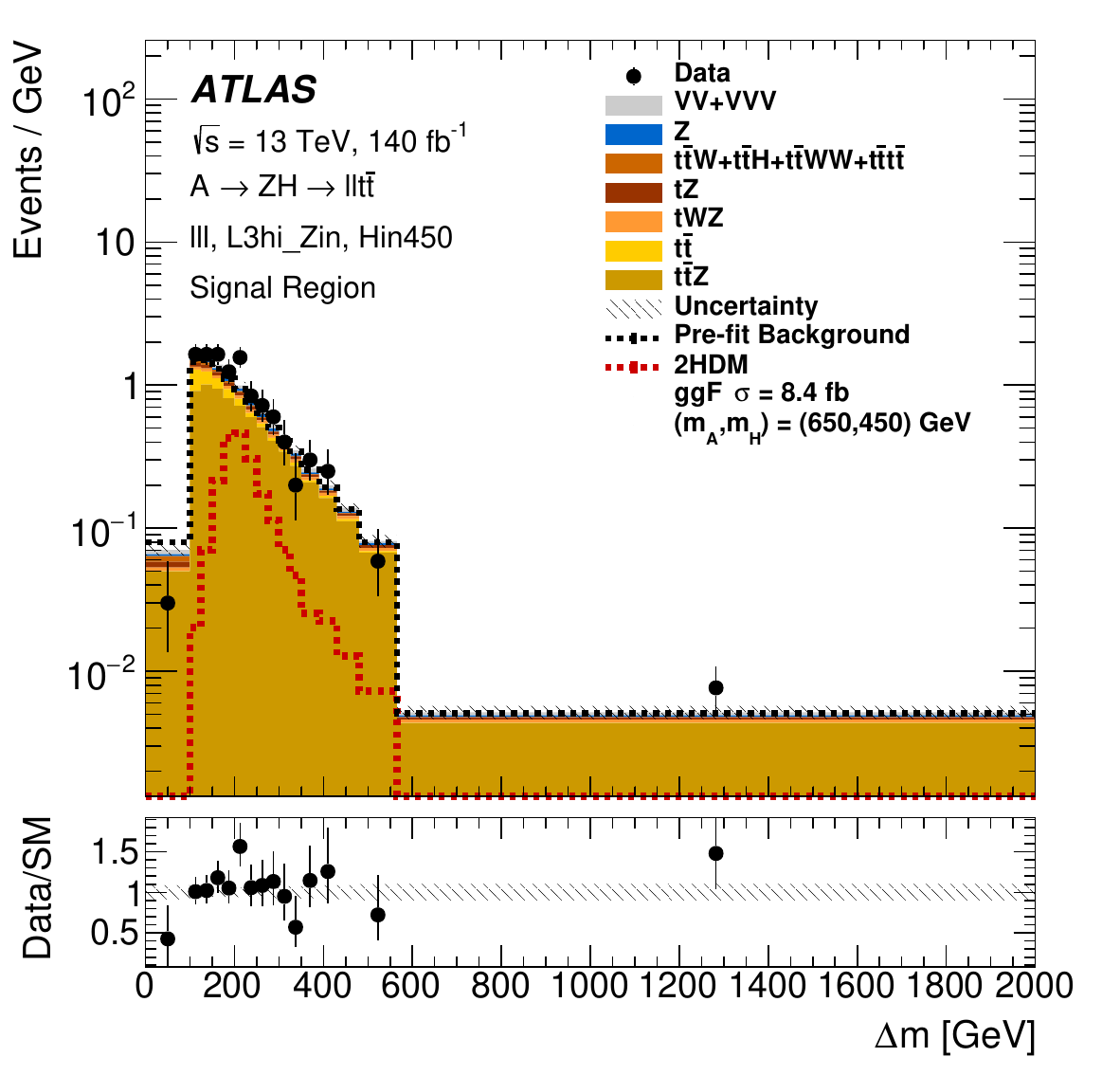}\label{fig:postfit_fitvar:lltt}}
\subfigure[]{\includegraphics[width=0.329\textwidth]{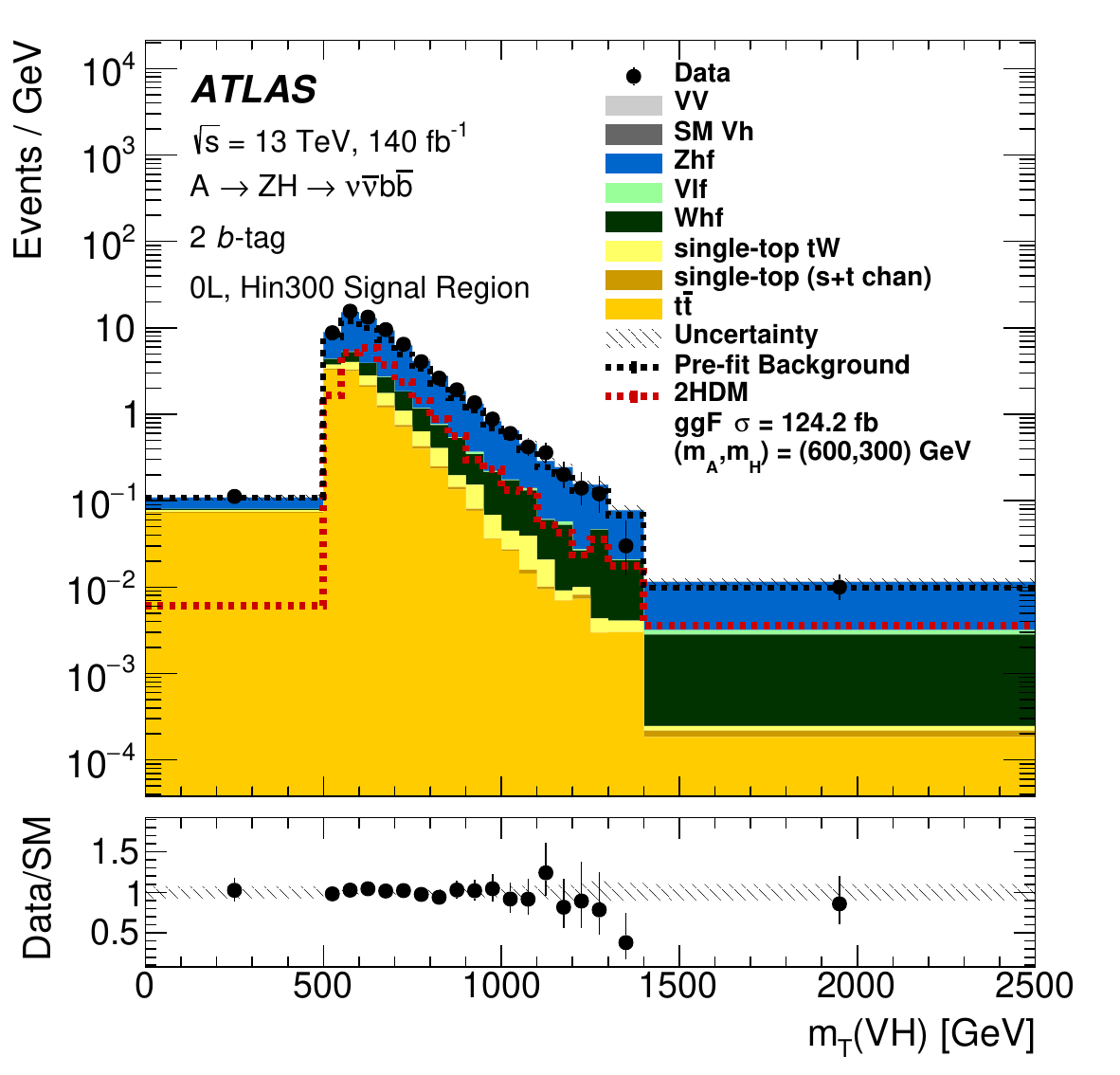}\label{fig:postfit_fitvar:vvbb:2tag}}
\subfigure[]{\includegraphics[width=0.329\textwidth]{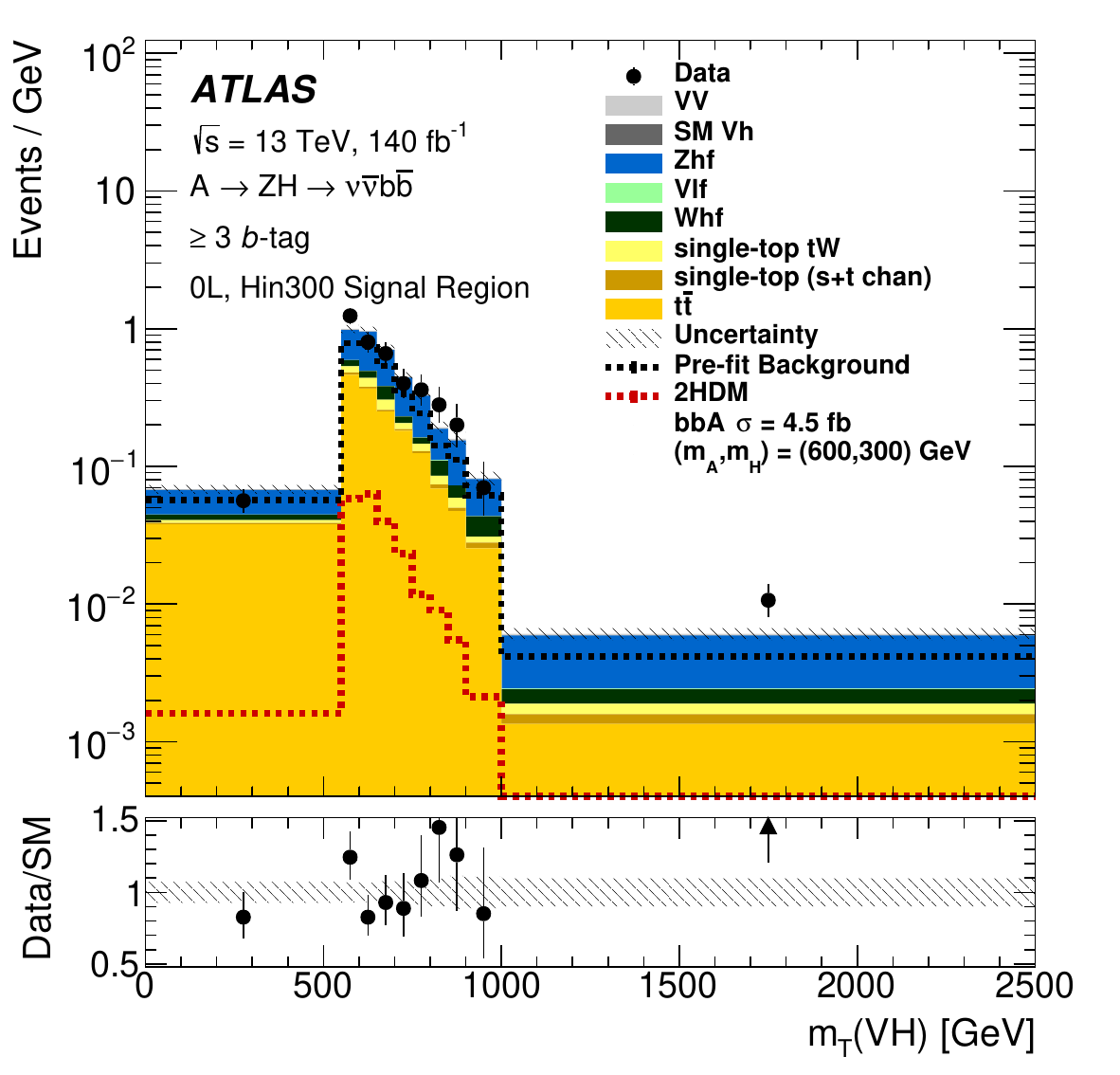}\label{fig:postfit_fitvar:vvbb:3ptag}}
\caption{The distribution of the fit discriminant \(\Delta m=m(\lltt)-m(\ttbar)\) in the SR of the \lltt channel for the \(\mHhypo=450\)~\GeV\ hypothesis (a). The distribution of the fit discriminant \mtvh in the SR of the \vvbb channel in the \twotag (b) and \threetag (c) region, for the \(\mHhypo=300\)~\GeV\ hypothesis. The background yields are obtained from a background-only fit to data. Signal distributions corresponding to ggF or bbA production normalised to the theory cross-section are shown for comparison. The data are represented as black points and the associated error bars represent the statistical uncertainty. The hatched band indicates the combined statistical and systematic uncertainty for the sum of backgrounds. The quantity on the vertical axis is the number of events divided by the bin width in \GeV.}
\label{fig:postfit_fitvar}
\end{figure}

\begin{figure}
\centering
\subfigure[]{\includegraphics[width=0.329\textwidth]{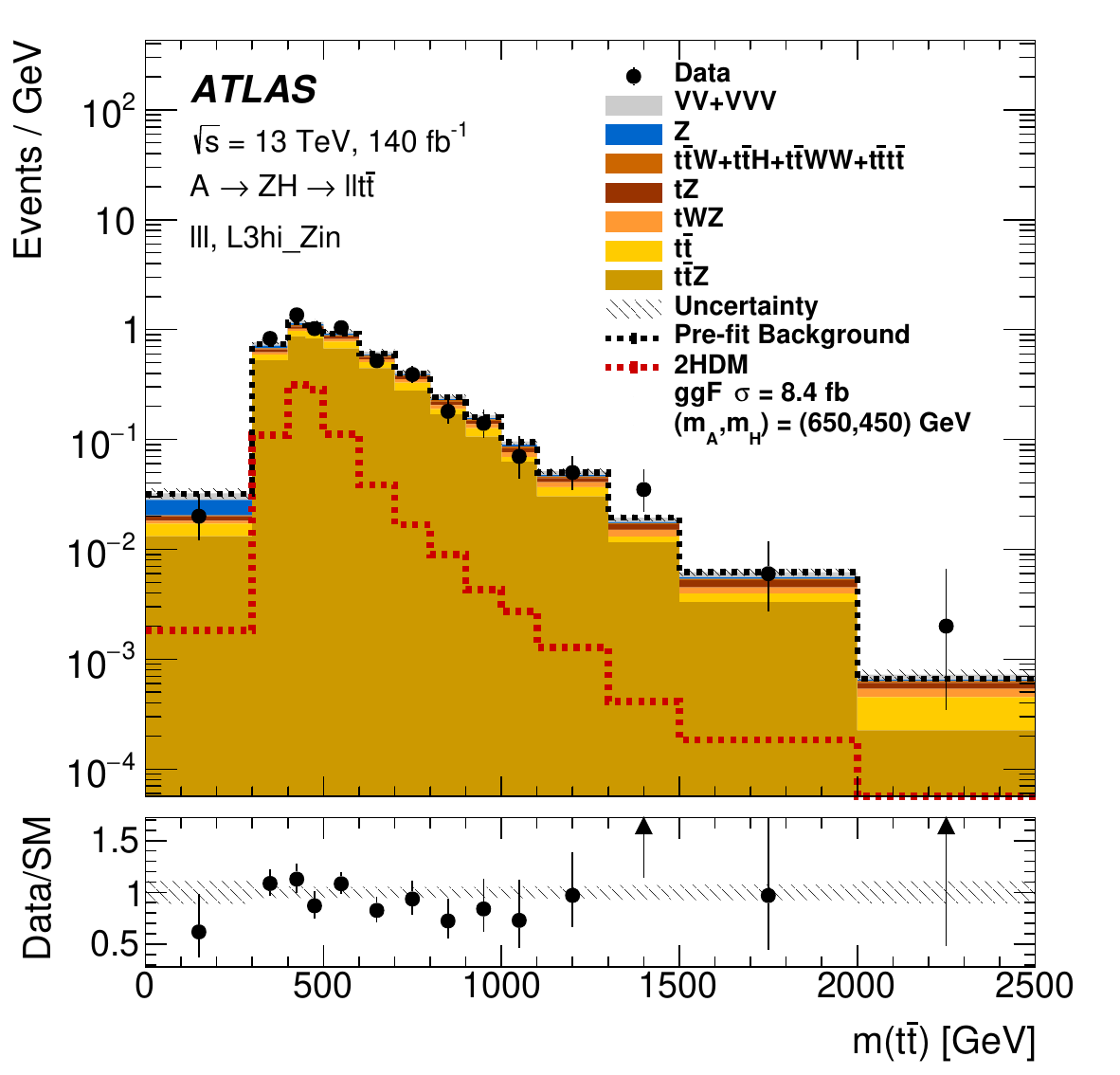}}
\subfigure[]{\includegraphics[width=0.329\textwidth]{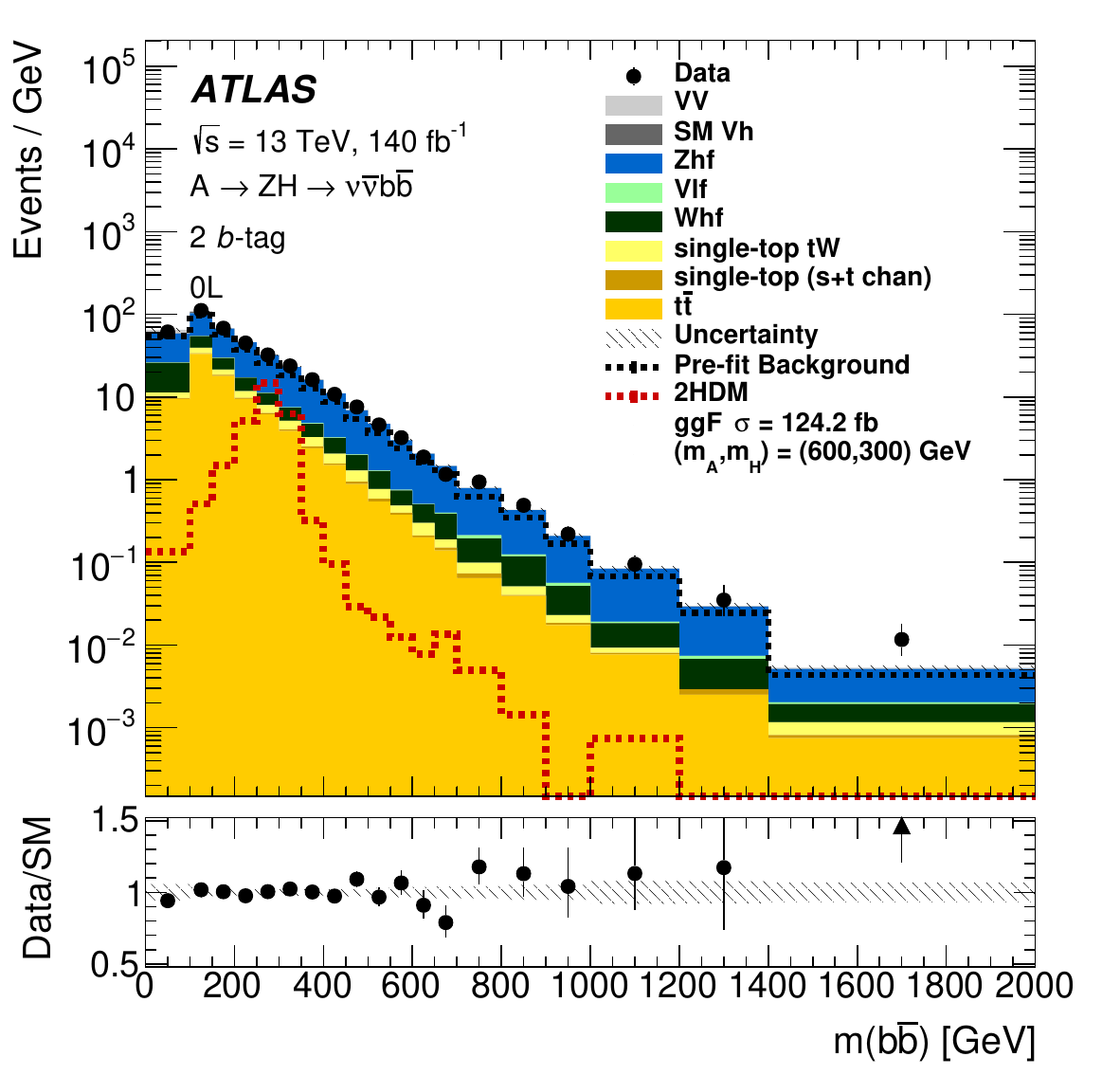}}
\subfigure[]{\includegraphics[width=0.329\textwidth]{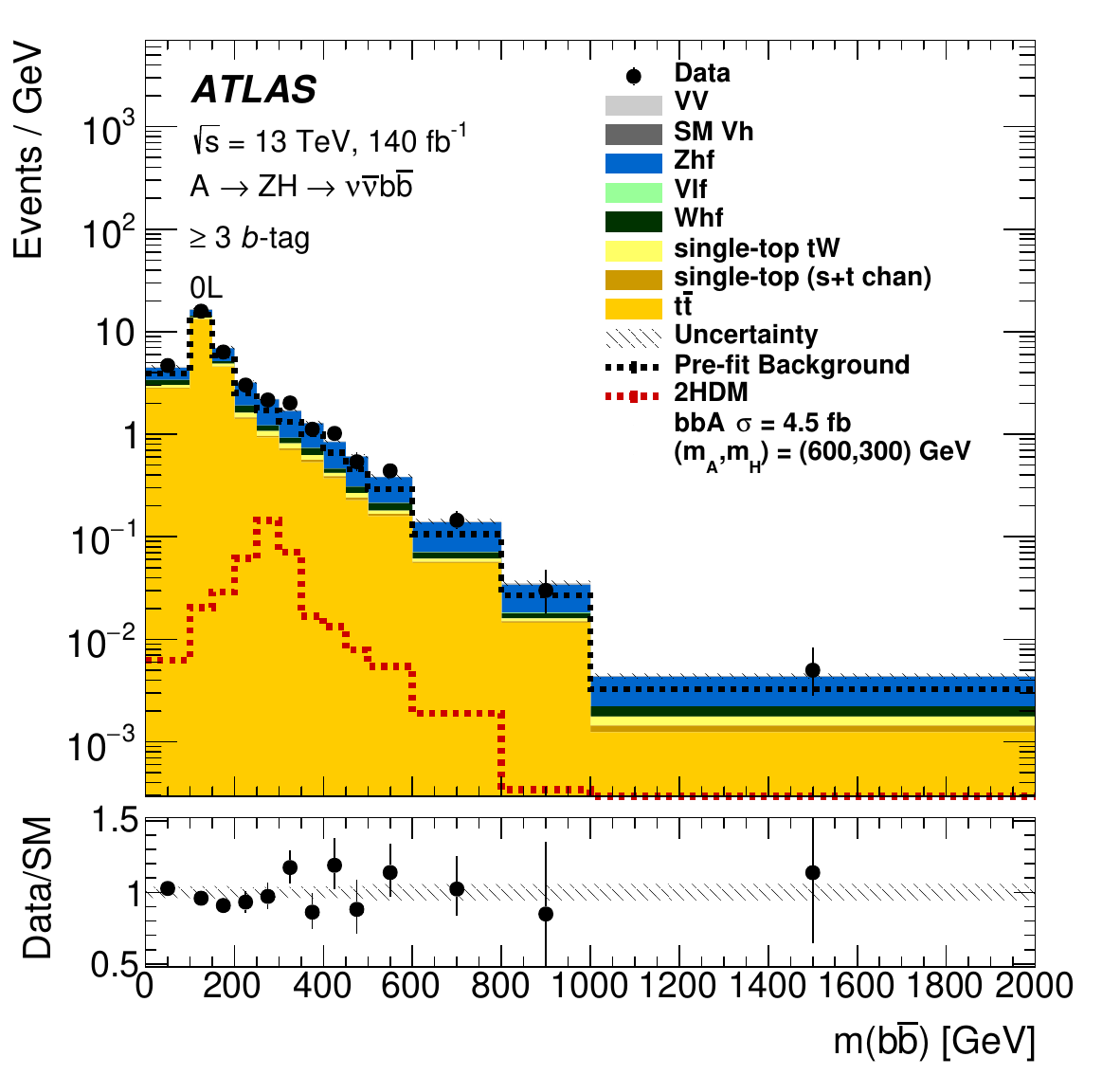}}
\caption{The \(m(\ttbar)\) distribution in the \Lhi\!\!\_\Zin region of the \lltt channel (a) and the \(m(\bbbar)\) distribution in the \vvbb channel in the \twotag (b) and \threetag (c) \zeroL region. The background yields are obtained from a background-only fit to data. Signal distributions corresponding to ggF or bbA production normalised to the theory cross-section are shown for comparison. The data are represented as black points and the associated error bars represent the statistical uncertainty. The hatched band indicates the combined statistical and systematic uncertainty for the sum of backgrounds. The quantity on the vertical axis is the number of events divided by the bin width in \GeV. }
\label{fig:postfit_mH}
\end{figure}

The yields and post-fit distributions presented here are obtained from background-only fits to the data in the SR and CRs.
The constraints from these fits are also applied to the VRs to gauge whether the fit can describe the data in regions that are kinematically close to the SR and CRs but are not included in the fits.
Representative yields in the SR, CRs and VRs are shown in \Cref{fig:yields} and \Cref{tab:results:lltt_yields,tab:results:vvbb_2tag_yields,tab:results:vvbb_3ptag_yields}.
Representative distributions of the fit discriminant and the mass of the \(H\) candidate in the SR are shown in \Cref{fig:postfit_fitvar,fig:postfit_mH}.

\begin{table}[ht]
\caption{Yields in the \lltt channel obtained from the background-only fit to data using \Hin{450} as the signal region. The indicated uncertainties include statistical and systematic components. The value next to the region name refers to the \mHhypo hypothesis.}\label{tab:results:lltt_yields}
\centering
\scalebox{0.9}{\begin{tabular}{|l|ccc|c|c|c|}
\hline
&                     &   \Lhi\!\!\_\Zin    &                     &                     &        (VR)         &       (VR)          \\
&     \Hlo{450} (CR)  &    \Hin{450} (SR)   &   \Hhi{450} (CR)    &       \fake (CR)    & \Lhi\!\!\_\Zout     & \Llo\!\!\_\Zin      \\
\hline
$\ttZ$ &    $5.1\pm0.9$ &   $200\pm22$ &    $113\pm13$ &      $2.3\pm0.6$ &    $29.0\pm3.2$\phantom{0} &   $30.8\pm3.4$\phantom{0} \\
$\ttbar$ &    $1.2\pm0.8$ &     $29\pm9$ &      $16\pm6$ &         $40\pm7$\phantom{0} &       $46\pm14$ &      $54\pm17$ \\
\(tWZ\) &  $0.40\pm0.14$ &     $12\pm4$ &      $10\pm4$ &    $0.13\pm0.04$ &     $1.6\pm0.5$ &    $2.2\pm0.7$ \\
\(tZq\) &    $0.6\pm0.4$ &     $13\pm8$ &      $10\pm6$ &  $0.046\pm0.032$ &     $1.8\pm1.2$ &    $2.4\pm1.6$ \\
\(VV\)+\;\!\(VVV\) &    $1.5\pm0.5$ &     $15\pm4$ &  $11.1\pm3.5$ &  $0.034\pm0.013$ &     $2.3\pm0.6$ &    $3.1\pm0.9$ \\
\(Z\) &    $1.5\pm1.1$ &     $11\pm4$ &   \phantom{0}$3.9\pm1.6$ &  $0.025\pm0.010$ &     $3.7\pm1.4$ &      $39\pm15$ \\
$t\bar{t}W$+\;\!$t\bar{t}H$\;\!+\;\!$t\bar{t}WW$+\;\!$t\bar{t}t\bar{t}$ &  $0.16\pm0.05$ &  \phantom{0}$6.8\pm0.9$ &   \phantom{0}$4.9\pm0.9$ &      $7.4\pm1.8$ &     $8.4\pm1.8$ &  $1.63\pm0.27$ \\
Total background &   $10.5\pm1.5$\phantom{0} &   $285\pm15$ &    $169\pm10$ &         $50\pm7$\phantom{0} &       $93\pm13$ &     $133\pm21$\phantom{0} \\
\hline
\hline
Data &              7 &          303 &           153 &               49 &              84 &            119 \\
\hline
\end{tabular}}
\end{table}


\begin{table}[ht]
\caption{Yields in the \twotag regions of the \vvbb channel obtained from the background-only fit to data using \Hin{300} as the signal region. The indicated uncertainties include statistical and systematic components. The value next to the region name refers to the \mHhypo hypothesis.}\label{tab:results:vvbb_2tag_yields}
\centering
\scalebox{0.85}{\begin{tabular}{|l|ccc|c|c|c|}
\hline
&                     &     \zeroL          &                     &                     &                     &                     \\
&   \Hlo{300} (CR)    &   \Hin{300} (SR)    &   \Hhi{300} (CR)    &      \twoL (CR)     &       \emu (CR)     &      \oneL (VR)     \\
\hline
\ttbar &   $3800\pm400$\phantom{0} &    $600\pm80$\phantom{0} &     $290\pm40$\phantom{0} &     $370\pm60$\phantom{0} &   $8700\pm500$\phantom{0} &  $336000\pm28000$\phantom{0} \\
Single-top (\(s\)-, \(t\)-chan) &      $93\pm18$ &      $26\pm4$\phantom{0} &   $16.7\pm3.1$\phantom{0} &  $0.70\pm0.15$ &   $12.4\pm0.7$\phantom{0} &    $16700\pm1100$\phantom{0} \\
Single-top \(tW\) &    $600\pm400$ &    $160\pm90$\phantom{0} &      $90\pm60$ &      $43\pm23$ &    $800\pm500$ &   $23000\pm12000$ \\
Whf &   $2800\pm900$\phantom{0} &   $330\pm100$ &     $230\pm70$\phantom{0} &    $2.8\pm1.0$ &       $29\pm9$\phantom{0} &    $21000\pm7000$\phantom{0} \\
Zhf &   $8500\pm900$\phantom{0} &  $2200\pm120$\phantom{0} &    $1620\pm90$\phantom{00} &   $5370\pm120$\phantom{0} &   $18.1\pm1.3$\phantom{0} &       $1070\pm90$\phantom{00} \\
Vlf &       $44\pm8$\phantom{0} &  $10.8\pm1.7$\phantom{0} &   $12.3\pm2.3$\phantom{0} &       $23\pm5$\phantom{0} &  $0.35\pm0.08$ &        $330\pm60$\phantom{0} \\
\(VHbb\) &    $210\pm130$ &   $0.8\pm0.5$ &  $0.48\pm0.31$ &      $60\pm40$ &  $0.37\pm0.23$ &       $350\pm220$ \\
\(VV\) &    $770\pm150$ &  $15.4\pm1.7$\phantom{0} &   $12.3\pm1.6$\phantom{0} &     $207\pm24$\phantom{0} &  $1.63\pm0.20$ &      $1260\pm140$\phantom{0} \\
Total background &  $16960\pm170$\phantom{00} &   $3350\pm50$\phantom{00} &    $2270\pm50$\phantom{00} &    $6080\pm90$\phantom{00} &   $9620\pm110$\phantom{0} &  $400000\pm26000$\phantom{0} \\
\hline
\hline
Data &          16961 &          3389 &           2266 &           6037 &           9618 &            415808 \\
\hline
\end{tabular}}
\end{table}
 
\begin{table}[ht]
\caption{Yields in the \threetag regions of the \vvbb channel obtained from the background-only fit to data using \Hin{300} as the signal region. The indicated uncertainties include statistical and systematic components. The value next to the region name refers to the \(m_H\) hypothesis.}\label{tab:results:vvbb_3ptag_yields}
\centering
\scalebox{0.85}{\begin{tabular}{|l|ccc|c|c|c|}
\hline
&                     &        \zeroL       &                     &                     &                     &                    \\
&   \Hlo{300} (CR)    &   \Hin{300} (SR)    &    \Hhi{300} (CR)   &      \twoL (CR)     &       \emu (CR)     &      \oneL (VR)    \\
\hline
\ttbar &    $1200\pm70$\phantom{00} &  $101\pm8$\phantom{00} &  $80\pm9$\phantom{0} &   $16.9\pm3.1$\phantom{0} &   $385\pm26$\phantom{0} &  $19300\pm1400$\phantom{0} \\
Single-top (\(s\)-, \(t\)-chan) &   $11.0\pm1.2$\phantom{0} &    $3.9\pm0.5$ &    $4.0\pm0.4$ &                 -- &    $0.28\pm0.10$ &      $310\pm27$\phantom{0} \\
Single-top \(tW\) &     $70\pm50$ &     $13\pm8$\phantom{0} &     $8\pm7$ &     $1.2\pm0.8$ &     $27\pm19$ &    $1000\pm700$\phantom{0} \\
Whf &     $82\pm28$ &     $18\pm6$\phantom{0} &     $14\pm5$\phantom{0} &    $0.13\pm0.04$ &    $1.2\pm0.4$ &     $530\pm170$ \\
Zhf &    $340\pm50$\phantom{0} &    $106\pm10$\phantom{0} &    $91\pm10$ &     $173\pm13$\phantom{0} &    $0.60\pm0.16$ &    $43\pm5$\phantom{0} \\
Vlf &  $0.73\pm0.33$ &  $0.14\pm0.05$ &  $0.17\pm0.04$ &  $0.0040\pm0.0020$ &               -- &     $6.9\pm2.8$ \\
\(VHbb\) &    $3.7\pm2.4$ &  $0.48\pm0.31$ &  $0.42\pm0.27$ &        $1.1\pm0.7$ &  $0.010\pm0.007$ &     $5.3\pm3.4$ \\
\(VV\) &       $21\pm4$\phantom{0} &    $3.7\pm0.5$ &    $3.3\pm0.4$ &        $6.6\pm0.9$ &  $0.037\pm0.018$ &        $35\pm4$\phantom{0} \\
Total background &    $1720\pm40$\phantom{00} &     $245\pm9$\phantom{00} &    $201\pm8$\phantom{00} &     $199\pm12$\phantom{0} &    $415\pm19$\phantom{0} &  $21200\pm1300$\phantom{0} \\
\hline
\hline
Data &           1702 &            251 &            203 &                198 &              428 &           21356 \\
 
\hline
\end{tabular}}
\end{table}


The normalisation factors in the \lltt channel are found to be compatible with unity for the \(\ttbar Z\) background, while for the \ttbar background they range from 1.5 to 1.8 for the different \mHhypo hypotheses with an uncertainty of 0.5. The \ttbar normalisation factor is higher than unity due to mismodelling of \ttbar events with a non-prompt or misidentified lepton; this has been verified with a data-driven method.
In the \vvbb channel, the normalisation factors for the \ttbar background are compatible with unity, while for the \(Z\)hf background they are in the range 1.2--1.3 with an uncertainty of 0.1 in the \twotag region and in the range 1.4--1.7 with an uncertainty of 0.2 in the \threetag regions. The Zhf normalisation factors are higher than unity due to mismodelling of the Zhf process in \SHERPA~\cite{ATLAS:2021shl,ATLAS:2020fcp,ATLAS:2022enb}.
In the \lltt channel, the \ttbar template corresponds to \ttbar events with two prompt leptons and one non-prompt or misidentified lepton, while in the \vvbb channel it corresponds to \ttbar events with  two prompt leptons, so the two normalisation factors obtained above are not comparable.

\subsection{Upper limits on the production cross-section for $A\to ZH\to \lltt/\vvbb$}\label{subsec:limits}
\begin{figure}
\centering
\subfigure[]{\includegraphics[width=0.48\textwidth]{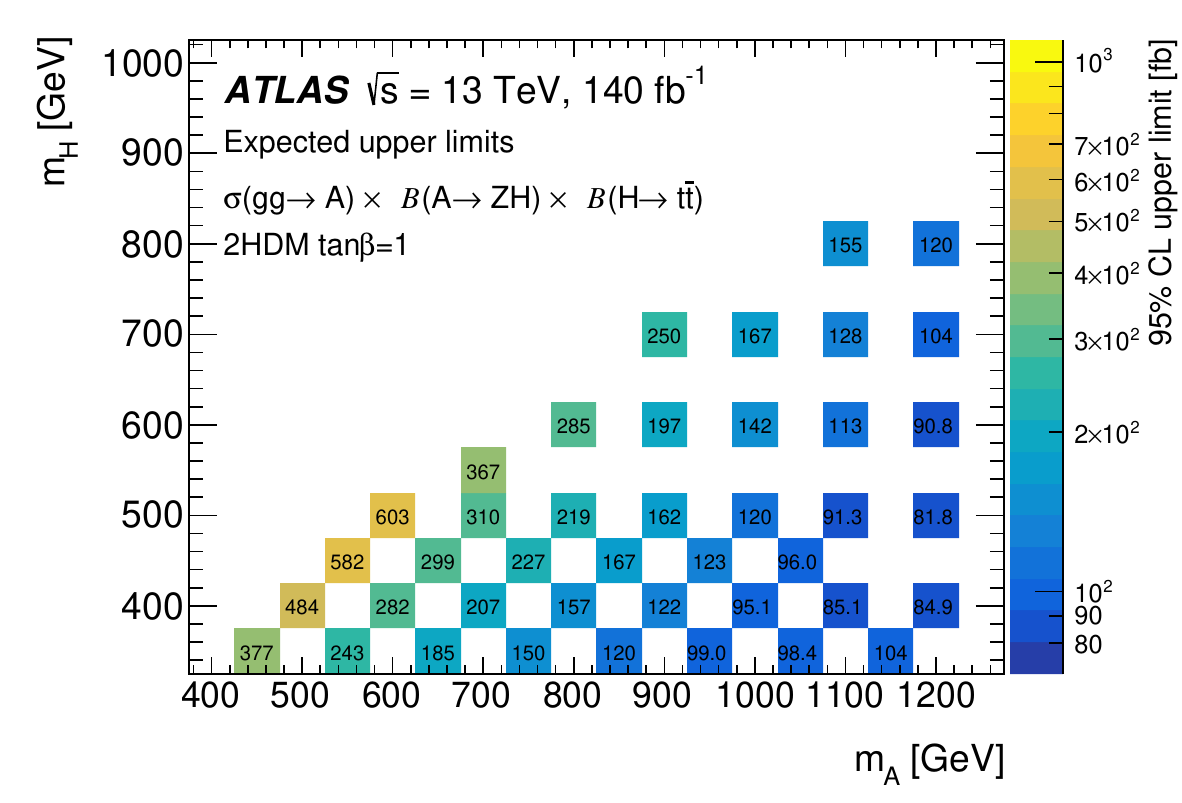}\label{fig:limits:lltt:exp}}
\subfigure[]{\includegraphics[width=0.48\textwidth]{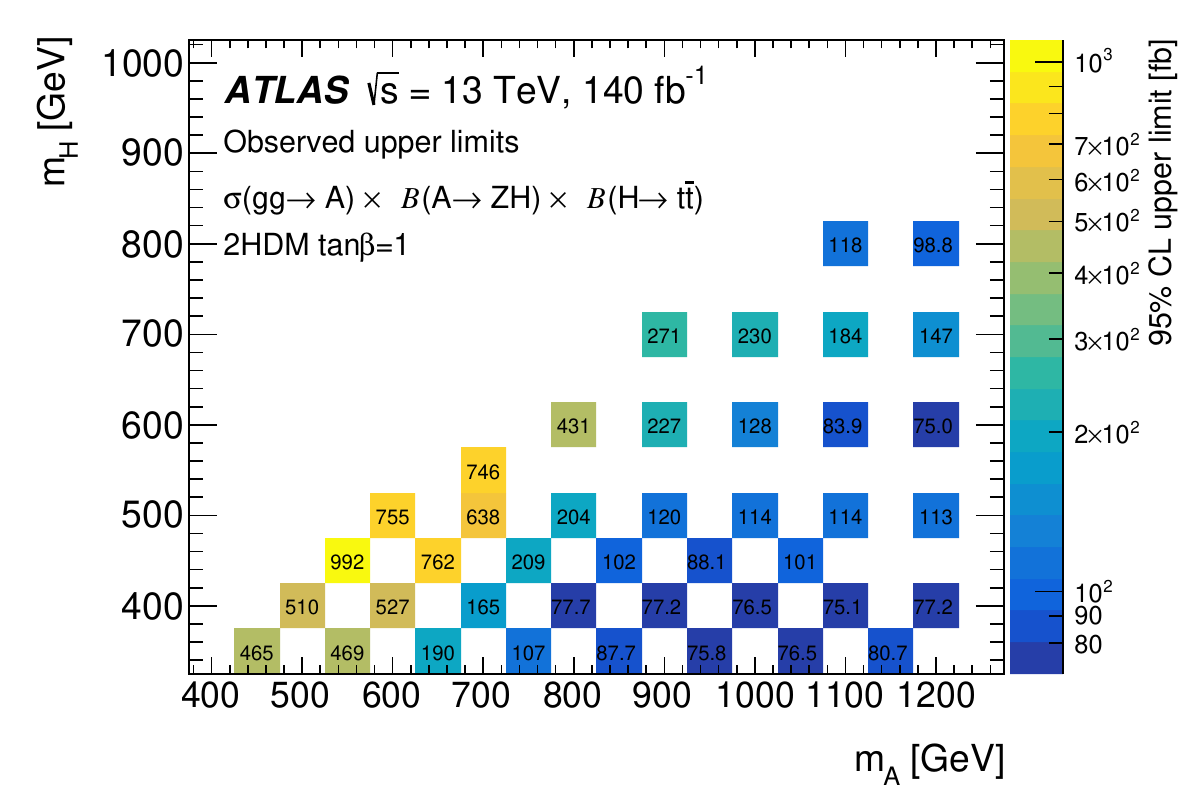}\label{fig:limits:lltt:obs}}
\subfigure[]{\includegraphics[width=0.48\textwidth]{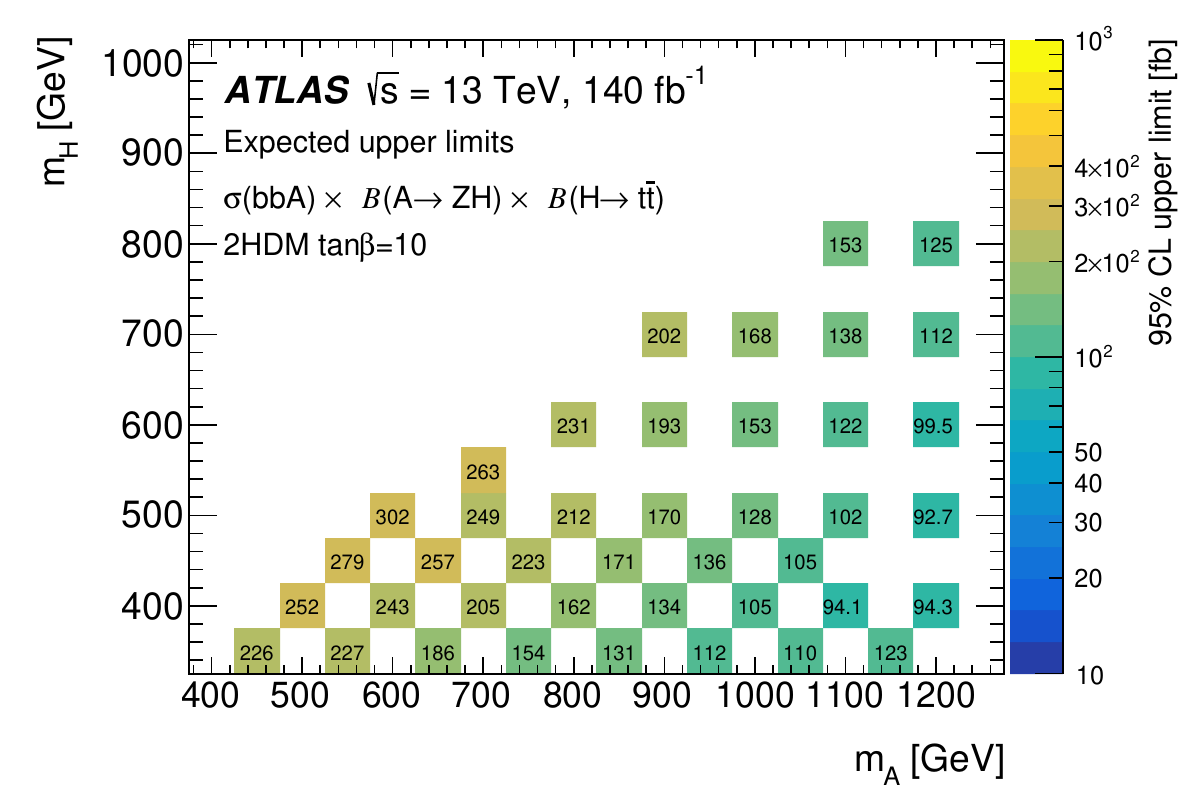}\label{fig:limits:lltt:bbA:exp}}
\subfigure[]{\includegraphics[width=0.48\textwidth]{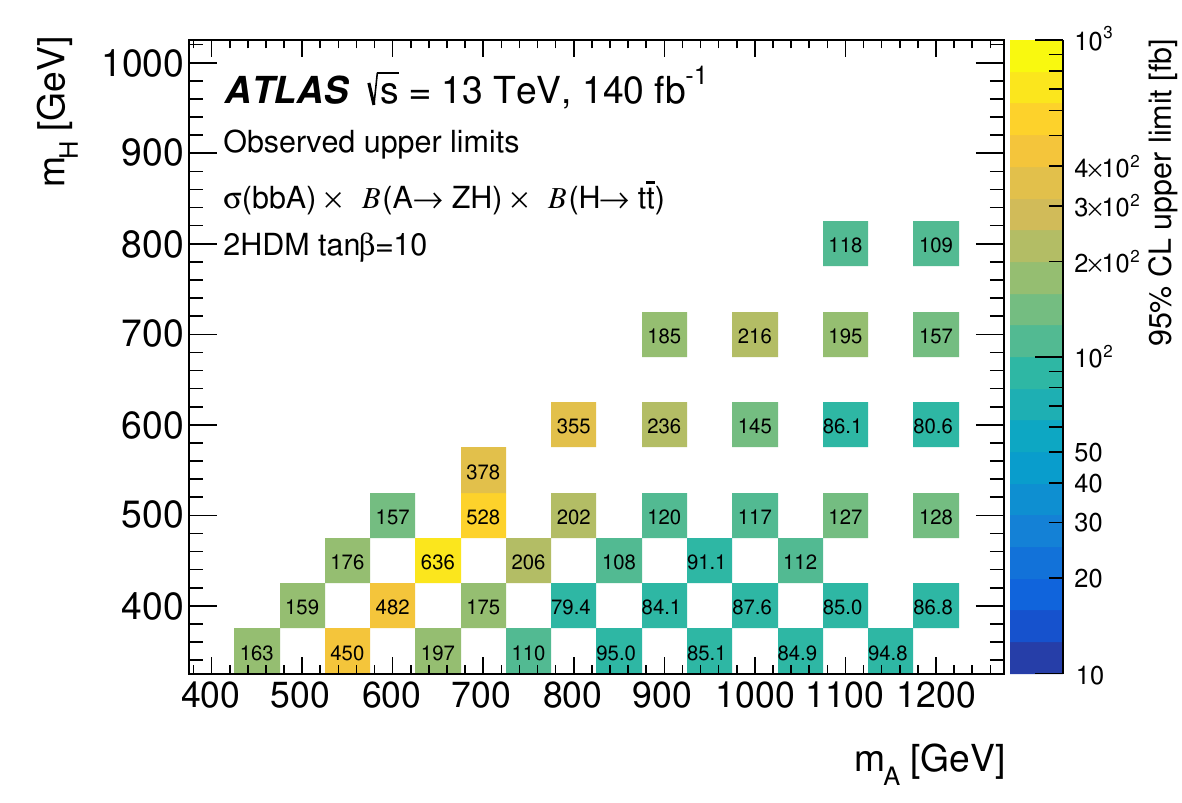}\label{fig:limits:lltt:bbA:obs}}
\caption{Expected (a,c) and observed (b,d) upper limits at 95\% CL on $\sigma(gg\to A)\times B(A\to ZH)\times B(H\to \ttbar)$ (a,b) and $\sigma(b\bbar A)\times B(A\to ZH)\times B(H\to \ttbar)$ (c,d) in the ($m_A, m_H$) plane. The limits are shown for either $\tan\beta=1$ or $\tan\beta=10$ in ggF or bbA production, respectively. The \(\tan\beta\) value is relevant only for the choice of $A$ boson width.}
\label{fig:xseclimits:lltt}
\end{figure}

\begin{figure}
\centering
\subfigure[]{\includegraphics[width=0.48\textwidth]{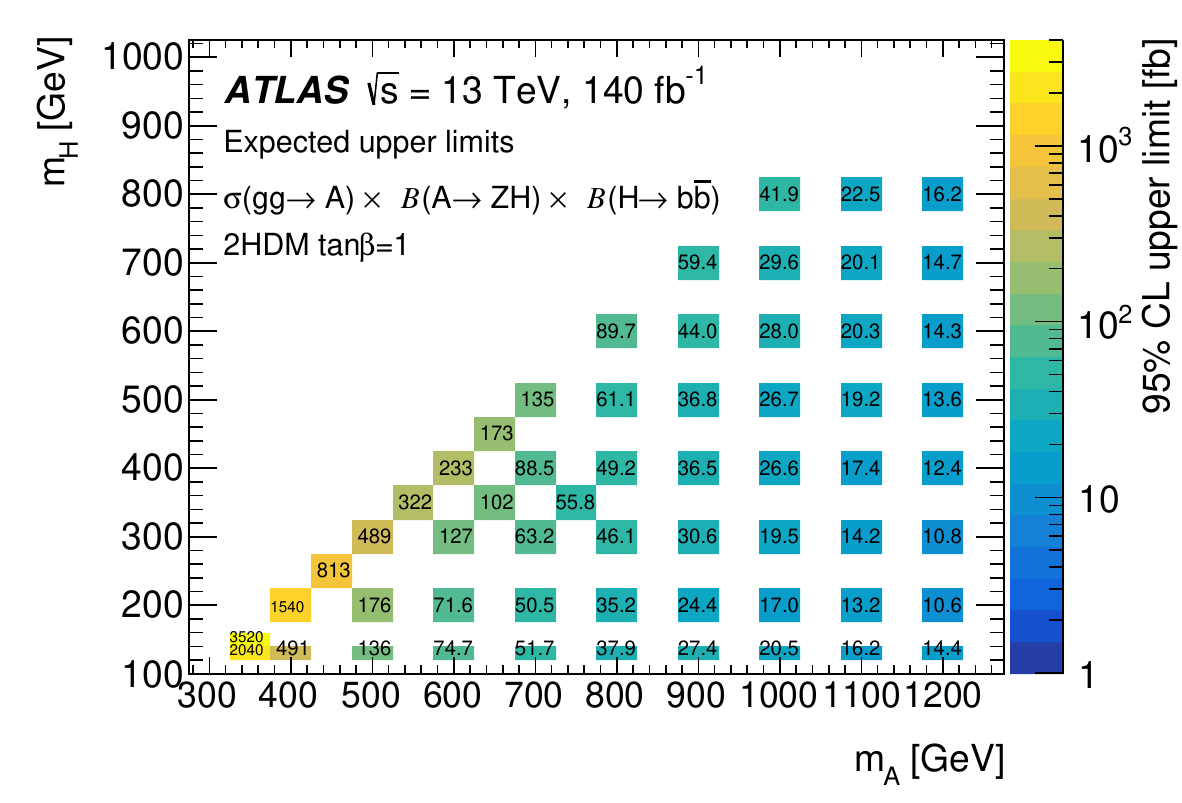}\label{fig:limits:vvbb:ggF:exp}}
\subfigure[]{\includegraphics[width=0.48\textwidth]{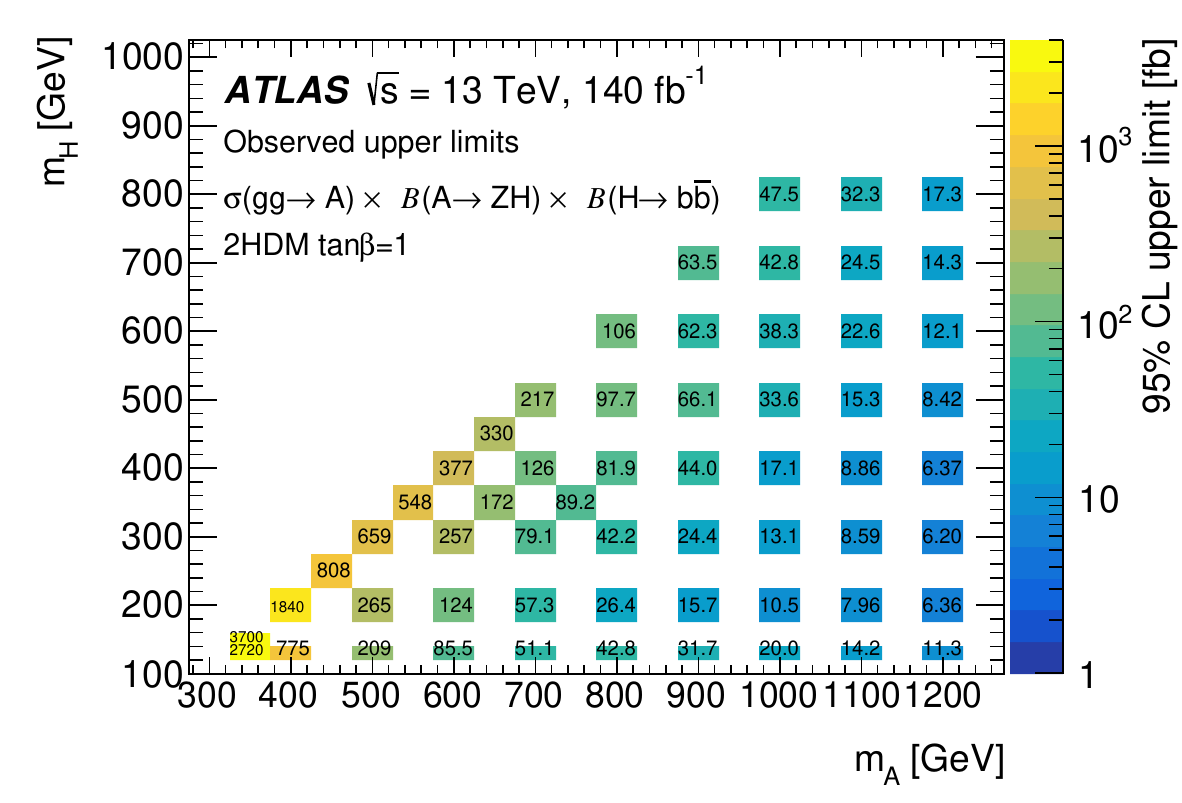}\label{fig:limits:vvbb:ggF:obs}}
\subfigure[]{\includegraphics[width=0.48\textwidth]{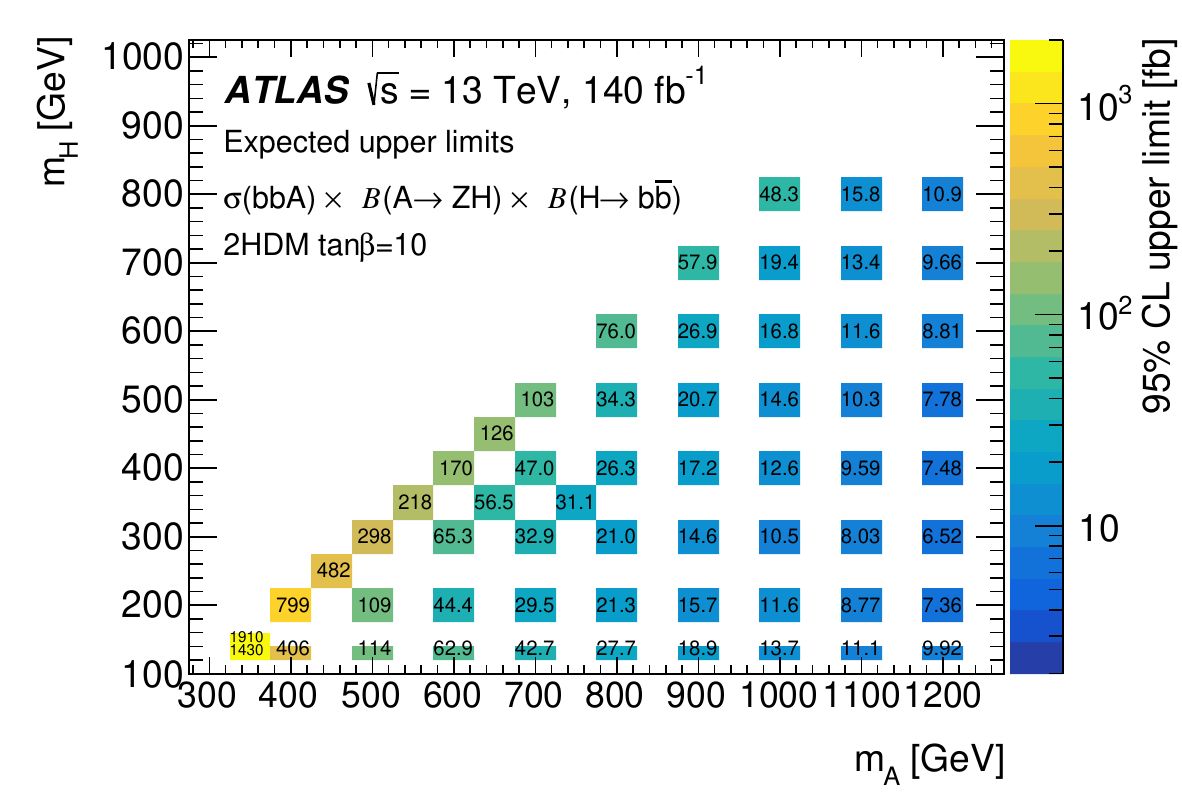}\label{fig:limits:vvbb:bbA:exp}}
\subfigure[]{\includegraphics[width=0.48\textwidth]{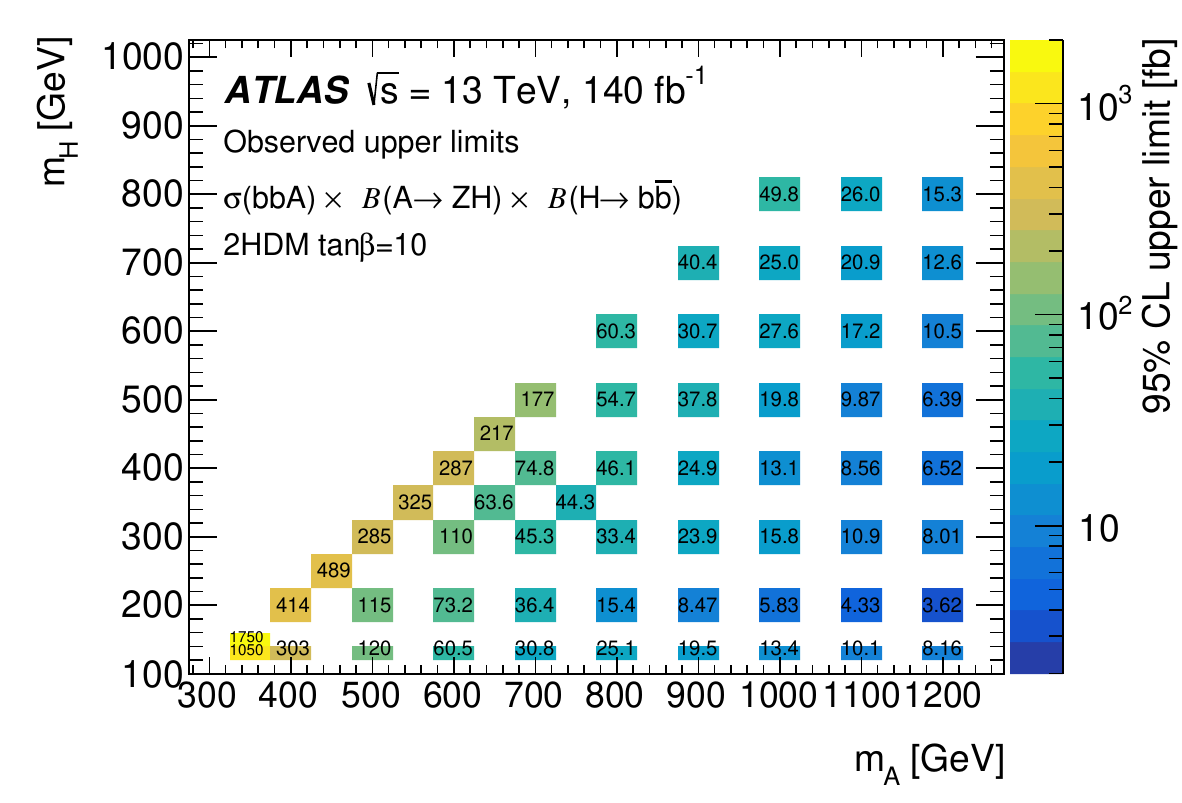}\label{fig:limits:vvbb:bbA:obs}}
\caption{Expected (a,c) and observed (b,d) upper limits at 95\% CL on $\sigma(gg\to A)\times B(A\to ZH)\times B(H\to b\bbar)$ (a,b) and $\sigma(b\bbar A)\times B(A\to ZH)\times B(H\to b\bbar)$ (c,d) in the ($m_A, m_H$) plane. The limits are shown for either $\tan\beta=1$ or $\tan\beta=10$ in ggF or bbA production, respectively. The \(\tan\beta\) value is relevant only for the choice of $A$ boson width.}
\label{fig:xseclimits:vvbb}
\end{figure}

Upper limits on the production cross-section for the $A$ boson times the decay branching ratios, $B(A\to ZH)\times B(H\to \ttbar)$ in the \lltt channel and $B(A\to ZH)\times B(H\to b\bbar)$ in the \vvbb channel, are derived for the ggF
and bbA production modes and are shown in \Cref{fig:xseclimits:lltt,fig:xseclimits:vvbb}.
While the limits generally depend on the natural width of the
$A$ and $H$ bosons, for the parameter space that is relevant for this search
only the width of the $A$ boson matters, with the width of the $H$ boson
always being very small compared to the experimental resolution.
The width of the $A$ boson increases as $m_A-m_H$ increases
and is roughly independent of $\tan\beta$ for $\tan\beta \gtrsim 5$  but becomes larger for
smaller values of $\tan\beta$ (i.e.\ $\tan\beta \lesssim 5$). This implies that the limits
provided in \Cref{fig:xseclimits:lltt,fig:xseclimits:vvbb} for $\tan\beta=10$ are generally applicable for
$\tan\beta \gtrsim 5$, while the limits provided for $\tan\beta=1$ are not applicable for other $\tan\beta$ values.
Upper limits for other $\tan\beta$ values are provided in Figures~\ref{fig:xseclimits:lltt:ggF} to \ref{fig:xseclimits:vvbb:bbA:tanbHi} in the Appendix. Limits for intermediate $\tan\beta$ values can be obtained by interpolating between the limits for the given $\tan\beta$ values.
 
To obtain more realistic limits, instead of using narrow-width $A$ bosons, the signals are generated with a natural width that corresponds to the prediction of the 2HDM for $\tan\beta=1$ in ggF production and $\tan\beta=10$ in bbA production.
For the 2HDM benchmarks considered and the parameter
space that is relevant for this search (see Section~\ref{subsec:2hdm}), the $\tan\beta$ and
Higgs boson mass values are enough to define the $A$ boson width.
The choice $\tan\beta=10$ is made for bbA production because at this value this
production mechanism is dominant
in the benchmark models discussed in Section~\ref{subsec:2hdm} (Type-II and flipped 2HDM).
In the 2HDM benchmarks considered here, the width of the $A$ boson relative
to its mass is a few percent for low $m_A$ values and increases at high $m_A$.
For example, for the $m_A$ range shown in Figures~\ref{fig:limits:lltt:exp} and \ref{fig:limits:lltt:obs}
for the $\lltt$ channel, the $A$ boson width ranges from 4.3\% to 37\% of its mass.

The observed upper limit in the \lltt channel in the case of ggF production varies from
75.0~fb for $(m_A, m_H)=(1200, 600)$~\GeV\ to
992~fb for $(m_A, m_H)=(550, 450)$~\GeV;
this is to be compared with the respective expected limits of 90.8~fb and 582~fb.
The observed upper limit in the \lltt channel in the case of bbA production varies from
79.4~fb for $(m_A, m_H)=(800, 400)$~\GeV\ to 636~fb for $(m_A, m_H)=(650, 450)$~\GeV;
this is to be compared with the respective expected limits of 162~fb and 257~fb.
Similarly, for the \vvbb channel, the observed upper limit in the case of ggF production varies from
6.2~fb for $(m_A, m_H)=(1200, 300)$~\GeV\ to
3700~fb for $(m_A,m_H)=(350, 150)$~\GeV;
this is to be compared with the respective expected limits of 10.6~fb and 3520~fb.
Finally, for the \vvbb channel, the observed upper limit in the case of bbA production varies from
3.62~fb for $(m_A, m_H)=(1200, 200)$~\GeV\ to
1750~fb for $(m_A, m_H)=(350, 150)$~\GeV;
this is to be compared with the respective expected limits of 9.92~fb and 1910~fb.

\subsection{Interpretation in the context of 2HDM} \label{subsec:2hdm}
 
The upper limits shown in Section~\ref{subsec:limits} are interpreted in the context of the CP-conserving 2HDM.
For this interpretation, several assumptions are made to reduce the number of free parameters in the model.
The $H^{\pm}$ bosons are assumed to have the same mass as the $A$ boson, whereas $m_H < m_A$ is assumed for the masses of the $A$ and $H$ bosons.
The 2HDM parameter $m^2_{12}$ is fixed to $m^2_{A}\tan\beta/(1+\tan^2\beta)$.
The $h$ boson is assumed to have a mass of $125~\GeV$ and its couplings to fermions and vector bosons are set to be the same as those of the SM Higgs boson at lowest order by choosing $\cos(\beta-\alpha)=0$.
The widths of the $A$ and $H$ bosons are taken from the predictions of the 2HDM.
These assumptions leave three free parameters: $m_A$, $m_H$ and $\tan\beta$.
In addition, there are four possible arrangements of the Yukawa couplings, which are known as type-I, type-II, lepton specific and flipped 2HDM.
For the parameter space that is relevant for this search, the widths of the $A$ and $H$ bosons differ very little across the different 2HDM types in comparison with the experimental mass resolution.
In the same parameter space, the $A$ boson width is larger than the $H$ boson width, so the quoted limits from this search cannot be interpreted as limits for the $H\to ZA$ process.
The cross-sections for $A$ boson production in the 2HDM are
calculated with corrections up to NNLO in QCD for ggF and
bbA production in the five-flavour scheme as implemented in
{\texttt SusHi}~\cite{Harlander:2012pb,Harlander:2005rq,Harlander:2003ai,Harlander:2002wh}.
For bbA production, a cross-section in the four-flavour
scheme is also calculated as described in Refs.~\cite{Dawson:2003kb,Dittmaier:2003ej} and the results are combined with the five-flavour scheme calculation following Ref.~\cite{Harlander:2011aa}.
The Higgs boson branching
ratios are calculated using {\texttt 2HDMC}~\cite{Eriksson:2009ws}.
The procedure used to calculate the cross-sections and branching ratios, as well as to choose the 2HDM parameter values, follows Ref.~\cite{deFlorian:2016spz}.
 
\begin{figure}[ht]
\centering
\subfigure[]{\includegraphics[width=0.48\textwidth]{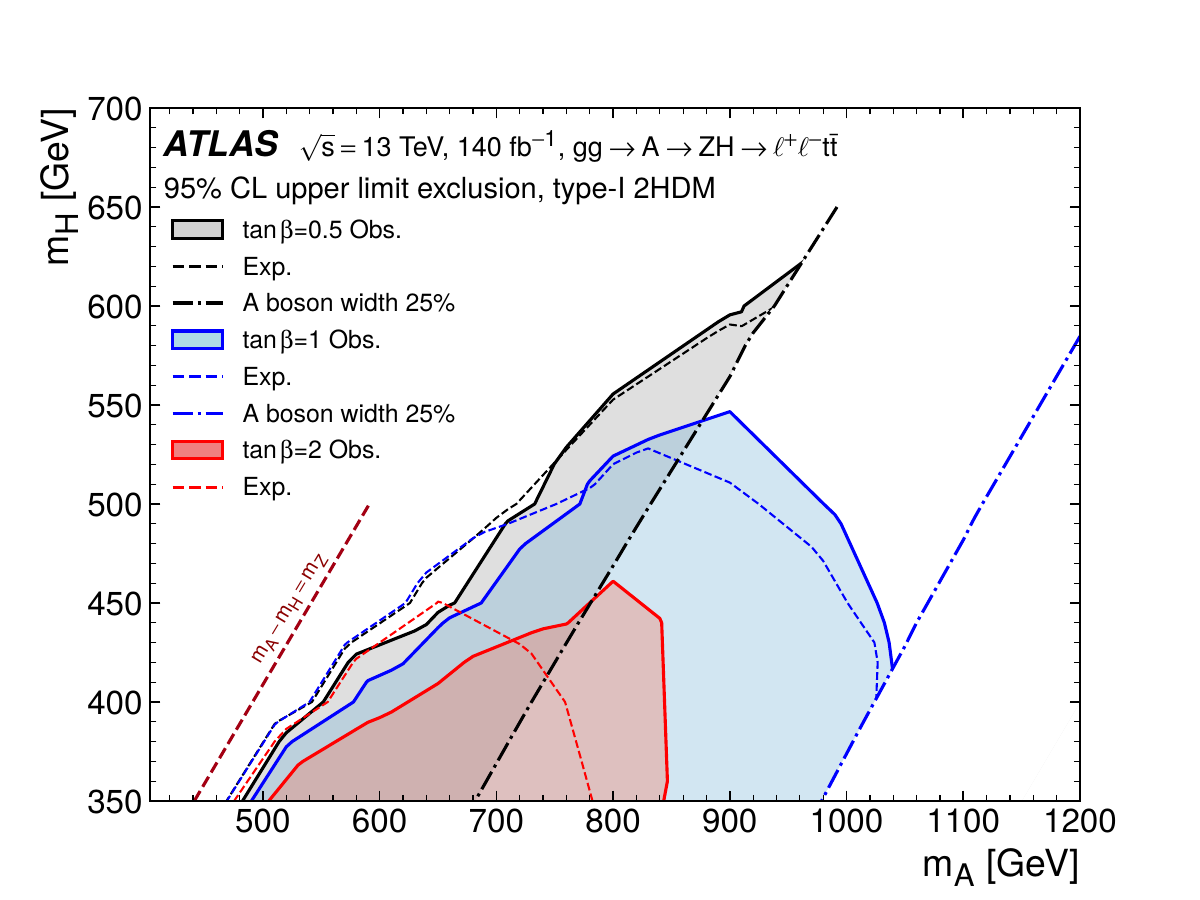}\label{fig:2HDM:lltt}}
\subfigure[]{\includegraphics[width=0.48\textwidth]{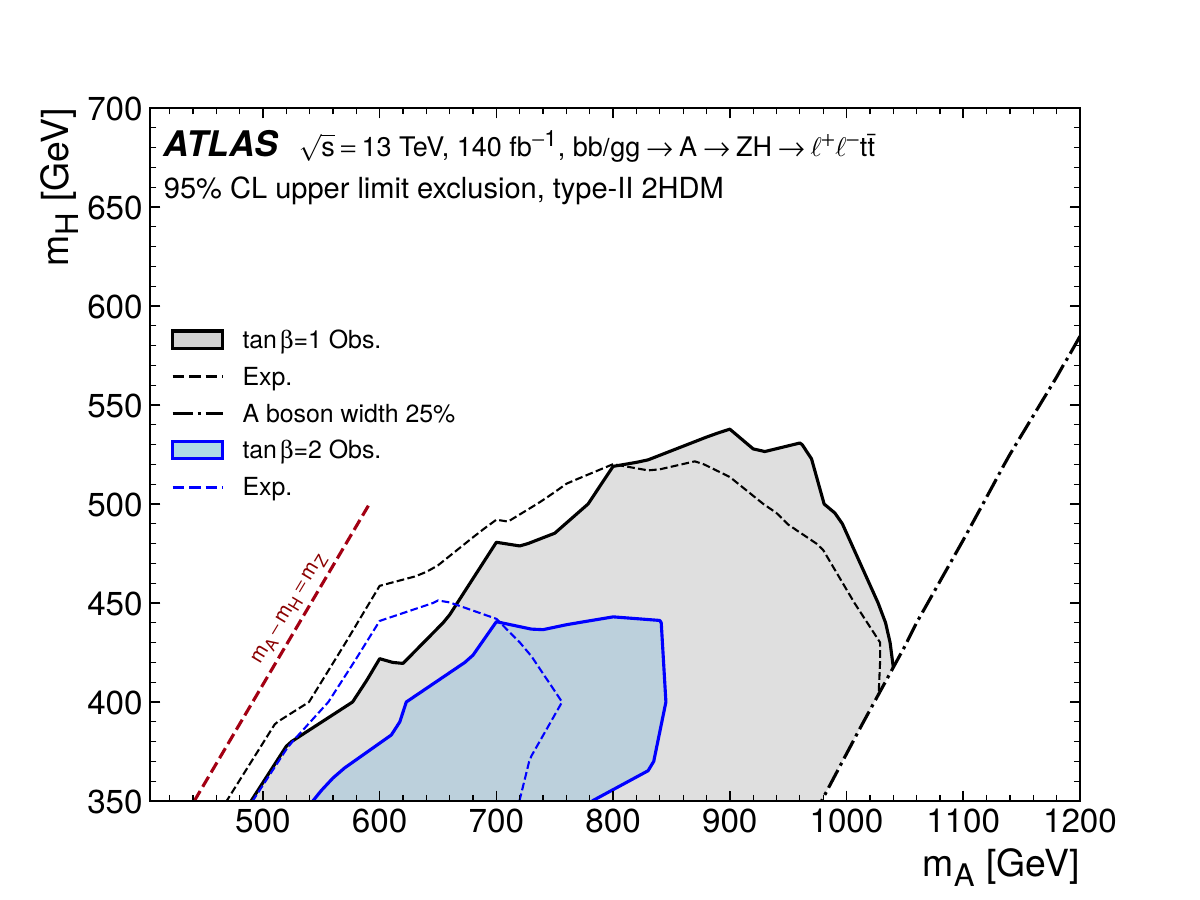}\label{fig:2HDM:lltt:t2}}
\subfigure[]{\includegraphics[width=0.48\textwidth]{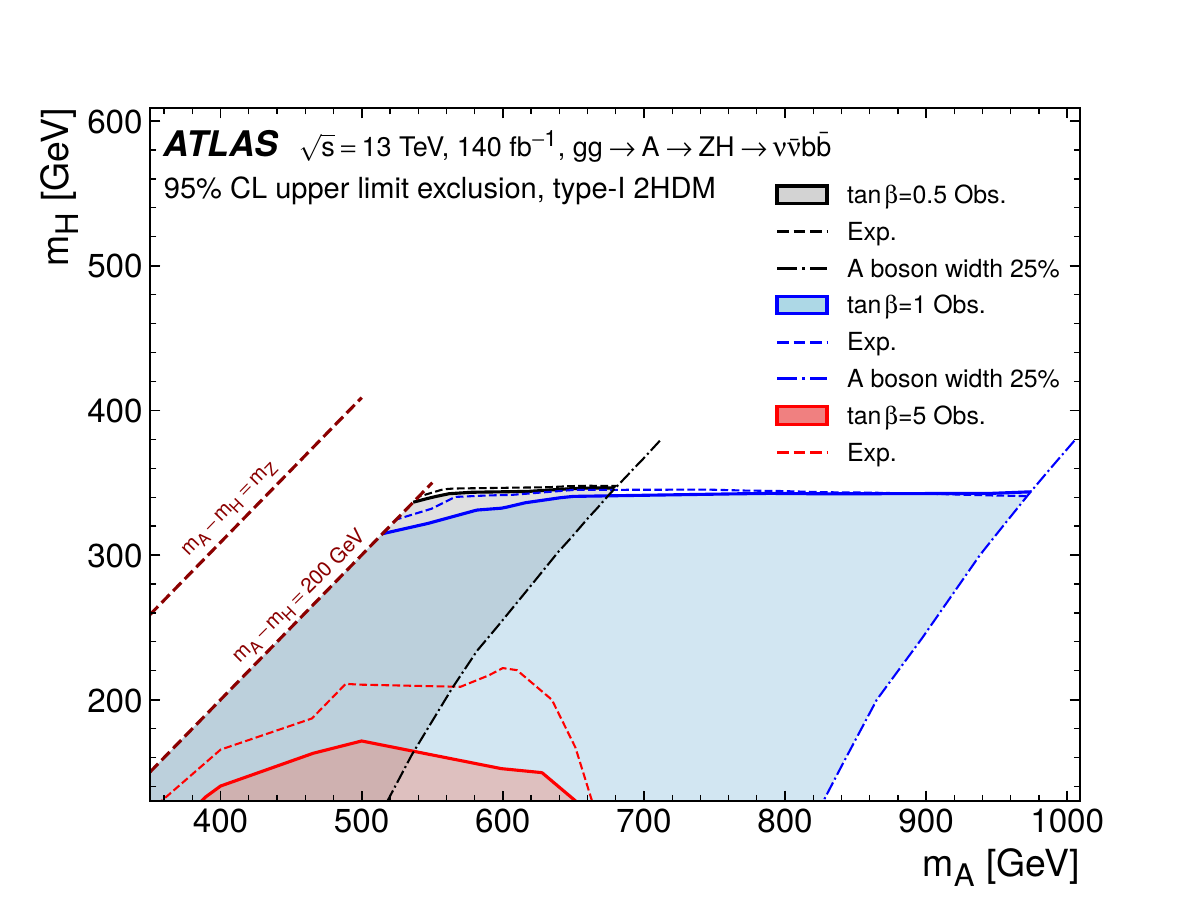}\label{fig:2HDM:vvbb:t1}}
\subfigure[]{\includegraphics[width=0.48\textwidth]{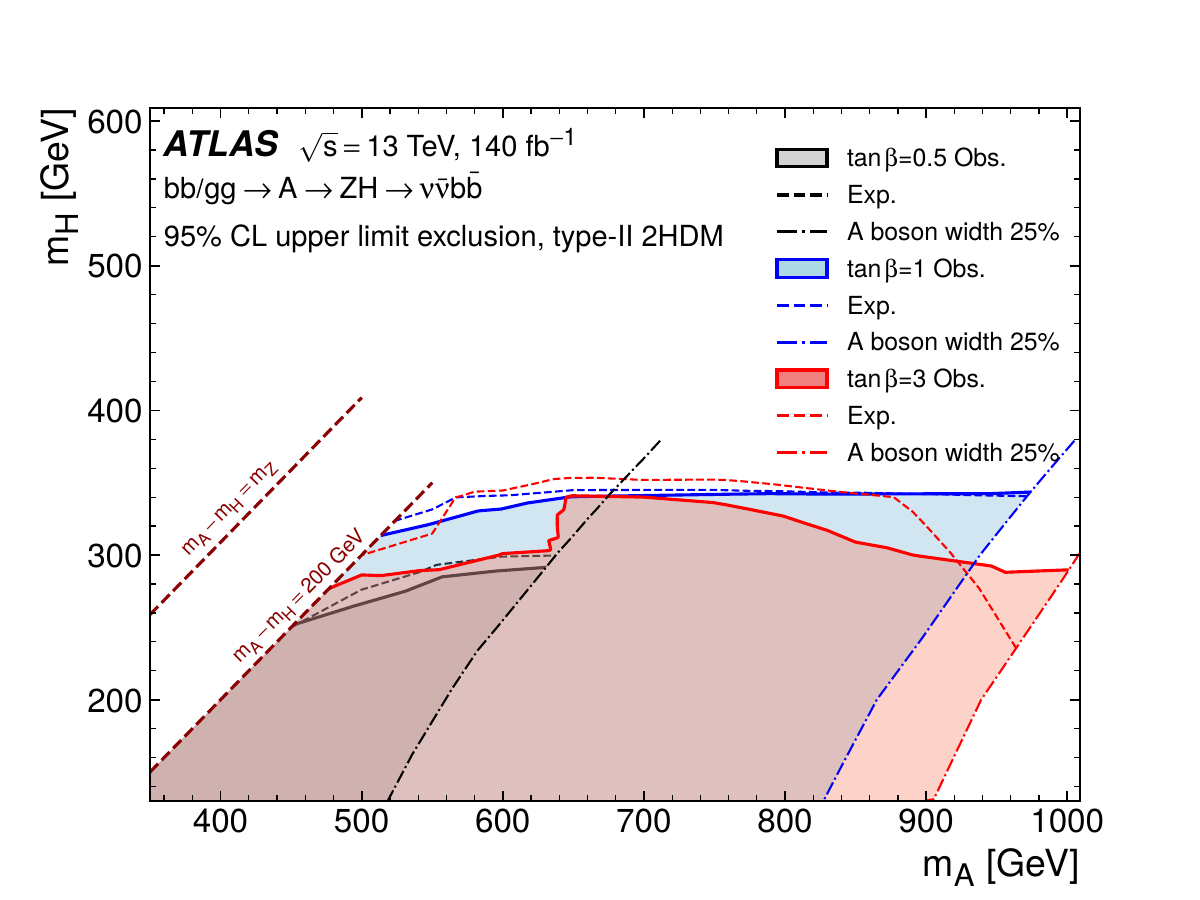}\label{fig:2HDM:vvbb:t2_tb_lo}}
\subfigure[]{\includegraphics[width=0.48\textwidth]{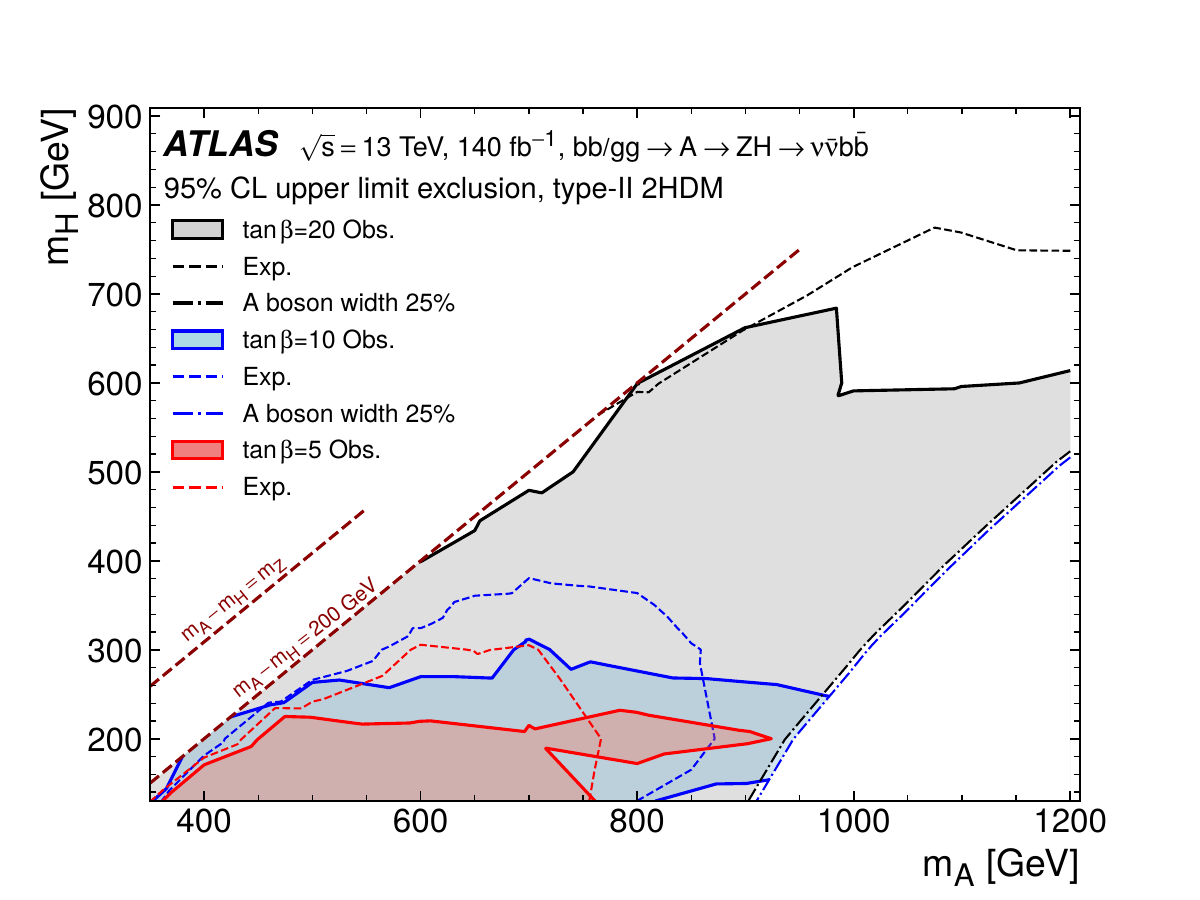}\label{fig:2HDM:vvbb:t2_tb_hi}}
\caption{Observed and expected 95\% CL exclusion regions in the $(m_A, m_H)$ plane for various $\tan\beta$ values for the \lltt channel, type-I (a) and type-II (b) 2HDM, and \vvbb channel, type-I (c) and type-II (d,e) 2HDM.
The line at $m_A-m_H=200$~\GeV\ shown in (c--e) corresponds to the edge of the analysis sensitivity due to the \met requirement.}
\label{fig:2HDM}
\end{figure}

The upper limits are interpreted as constraints in the $(m_A,m_H)$ plane for several $\tan\beta$
values. The widths of the $A$ and $H$ bosons change as a function of  $\tan\beta$ and these
variations are taken into account when calculating the constraints. The results are quoted
only for cases in which the width of the $A$ boson is no more than 25\% of $m_A$.
Figures~\ref{fig:2HDM:lltt} and \ref{fig:2HDM:lltt:t2} show
the constraints from the \lltt channel for the type-I and type-II 2HDM, respectively. Constraints from this channel
for the lepton-specific 2HDM are very similar to type-I. Constraints from the \vvbb
channel are shown in Figure~\ref{fig:2HDM:vvbb:t1} for the type-I 2HDM, and in Figures~\ref{fig:2HDM:vvbb:t2_tb_lo} and \ref{fig:2HDM:vvbb:t2_tb_hi} for the type-II 2HDM.
These results extend the reach of the $A\to ZH\to \ell\ell b\bar{b}$ search reported in Ref.~\cite{HDBS-2018-13}, especially in parts of the parameter space with $m_H > 350$ GeV and $m_A > 800$ GeV.
 
The fact that the observed exclusion limit for the \lltt channel is smaller than the expected exclusion limit in the region around $(m_A,m_H)=(650,450)$~\GeV\ in \Cref{fig:2HDM:lltt,fig:2HDM:lltt:t2} is a consequence of the small data excess observed in \Cref{fig:postfit_fitvar:lltt}.
Similarly, in the \vvbb channel, at low $\tan\beta$ in \Cref{fig:2HDM:vvbb:t2_tb_lo} the expected and observed limits diverge because of a small excess of events for $550~\GeV \lesssim \mtvh \lesssim 650~\GeV$.
Finally, differences seen between the limits at high $\tan\beta$ are affected by small excesses or deficits in the data; in particular, the smaller observed exclusion limit in the region with $(m_A,m_H)=(600,300)$~\GeV\ is due to the excess observed in \Cref{fig:postfit_fitvar:vvbb:3ptag}.
 
\FloatBarrier
 
\subsection{Model-independent limits}
 
The upper limits presented above are based on the assumption that the $Z$ and $H$ candidates in the final state are produced resonantly and therefore have a significant model dependence. In addition to the 2HDM interpretation, limits with less model dependence are derived with a slight modification of the fit model as follows. Assuming that a resonance \(X\) that decays into a $\ttbar$ or $b\bar{b}$ pair is produced in association with a $Z$ boson, the number of signal events that will be recorded by the detector in a given bin of the reconstructed $m(\ttbar)$ or $m(b\bar{b})$ distribution will be equal to the integrated luminosity times the `visible cross-section' $\sigma_{\mathrm{vis}}(\ZllXtt)$ or $\sigma_{\mathrm{vis}}(\ZvvXbb)$.
 
\begin{figure}[ht]
\centering
\subfigure[]{\includegraphics[width=0.4\textwidth]{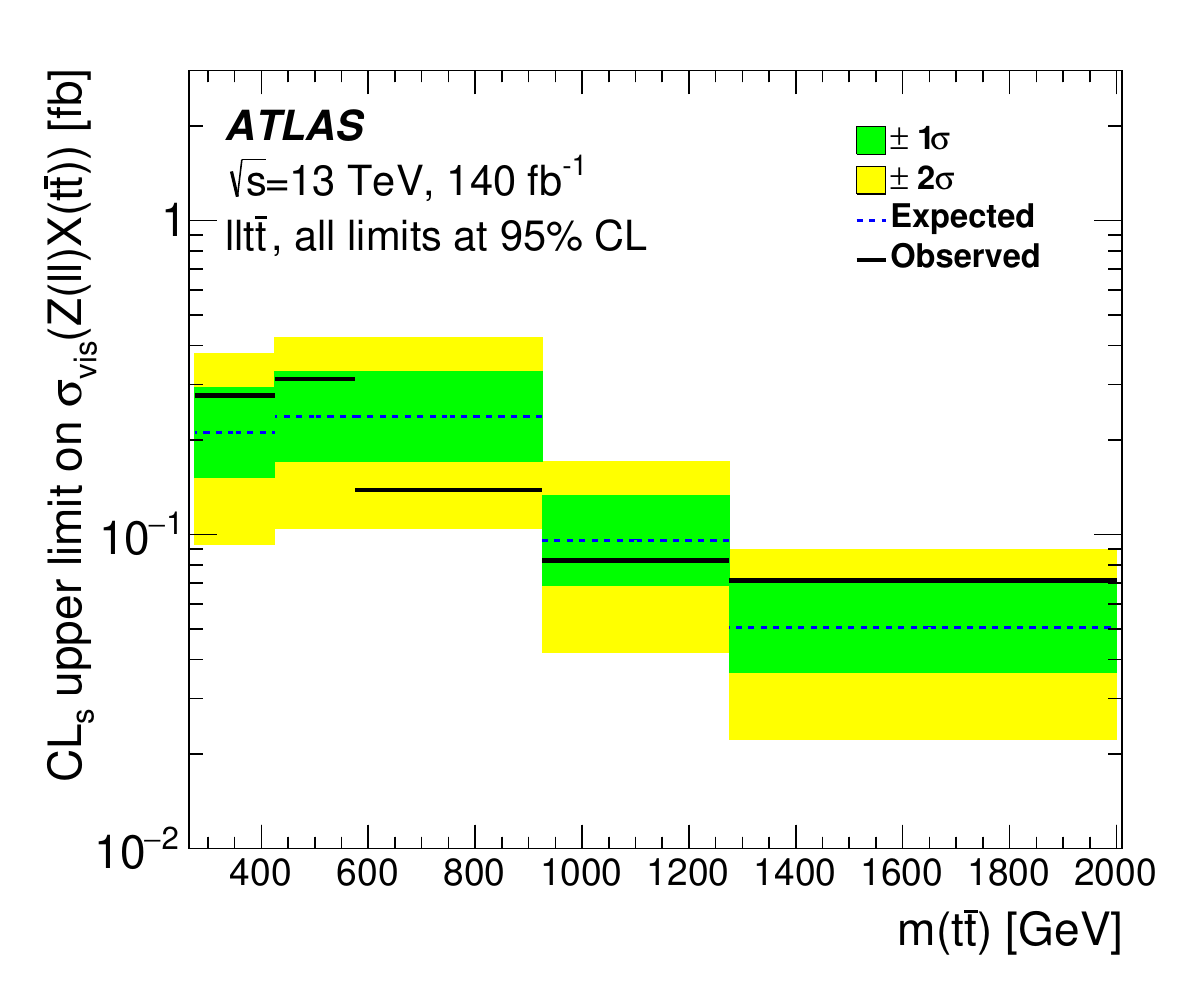}\label{fig:MIL:lltt}}
\subfigure[]{\includegraphics[width=0.4\textwidth]{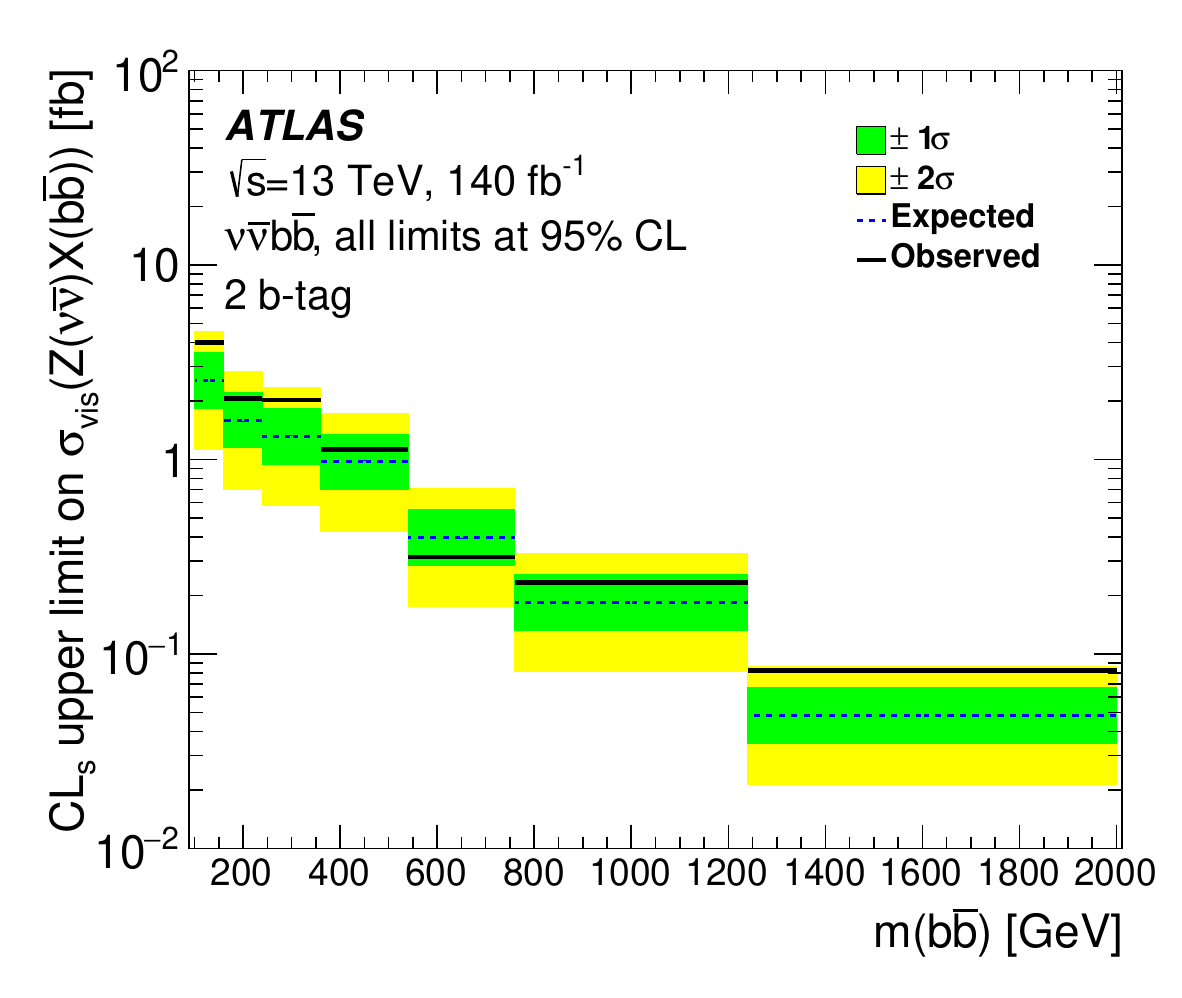}\label{fig:MIL:vvbb_2b}}
\subfigure[]{\includegraphics[width=0.4\textwidth]{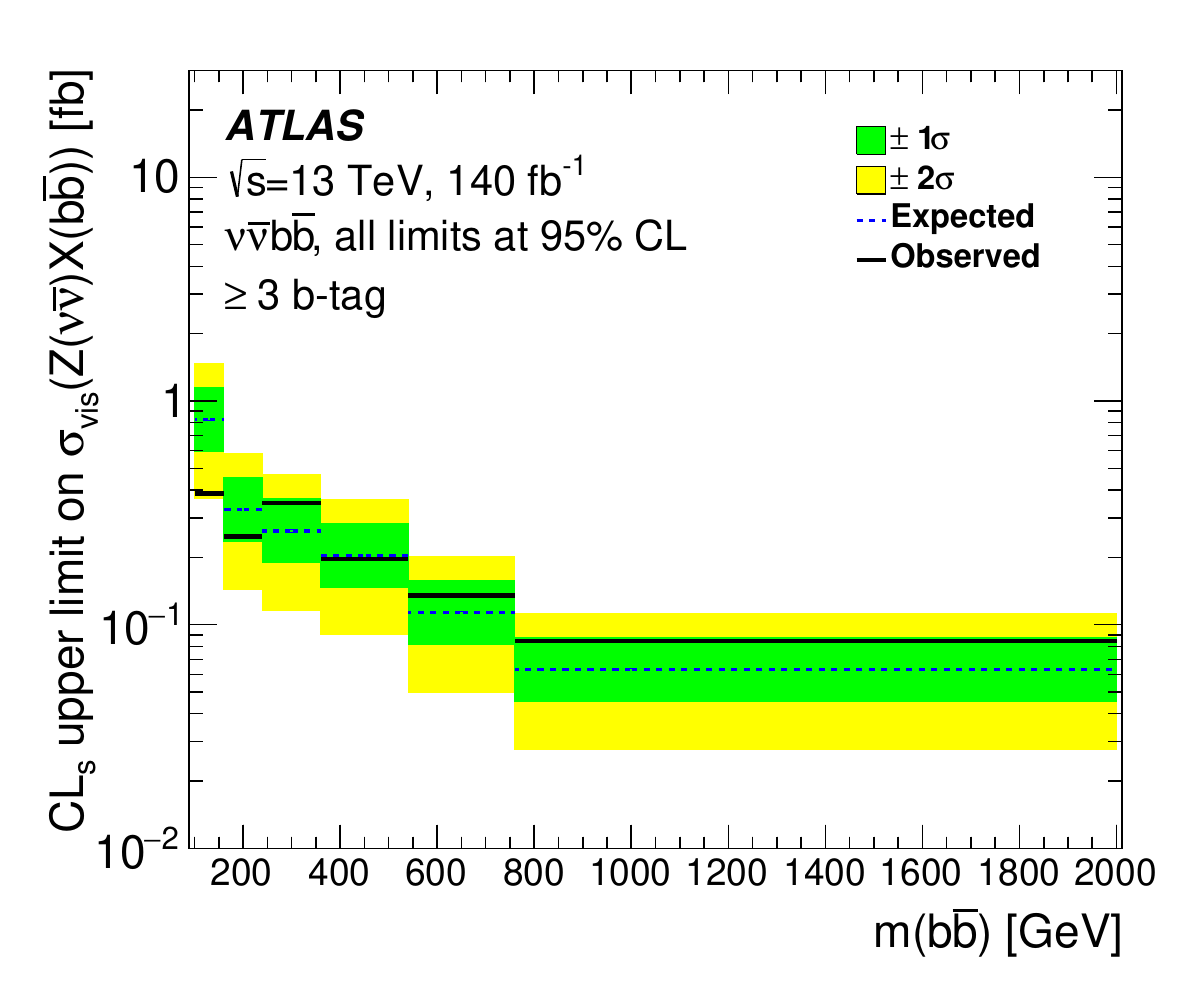}\label{fig:MIL:vvbb_3b}}
\caption{Observed (solid line) and expected (dashed line) 95\% CL upper limits on \(\sigma_{\mathrm{vis}}(\ZllXtt)\) (a) and \(\sigma_{\mathrm{vis}}(\ZvvXbb)\) obtained from the \twotag (b) and \threetag (c) regions. The limits are obtained in bins of reconstructed \(m(\ttbar)\) and \(m(\bbbar)\). The green and yellow bands correspond to the \(\pm1\,\sigma\) and \(\pm2\,\sigma\) intervals for the expected upper limits, respectively.}
\label{fig:MIL}
\end{figure}

Upper limits at 95\% CL are placed on $\sigma_{\mathrm{vis}}(\ZllXtt)$ and $\sigma_{\mathrm{vis}}(\ZvvXbb)$ using the fit model described in Section~\ref{sec:fit}, except that the $m(\ttbar)$ or $m(b\bar{b})$ distributions for the events that pass the SR selection are fitted using three bins. The signal template is constructed by adding a single signal event in the central bin of the $m(\ttbar)$ or $m(b\bar{b})$ distribution, with the adjacent bins used as CRs. The corresponding \fake region or \emu and \twoL regions are also used as CRs in this fit without any modification. Using the $m(\ttbar)$ or $m(b\bar{b})$ distributions instead of the $\Delta m$ or \mtvh distributions in the fit ensures that the limit thus obtained is independent of how the $Z$ and $H$ candidates are produced, and using a large bin that contains all of the signal events ensures that the limit does not depend strongly on the lineshape of the \ttbar or $b\bar{b}$ resonance.\footnote{For signals that predict a \ttbar or $b\bar{b}$ resonance with a mass that lies between the bin edges in Figure~\ref{fig:MIL}, the limits from all bins in which the signal contributes have to be combined, taking into account their respective acceptances.}
 
To obtain the upper limits on the visible cross-section as functions of $m(\ttbar)$ and $m(b\bar{b})$, multiple independent fits are performed, each time using different SRs, defined by the bin edges shown in \Cref{fig:MIL}.
These limits can be used to estimate sensitivities for theories involving high-mass $t\bar{t}$ or $b\bar{b}$ resonances, by comparing the upper limit shown in \Cref{fig:MIL} with the visible cross-section predicted by a specific theory, given by
\begin{equation*}
\sigma_{\text{vis}}^{\mathrm{theory}}=\sigma_{\text{theory}}\times B \times (\mathcal{A}\cdot\epsilon)_{m(\ttbar)/m(b\bar{b})},
\end{equation*}
where $\sigma_{\text{theory}}$ is the inclusive signal cross-section, $B$ is the product of the branching ratios for the decay chain and $(\mathcal{A}\cdot\epsilon)_{m(\ttbar)/m(b\bar{b})}$ is the acceptance times efficiency for reconstructing a signal-model event in a given bin of the $m(b\bar{b})$ or $m(\ttbar)$ distribution.
The use of the limits shown in \Cref{fig:MIL} requires the value of $(\mathcal{A}\cdot\epsilon)_{m(\ttbar)/m(b\bar{b})}$ for a given signal model to be obtained from a Monte Carlo `truth'-level analysis that reproduces the event selection described in Section~\ref{sec:selection}, incorporating the detector effects, for example via fast simulation packages such as \textsc{Delphes}~\cite{deFavereau:2013fsa} or smearing routines such as the ones provided in the \textsc{Rivet} framework~\cite{Buckley:2019stt,Bierlich:2019rhm}.
The applicability of these limits is not guaranteed outside the phase space in which $(\mathcal{A}\cdot\epsilon)$ values are provided, or when the derived $(\mathcal{A}\cdot\epsilon)$ values differ significantly from the values provided in this publication.
 
\FloatBarrier
 
\section{Conclusion}
\label{sec:conclusion}

A search for the decay of a heavy CP-odd Higgs boson $A$ into a heavy CP-even Higgs boson $H$ is presented, using the full LHC Run~2 dataset of 13~\TeV\ proton--proton collisions corresponding to an integrated luminosity of 140~fb$^{-1}$ recorded by the ATLAS detector. Two channels are considered in the present search. The \lltt channel, which probes $H$ boson decays into a pair of top quarks and $Z$ boson decays into a pair of charged leptons, provides sensitivity in the $m_H>350$~\GeV\ region. The \vvbb channel, which probes $H$ boson decays into a pair of $b$-quarks and $Z$ boson decays into neutrinos, extends the sensitivity of previous searches in parts of the parameter space with $m_A>800$ GeV and $m_H<350$ GeV.
 
No significant excess over the SM background prediction is observed.
Upper limits are set at the 95\% confidence level on the production cross-section times the decay branching ratios.
These results are interpreted in the context of the two-Higgs-doublet models and yield an exclusion range significantly larger than those obtained by previous searches sensitive to this parameter space.
In addition, upper limits are quoted  as a function of reconstructed $m(\ttbar)$ and $m(\bbbar)$, assuming the production of a \ttbar or \bbbar resonance.


\section*{Acknowledgements}


We thank CERN for the very successful operation of the LHC, as well as the
support staff from our institutions without whom ATLAS could not be
operated efficiently.
 
We acknowledge the support of
ANPCyT, Argentina;
YerPhI, Armenia;
ARC, Australia;
BMWFW and FWF, Austria;
ANAS, Azerbaijan;
CNPq and FAPESP, Brazil;
NSERC, NRC and CFI, Canada;
CERN;
ANID, Chile;
CAS, MOST and NSFC, China;
Minciencias, Colombia;
MEYS CR, Czech Republic;
DNRF and DNSRC, Denmark;
IN2P3-CNRS and CEA-DRF/IRFU, France;
SRNSFG, Georgia;
BMBF, HGF and MPG, Germany;
GSRI, Greece;
RGC and Hong Kong SAR, China;
ISF and Benoziyo Center, Israel;
INFN, Italy;
MEXT and JSPS, Japan;
CNRST, Morocco;
NWO, Netherlands;
RCN, Norway;
MEiN, Poland;
FCT, Portugal;
MNE/IFA, Romania;
MESTD, Serbia;
MSSR, Slovakia;
ARRS and MIZ\v{S}, Slovenia;
DSI/NRF, South Africa;
MICINN, Spain;
SRC and Wallenberg Foundation, Sweden;
SERI, SNSF and Cantons of Bern and Geneva, Switzerland;
MOST, Taiwan;
TENMAK, T\"urkiye;
STFC, United Kingdom;
DOE and NSF, United States of America.
In addition, individual groups and members have received support from
BCKDF, CANARIE, Compute Canada and CRC, Canada;
PRIMUS 21/SCI/017 and UNCE SCI/013, Czech Republic;
COST, ERC, ERDF, Horizon 2020, ICSC-NextGenerationEU and Marie Sk{\l}odowska-Curie Actions, European Union;
Investissements d'Avenir Labex, Investissements d'Avenir Idex and ANR, France;
DFG and AvH Foundation, Germany;
Herakleitos, Thales and Aristeia programmes co-financed by EU-ESF and the Greek NSRF, Greece;
BSF-NSF and MINERVA, Israel;
Norwegian Financial Mechanism 2014-2021, Norway;
NCN and NAWA, Poland;
La Caixa Banking Foundation, CERCA Programme Generalitat de Catalunya and PROMETEO and GenT Programmes Generalitat Valenciana, Spain;
G\"{o}ran Gustafssons Stiftelse, Sweden;
The Royal Society and Leverhulme Trust, United Kingdom.
 
The crucial computing support from all WLCG partners is acknowledged gratefully, in particular from CERN, the ATLAS Tier-1 facilities at TRIUMF (Canada), NDGF (Denmark, Norway, Sweden), CC-IN2P3 (France), KIT/GridKA (Germany), INFN-CNAF (Italy), NL-T1 (Netherlands), PIC (Spain), ASGC (Taiwan), RAL (UK) and BNL (USA), the Tier-2 facilities worldwide and large non-WLCG resource providers. Major contributors of computing resources are listed in Ref.~\cite{ATL-SOFT-PUB-2023-001}.


\clearpage
\appendix
\part*{Upper limits on the production cross sections for additional $\tan\beta$ values}
\addcontentsline{toc}{part}{Upper limits on the production cross sections for additional $\tan\beta$ values}

\begin{figure}[ht]
\centering
\subfigure[]{\includegraphics[width=0.48\textwidth]{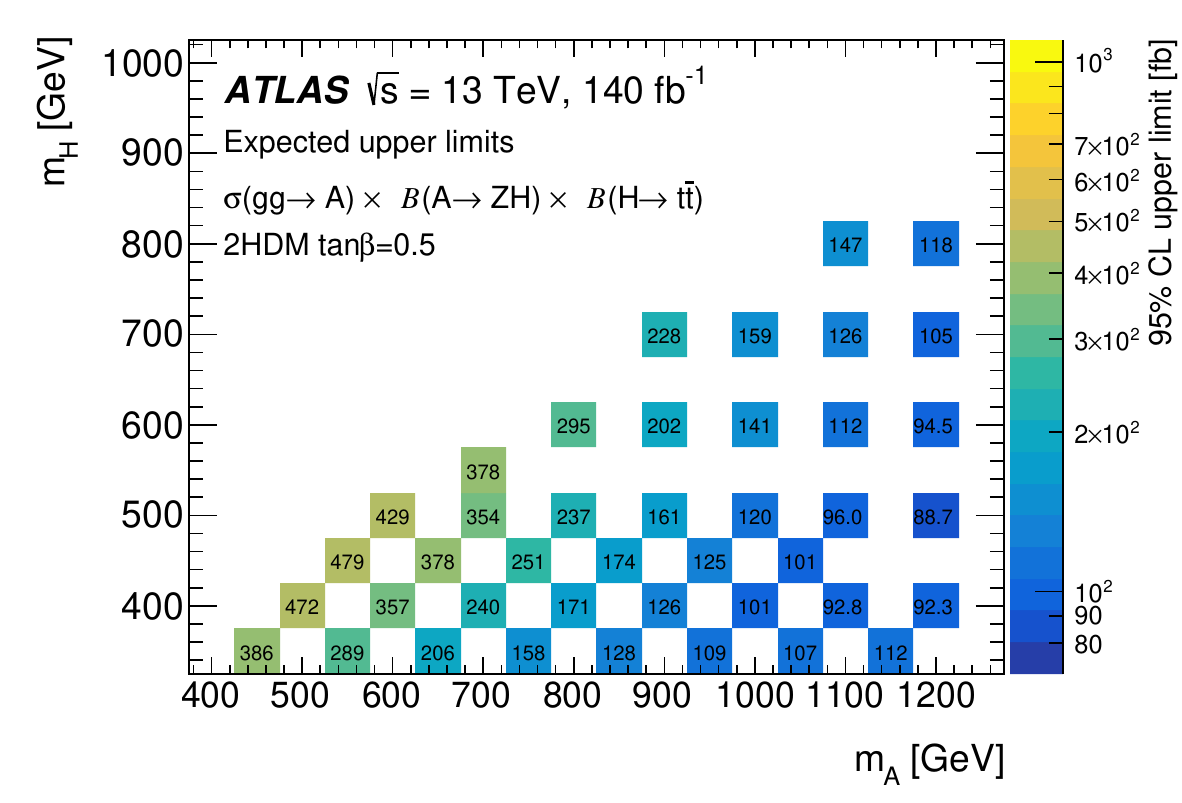}}
\subfigure[]{\includegraphics[width=0.48\textwidth]{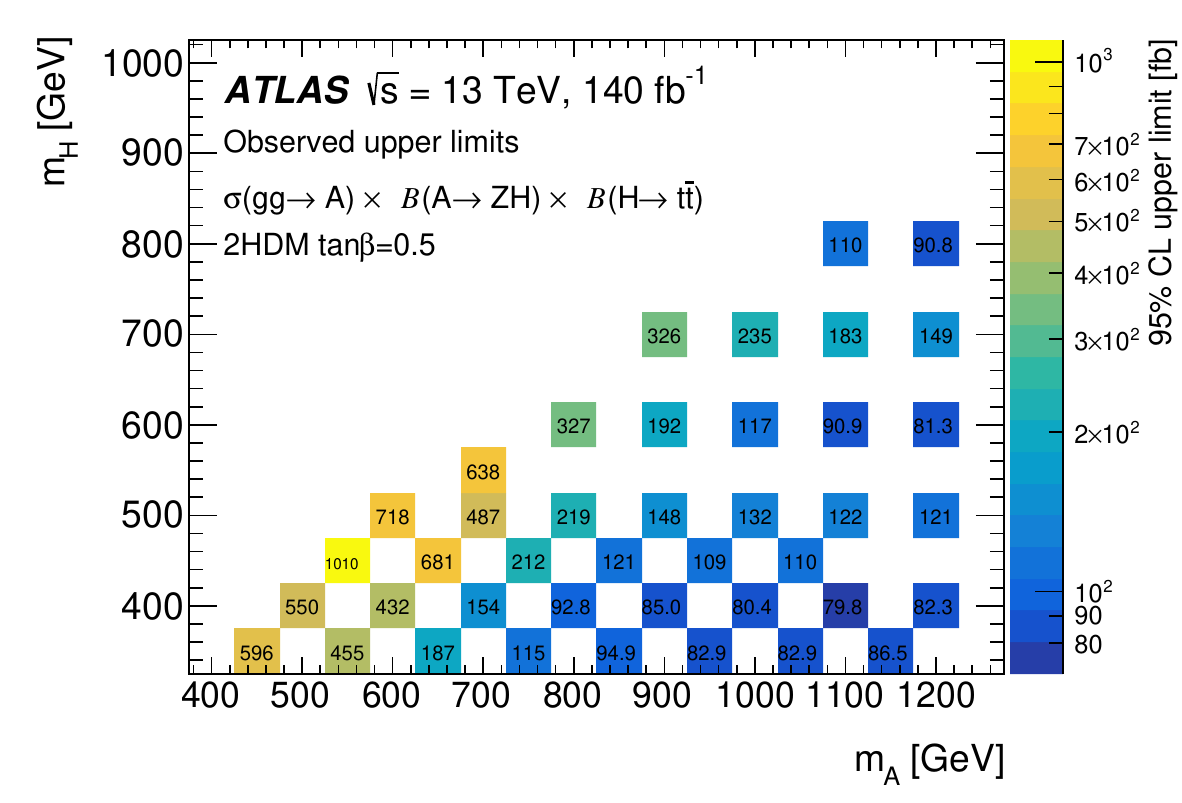}}
\subfigure[]{\includegraphics[width=0.48\textwidth]{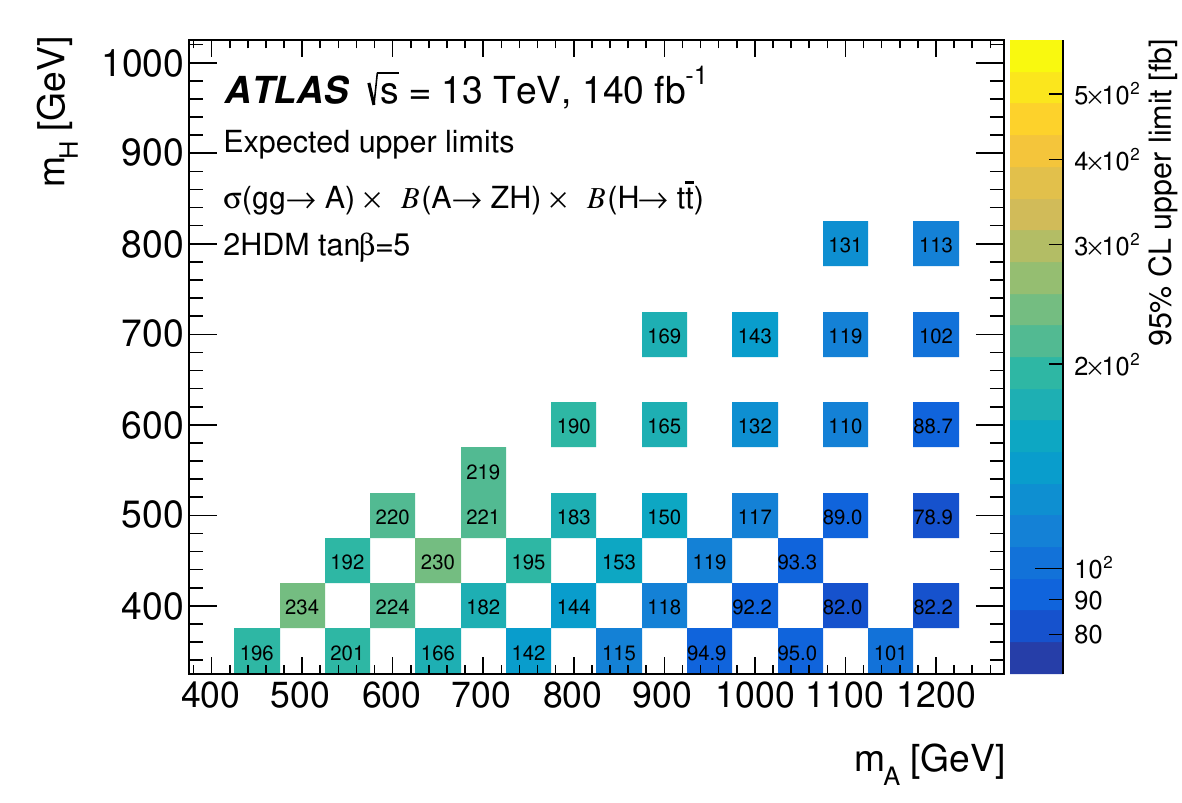}}
\subfigure[]{\includegraphics[width=0.48\textwidth]{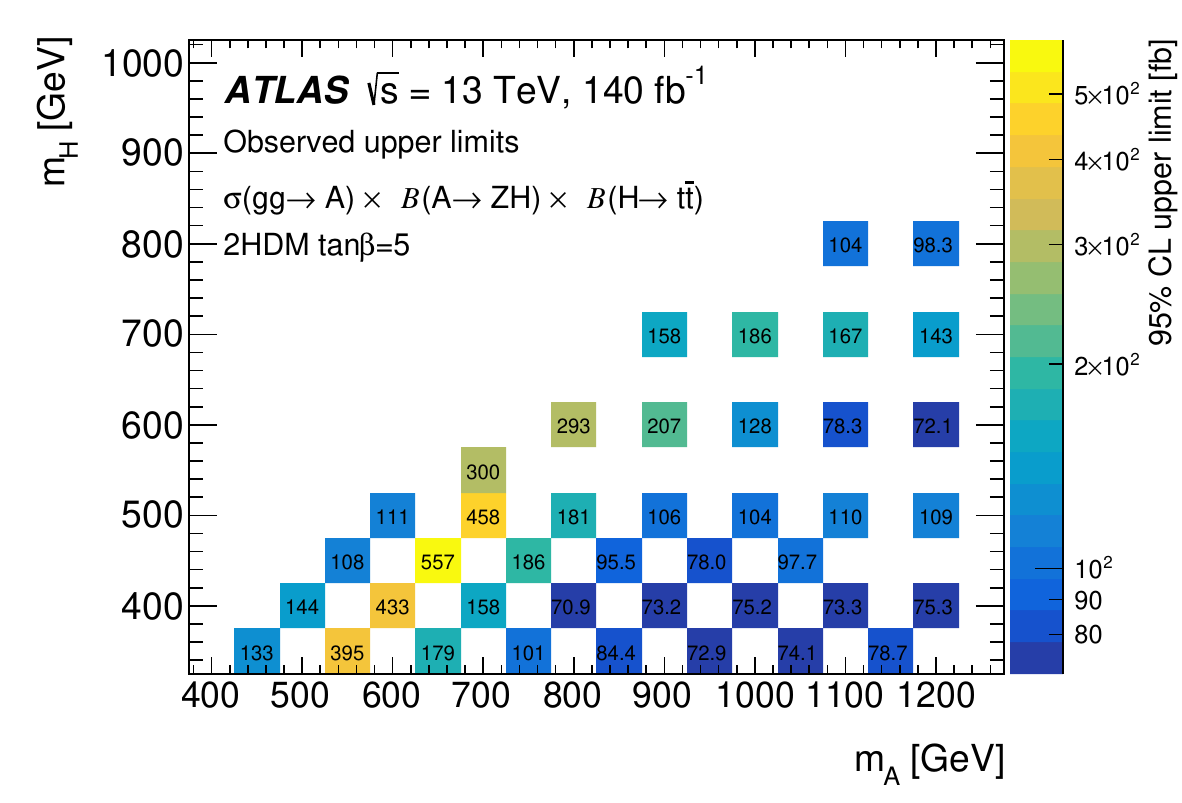}}
\caption{Expected and observed upper limits at 95\% CL on $\sigma(gg\to A)\times B(A\to ZH)\times B(H\to \ttbar)$. The limits are shown for $\tan\beta=0.5$ (a,b) and $\tan\beta=5$ (c,d). The \(\tan\beta\) value is relevant only for the choice of $A$ boson width.}
\label{fig:xseclimits:lltt:ggF}
\end{figure}
 
\begin{figure}[ht]
\centering
\subfigure[]{\includegraphics[width=0.48\textwidth]{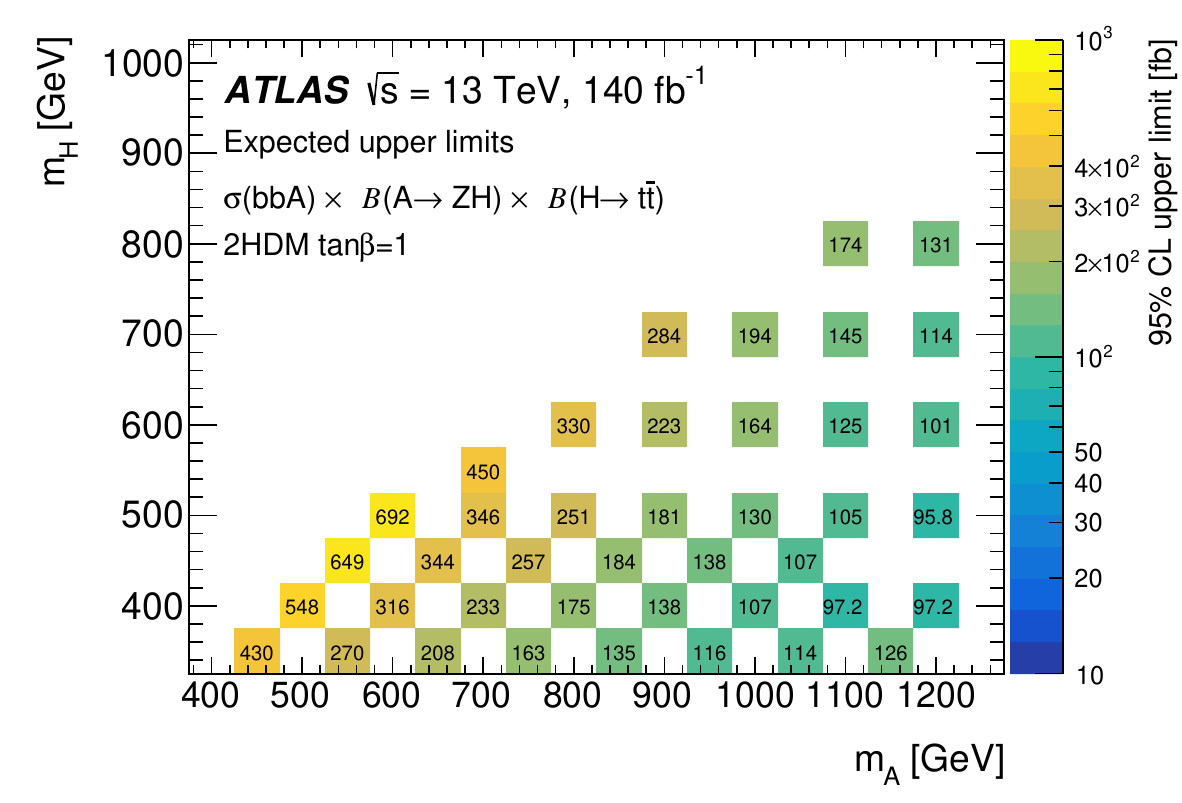}}
\subfigure[]{\includegraphics[width=0.48\textwidth]{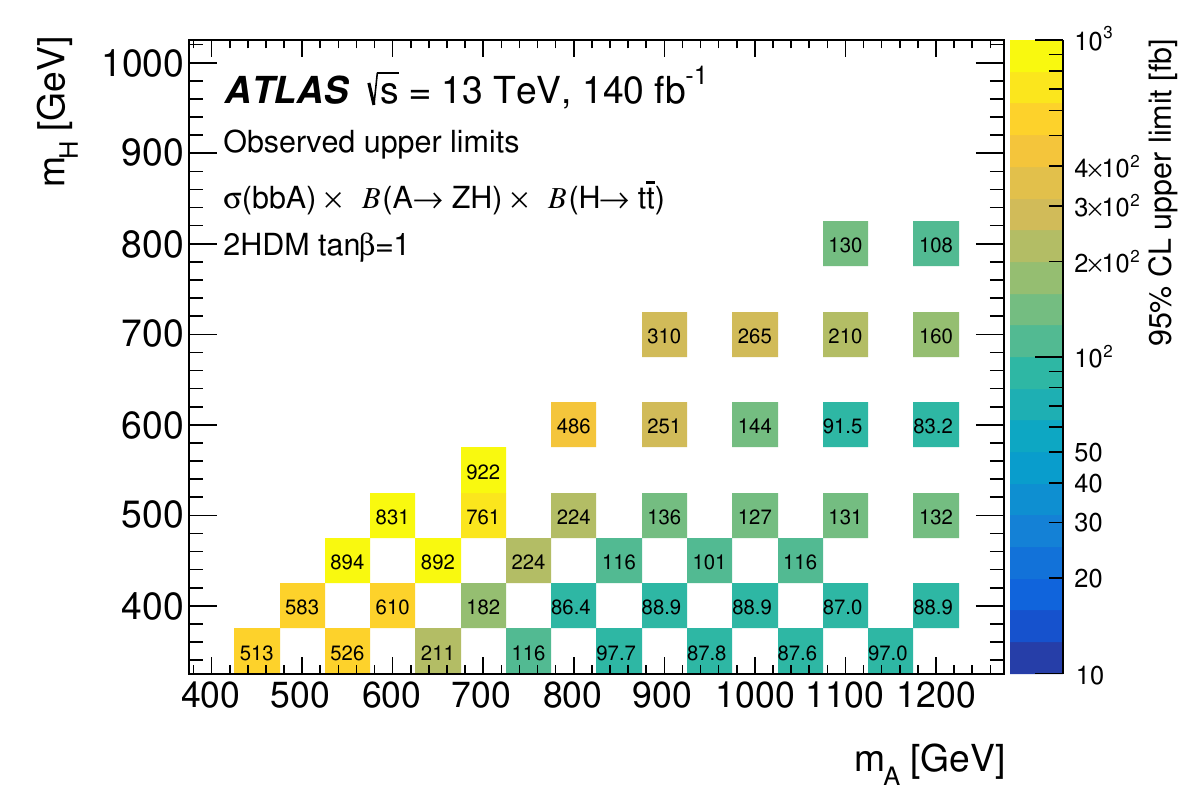}}
\subfigure[]{\includegraphics[width=0.48\textwidth]{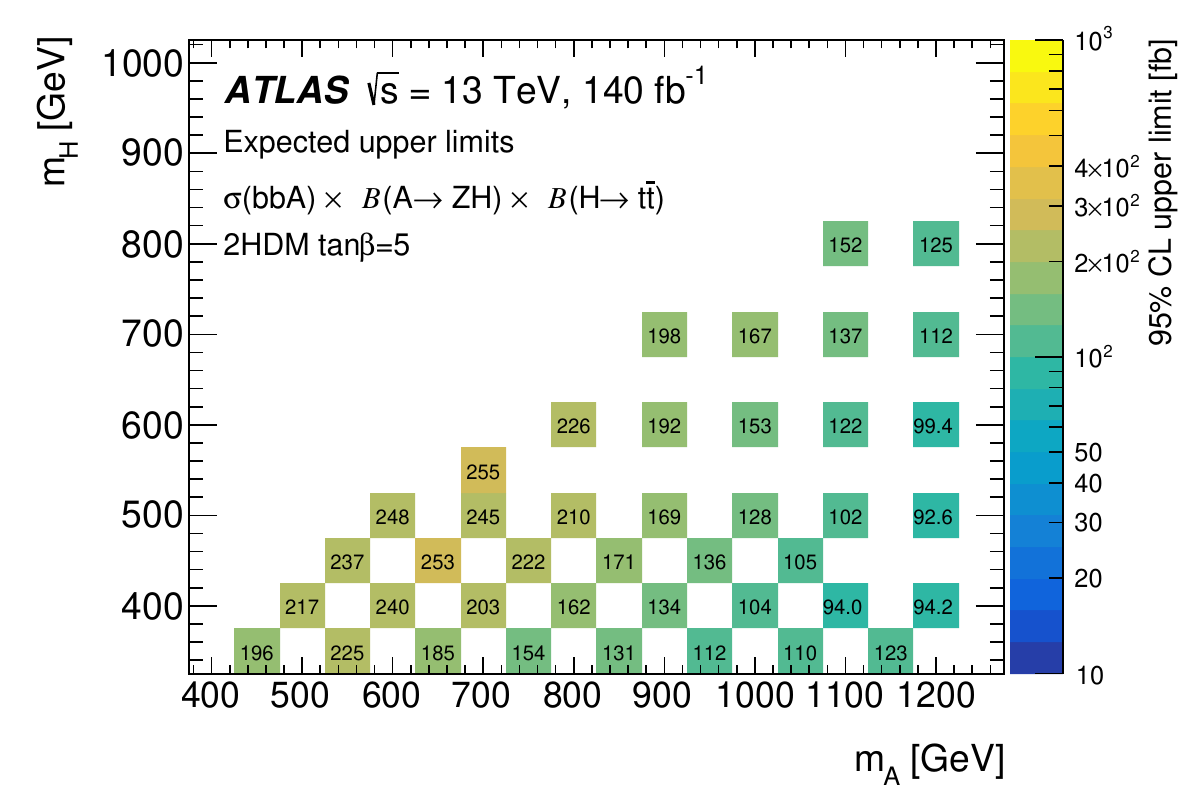}}
\subfigure[]{\includegraphics[width=0.48\textwidth]{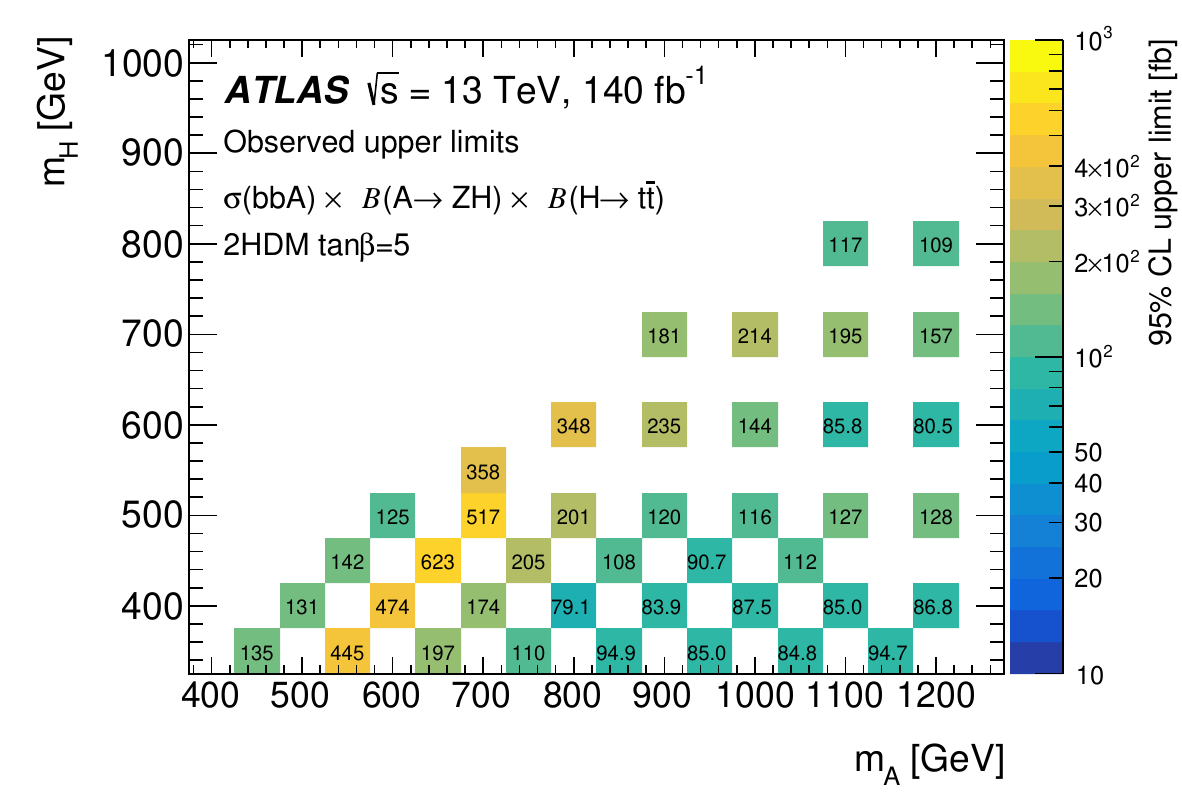}}
\caption{Expected and observed upper limits at 95\% CL on $\sigma(b\bbar A)\times B(A\to ZH)\times B(H\to \ttbar)$. The limits are shown for $\tan\beta=1$ (a,b) and $\tan\beta=5$ (c,d). The \(\tan\beta\) value is relevant only for the choice of $A$ boson width.}
\label{fig:xseclimits:lltt:bbA:tanbHi}
\end{figure}
 
\begin{figure}[ht]
\centering
\subfigure[]{\includegraphics[width=0.48\textwidth]{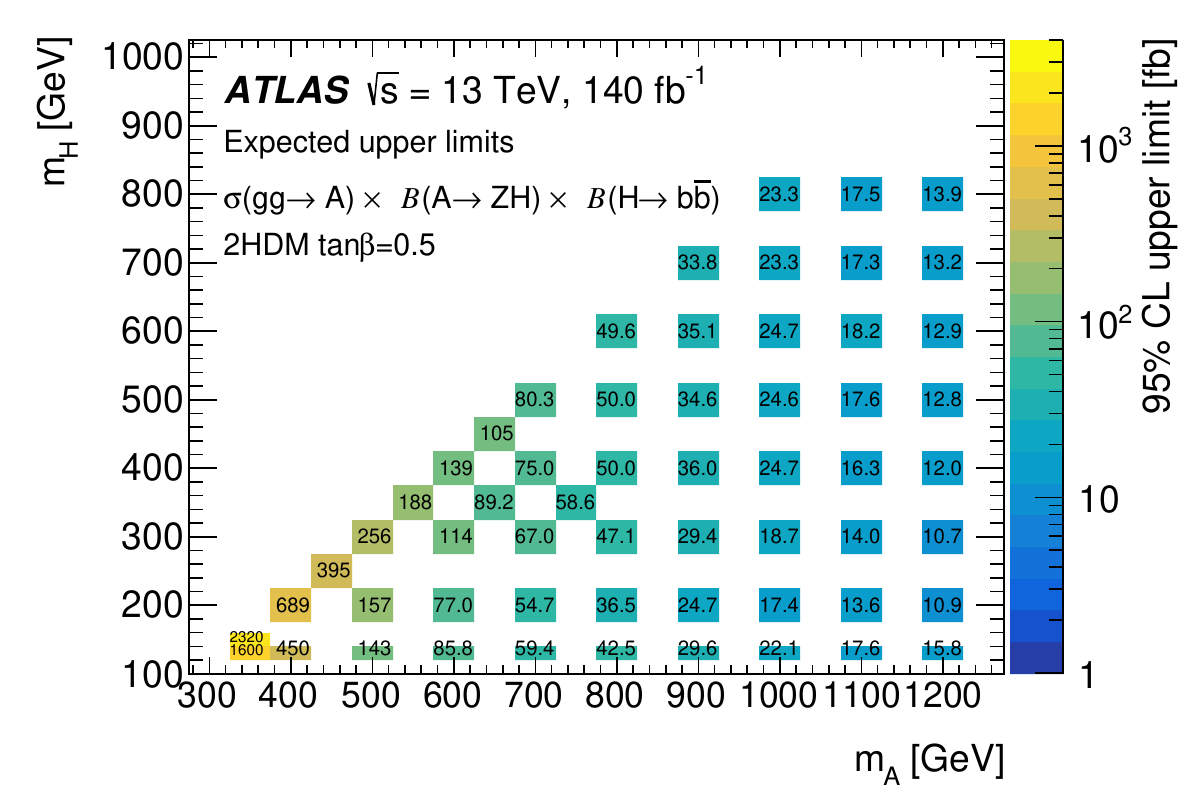}}
\subfigure[]{\includegraphics[width=0.48\textwidth]{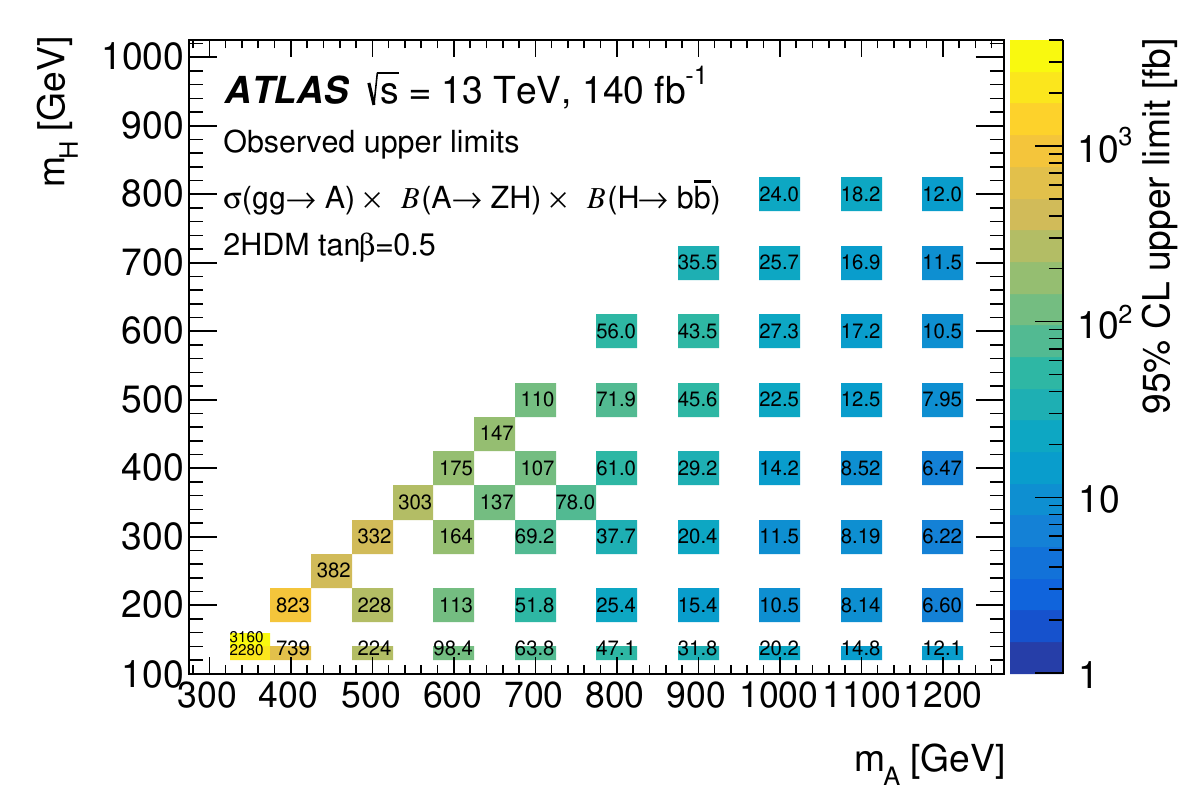}}
\subfigure[]{\includegraphics[width=0.48\textwidth]{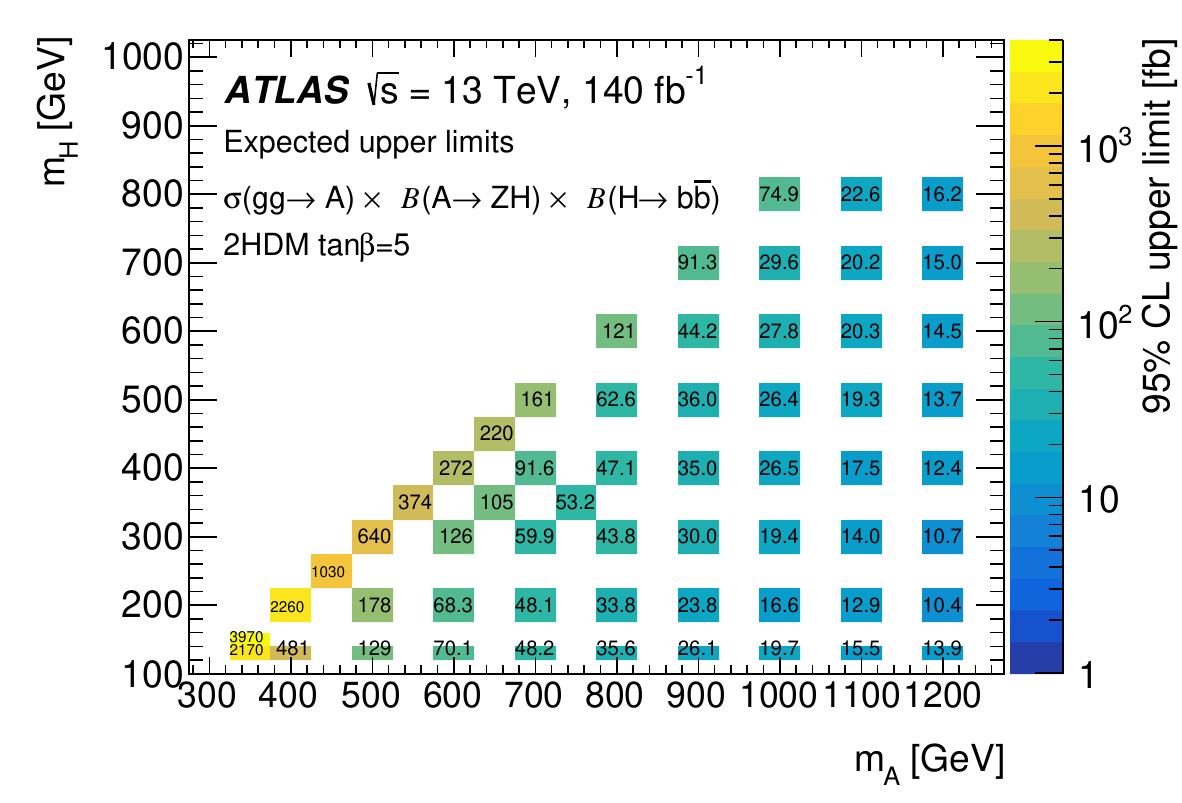}}
\subfigure[]{\includegraphics[width=0.48\textwidth]{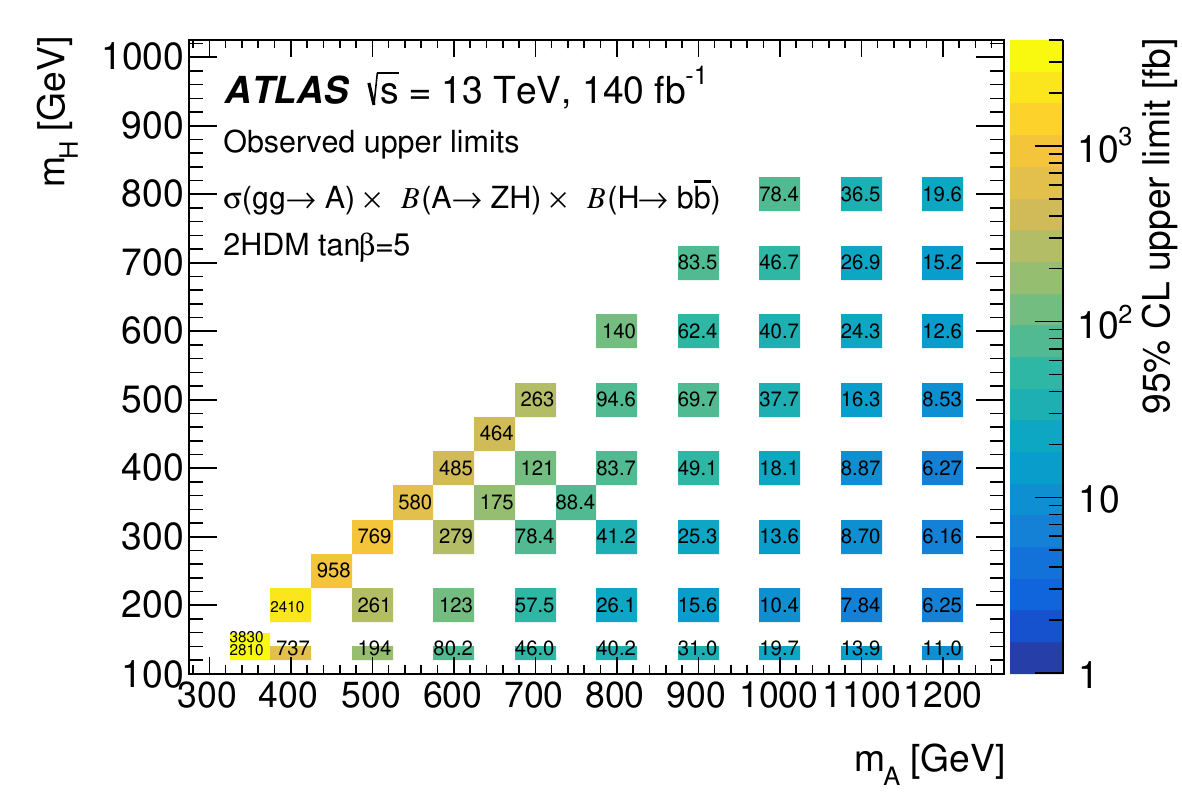}}
 
\caption{Expected and observed upper limits at 95\% CL on $\sigma(gg\to A)\times B(A\to ZH)\times B(H\to b\bbar)$. The limits are shown for $\tan\beta=0.5$ (a,b) and $\tan\beta=5$ (c,d). The \(\tan\beta\) value is relevant only for the choice of $A$ boson width.}
\label{fig:xseclimits:vvbb:ggF}
\end{figure}
 
\begin{figure}[ht]
\centering
\subfigure[]{\includegraphics[width=0.48\textwidth]{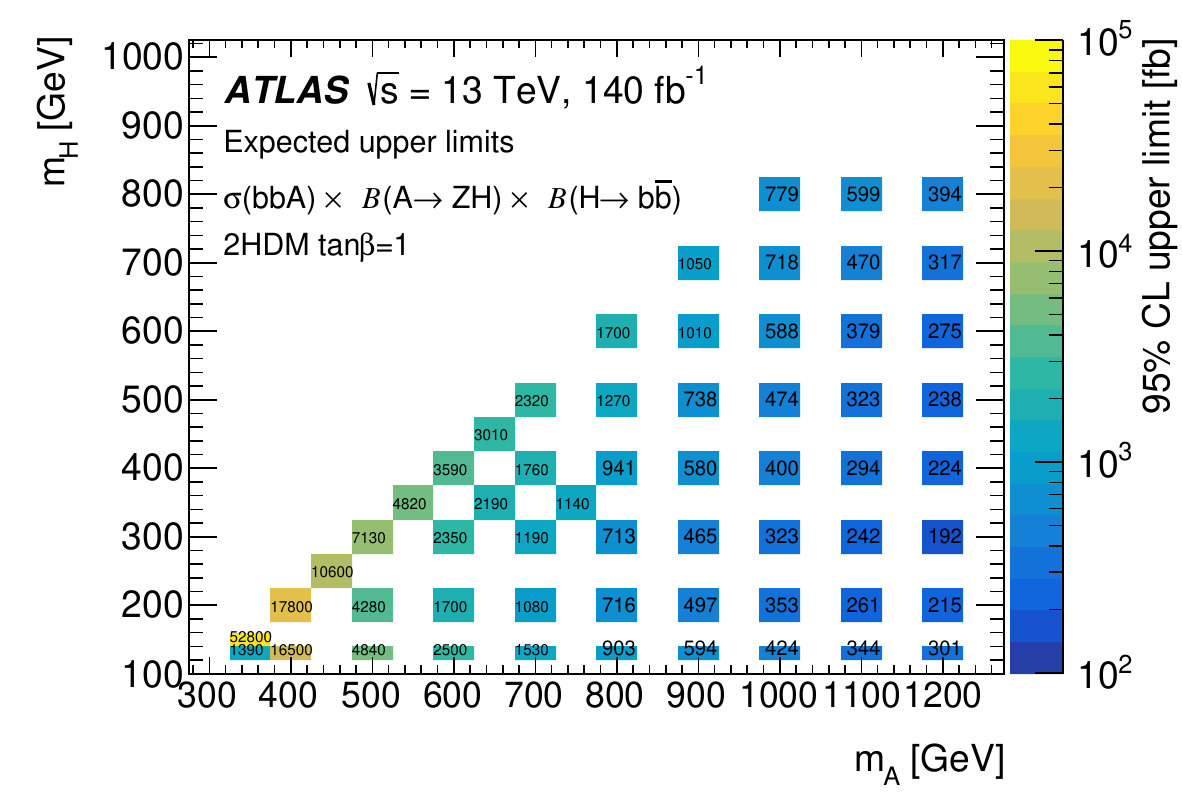}}
\subfigure[]{\includegraphics[width=0.48\textwidth]{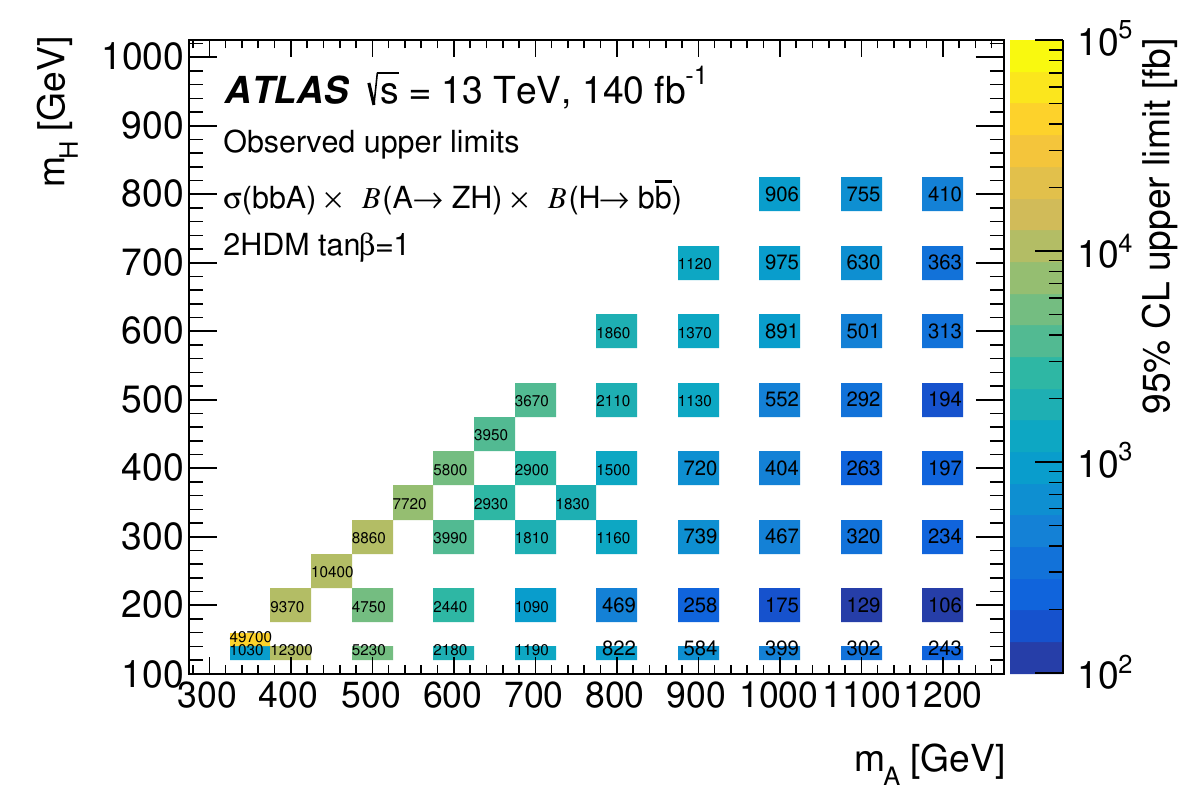}}
\subfigure[]{\includegraphics[width=0.48\textwidth]{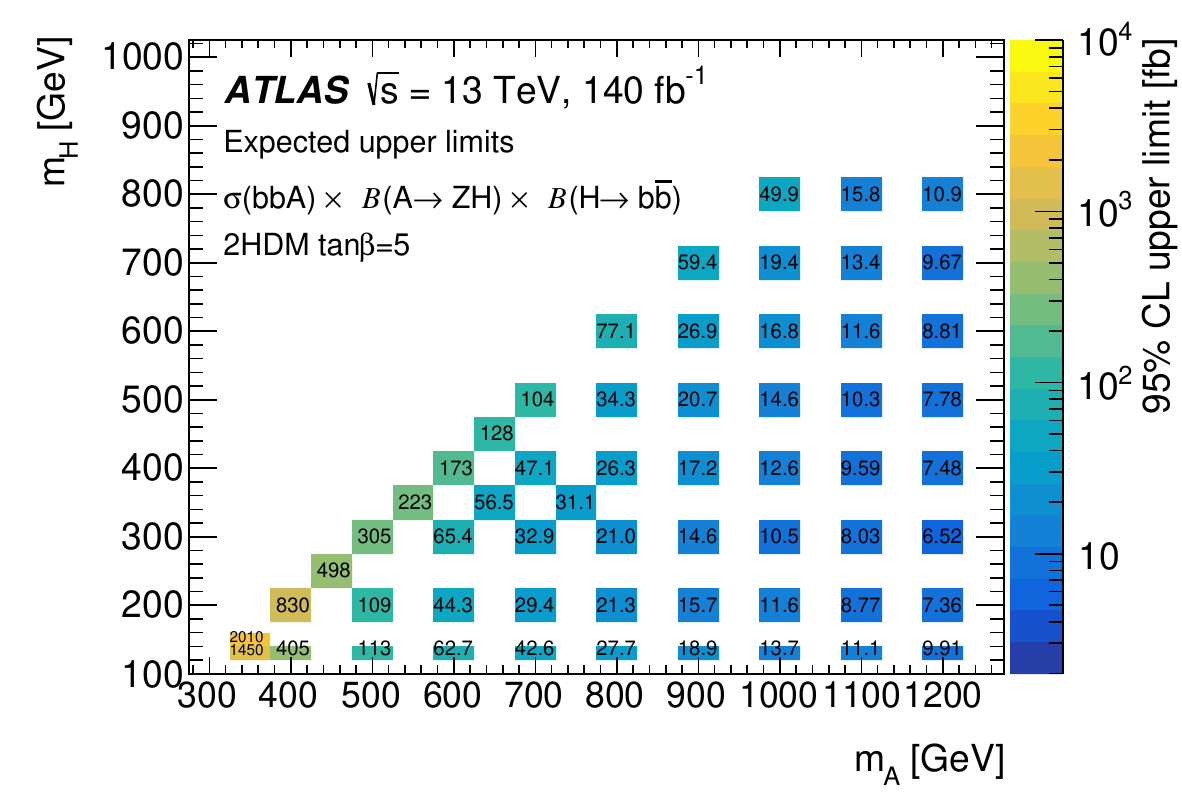}}
\subfigure[]{\includegraphics[width=0.48\textwidth]{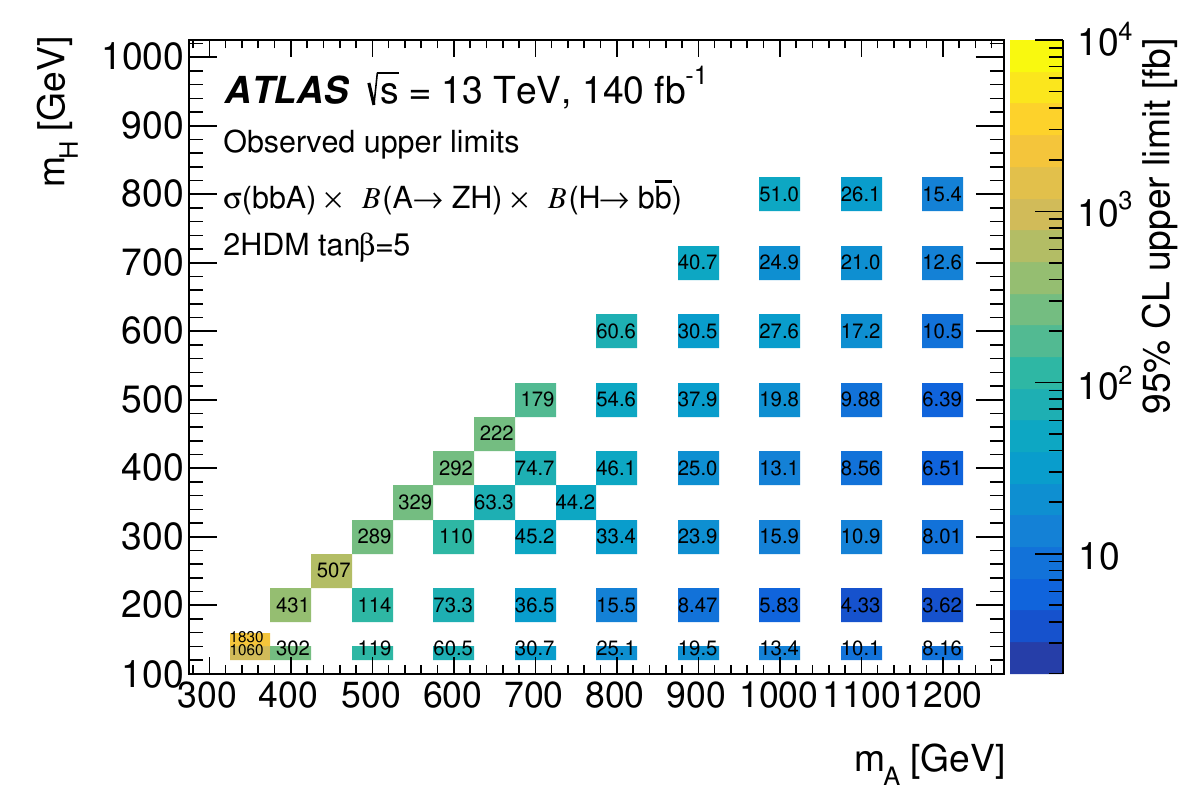}}
\subfigure[]{\includegraphics[width=0.48\textwidth]{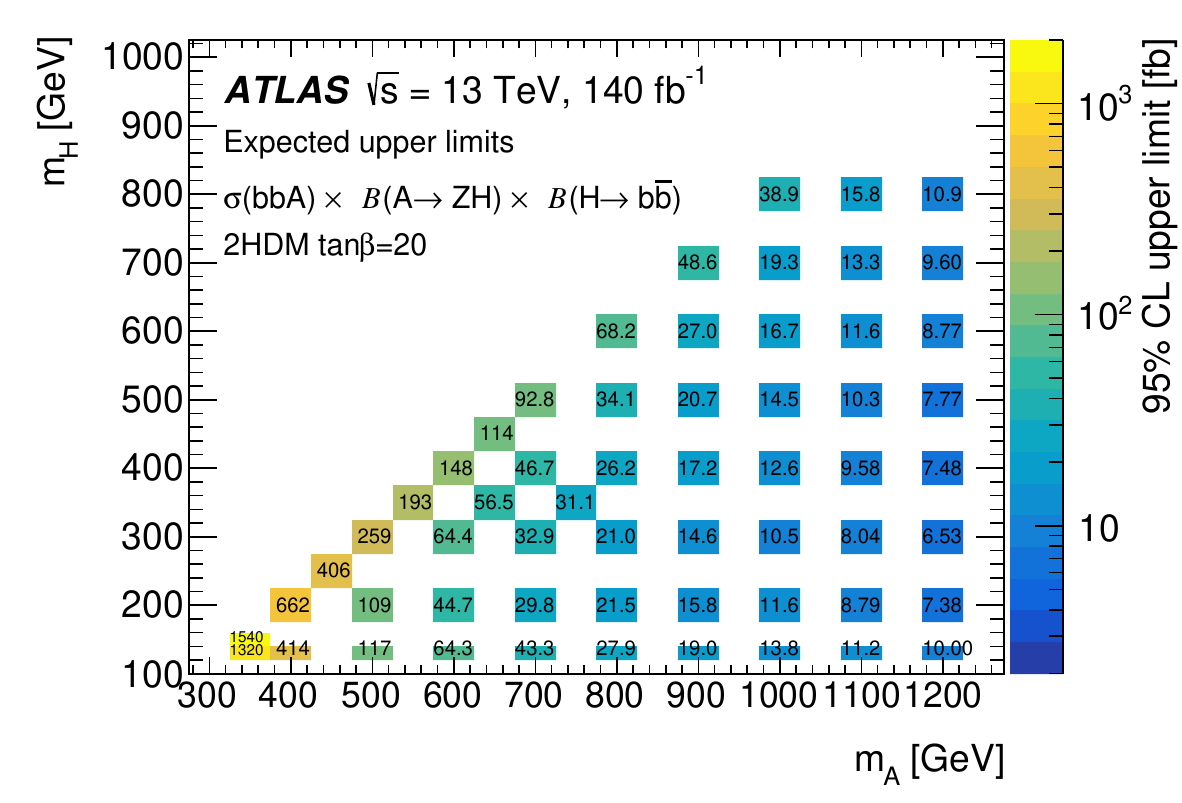}}
\subfigure[]{\includegraphics[width=0.48\textwidth]{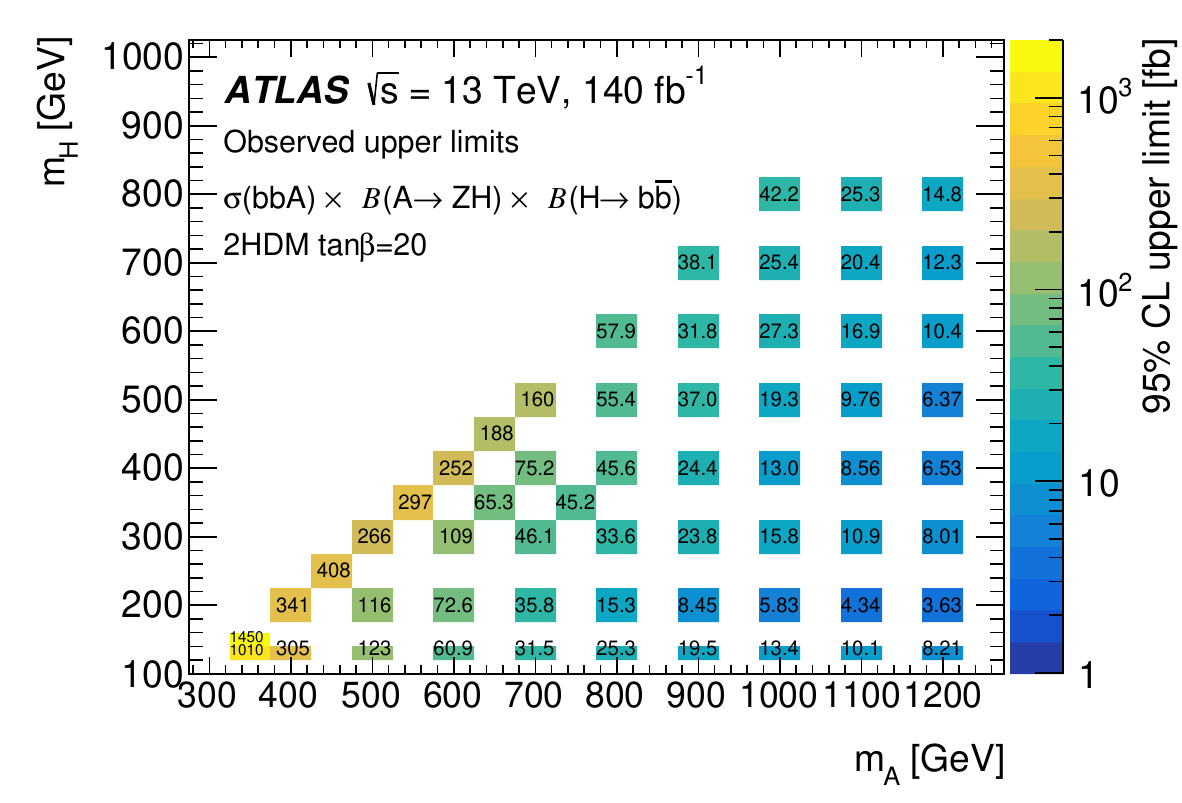}}
\caption{Expected and observed upper limits at 95\% CL on $\sigma(b\bbar A)\times B(A\to ZH)\times B(H\to b\bbar)$. The limits are shown for $\tan\beta=1$ (a,b), $\tan\beta=5$ (c,d) and $\tan\beta=20$ (e,f). The \(\tan\beta\) value is relevant only for the choice of $A$ boson width.}
\label{fig:xseclimits:vvbb:bbA:tanbHi}
\end{figure}


\FloatBarrier
 
\printbibliography
 
\clearpage
 
\begin{flushleft}
\hypersetup{urlcolor=black}
{\Large The ATLAS Collaboration}

\bigskip

\AtlasOrcid[0000-0002-6665-4934]{G.~Aad}$^\textrm{\scriptsize 102}$,
\AtlasOrcid[0000-0002-5888-2734]{B.~Abbott}$^\textrm{\scriptsize 120}$,
\AtlasOrcid[0000-0002-1002-1652]{K.~Abeling}$^\textrm{\scriptsize 55}$,
\AtlasOrcid[0000-0001-5763-2760]{N.J.~Abicht}$^\textrm{\scriptsize 49}$,
\AtlasOrcid[0000-0002-8496-9294]{S.H.~Abidi}$^\textrm{\scriptsize 29}$,
\AtlasOrcid[0000-0002-9987-2292]{A.~Aboulhorma}$^\textrm{\scriptsize 35e}$,
\AtlasOrcid[0000-0001-5329-6640]{H.~Abramowicz}$^\textrm{\scriptsize 151}$,
\AtlasOrcid[0000-0002-1599-2896]{H.~Abreu}$^\textrm{\scriptsize 150}$,
\AtlasOrcid[0000-0003-0403-3697]{Y.~Abulaiti}$^\textrm{\scriptsize 117}$,
\AtlasOrcid[0000-0002-8588-9157]{B.S.~Acharya}$^\textrm{\scriptsize 69a,69b,m}$
% ...
\bigskip
\\

% ...
$^{29}$Physics Department, Brookhaven National Laboratory, Upton NY; United States of America.\\
% ...
$^{35}$$^{(a)}$Facult\'e des Sciences Ain Chock, R\'eseau Universitaire de Physique des Hautes Energies - Universit\'e Hassan II, Casablanca;$^{(b)}$Facult\'{e} des Sciences, Universit\'{e} Ibn-Tofail, K\'{e}nitra;$^{(c)}$Facult\'e des Sciences Semlalia, Universit\'e Cadi Ayyad, LPHEA-Marrakech;$^{(d)}$LPMR, Facult\'e des Sciences, Universit\'e Mohamed Premier, Oujda;$^{(e)}$Facult\'e des sciences, Universit\'e Mohammed V, Rabat;$^{(f)}$Institute of Applied Physics, Mohammed VI Polytechnic University, Ben Guerir; Morocco.\\
% ...
$^{49}$Fakult\"{a}t Physik , Technische Universit{\"a}t Dortmund, Dortmund; Germany.\\
% ...
$^{55}$II. Physikalisches Institut, Georg-August-Universit\"{a}t G\"ottingen, G\"ottingen; Germany.\\
% ...
$^{69}$$^{(a)}$INFN Gruppo Collegato di Udine, Sezione di Trieste, Udine;$^{(b)}$ICTP, Trieste;$^{(c)}$Dipartimento Politecnico di Ingegneria e Architettura, Universit\`a di Udine, Udine; Italy.\\
% ...
$^{102}$CPPM, Aix-Marseille Universit\'e, CNRS/IN2P3, Marseille; France.\\
% ...
$^{117}$Department of Physics, New York University, New York NY; United States of America.\\
% ...
$^{120}$Homer L. Dodge Department of Physics and Astronomy, University of Oklahoma, Norman OK; United States of America.\\
% ...
$^{150}$Department of Physics, Technion, Israel Institute of Technology, Haifa; Israel.\\
$^{151}$Raymond and Beverly Sackler School of Physics and Astronomy, Tel Aviv University, Tel Aviv; Israel.\\
% ...
$^{a}$ Also Affiliated with an institute covered by a cooperation agreement with CERN.\\
$^{b}$ Also at An-Najah National University, Nablus; Palestine.\\
$^{c}$ Also at Borough of Manhattan Community College, City University of New York, New York NY; United States of America.\\
$^{d}$ Also at Center for High Energy Physics, Peking University; China.\\
$^{e}$ Also at Center for Interdisciplinary Research and Innovation (CIRI-AUTH), Thessaloniki; Greece.\\
$^{f}$ Also at Centro Studi e Ricerche Enrico Fermi; Italy.\\
% ...
$^{m}$ Also at Department of Physics, King's College London, London; United Kingdom.\\
% ...

\end{flushleft}


\end{document}